\documentclass[11pt]{article}

\usepackage{amsmath}
\usepackage{amssymb}
\usepackage{amsfonts}
\usepackage{latexsym}
\usepackage{color}
\usepackage{cite}
\usepackage{makeidx}

\normalbaselines
\catcode `\@=11
\@addtoreset{equation}{section}

\catcode`\@=12

\newtheorem{theo}{Theorem}[section]
\newtheorem{lem}[theo]{Lemma}
\newtheorem{propaat}[theo]{Proposition}
\newtheorem{cor}[theo]{Corollary}

\newtheorem{definitionat}[theo]{Definition}

\newtheorem{Example}[theo]{Example}
 
 \numberwithin{equation}{section}

\newtheorem{remarkat}[theo]{Remark}

\newcommand{\betheo}{\begin{theo}$\!\!\!${\bf } }
\newcommand{\entheo}{\end{theo}}

\newcommand{\becor}{\begin{cor}$\!\!\!$  }
\newcommand{\encor}{\end{cor}}

\newcommand{\belem}{\begin{lem}$\!\!\!$  }
\newcommand{\enlem}{\end{lem}}

\newcommand{\beprop}{\begin{propaat}} 
\newcommand{\enprop}{\end{propaat}}

\newcommand{\bedefi}{\begin{definitionat}$\!\!$ \rm }
\newcommand{\findefi}{ \end{definitionat}}

\newcommand{\beex}{\begin{Example}$\!\!$ \rm }
\newcommand{\enex}{ \end{Example}}

\newcommand{\berem}{\begin{remarkat}$\!$ \rm }
\newcommand{\enrem}{ \end{remarkat}}

\newcommand{\be}{\begin{equation}}
\newcommand{\en}{\end{equation}}

\newcommand{\bea}{\begin{eqnarray}}
\newcommand{\ena}{\end{eqnarray}}

\newcommand{\beano}{\begin{eqnarray*}}
\newcommand{\enano}{\end{eqnarray*}}

\newcommand{\bee}{\begin{enumerate}}
\newcommand{\ene}{\end{enumerate}}

\newcommand{\bei}{\begin{itemize}}
\newcommand{\eni}{\end{itemize}}

\newcommand{\betab}{\begin{tabular}}
\newcommand{\entab}{\end{tabular}}

\newcommand{\uph}{\raisebox{0.7mm}{$\upharpoonright $}}%

\newcommand{\nn}{\nonumber}
\newcommand{\subn}[1]{_{\scriptscriptstyle #1}}
\newcommand{\ha}{^{\rm\textstyle *}}

\newcommand{\ov}[1]{\overline{#1}}

\newcommand{\mc}{\mathcal}
\newcommand{\mb}{\mathbb}
\newcommand{\RN}{\mb R}
\newcommand{\CN}{\mb C}
\def\NN{{\mathbb N}}
\def\ZN{{\mathbb Z}}

\def\t{{\mathfrak t}}
\def\D{{\mathcal D}}
\def\F{{\mathcal F}}
\def\G{{\mathcal G}}
\def\H{{\mathcal H}}
\def\I{{\mathcal I}}
\def\J{{\mathcal J}}
\def\K{{\mathcal K}}
\def\M{{\mathcal M}}
\def\O{{\mathcal O}}
\def\P{{\mathcal P}}
\def\R{{\mathcal R}}
\def\SS{{\mathcal S}}
\def\T{{\mathcal T}}

\newcommand{\balpha}{\boldsymbol{\alpha}}

\newcommand{\bbeta}{\boldsymbol{\beta}}    
\newcommand{\balphabar}{\boldsymbol{\overline\alpha}}   

\newcommand{\norm}[2]{\left\| #2 \right\|_{#1}}
\newcommand{\dis}{\displaystyle}

\newcommand{\noi}{\noindent}
\newcommand{\ud}{\,\mathrm{d}}
\renewcommand{\leq}{\leqslant}
\renewcommand{\geq}{\geqslant}

\newcommand{\BH}{{\mc B}(\H)}

\def\hs{Hilbert space}

\def\dag{\dagger}

\newcommand{\vp}{\varphi}
\newcommand{\ip}[2]{\langle {#1} |{#2} \rangle}

\def\OL{\relax\ifmmode {\sf L}\else{\textsf L}\fi}
\def\OR{\relax\ifmmode {\sf R}\else{\textsf R}\fi}

\newcommand{\ta}{^\times}
\newcommand{\taa}{^{\times\times}}
\newcommand{\bdim}{ {\bf Proof. }}
 \newcommand{\edim}{\qed}
\newcommand{\pip}{{\sc pip}-space}
\newcommand{\ipip}{indexed {\sc pip}-space}
\newcommand{\spip}{{\sc pip}-subspace}
\newcommand{\dashvv}{\dashv \!\! \dashv}
 \newcommand{\co}{^{\#}}

\newcommand{\com}{{\scriptstyle\#}}
\newcommand{\HG}{\H(G)}
\newcommand{\qed}{\hfill $\square$}

\begin{document}

%
%

\begin{flushleft}
{\Large \sc Metric Operators, Generalized Hermiticity  \\[2mm] and Lattices of Hilbert Spaces}
\vspace*{7mm}

{\large\sf   Jean-Pierre Antoine $\!^{\rm a}$  and
Camillo Trapani $\!^{\rm b}$
}
\\[3mm]
$^{\rm a}$ \emph{\small Institut de Recherche en Math\'ematique et  Physique\\
\hspace*{3mm}Universit\'e catholique de Louvain \\
\hspace*{3mm}B-1348   Louvain-la-Neuve, Belgium\\
\hspace*{3mm}E-mail address: jean-pierre.antoine@uclouvain.be}
\\[2mm]
$^{\rm b}$ \emph{\small Dipartimento di Matematica e Informatica, Universit\`a di Palermo\\
\hspace*{3mm}I-90123, Palermo, Italy \\
\hspace*{3mm}E-mail address: camillo.trapani@unipa.it
}
\end{flushleft}

\thispagestyle{empty}

\begin{abstract}
A quasi-Hermitian operator  is an operator that is similar to its adjoint in some sense, via a
metric operator, i.e., a strictly positive self-adjoint operator. Whereas those metric operators are in general assumed to be bounded, we analyze the structure generated by unbounded metric operators in a Hilbert space.  It turns out that such operators generate a canonical lattice of Hilbert spaces, that is, the simplest case of a partial inner product space (PIP-space). 
We introduce several generalizations of the notion of similarity between operators, in particular, the notion of quasi-similarity, and we explore to what extend they preserve spectral properties. 
Then we apply some of the previous results to operators on a particular PIP-space, namely, a scale of  Hilbert spaces generated by a metric operator. Finally,  motivated by the recent developments of pseudo-Hermitian quantum mechanics, we  reformulate the notion of  pseudo-Hermitian operators in the preceding formalism.

\end{abstract}

\section{Introduction}
\label{sect_intro}

Pseudo-Hermitian quantum mechanics (QM) is a recent, unconventional, approach to QM, based on the use of non-self-adjoint Hamiltonians, whose self-adjointness can be restored by changing the ambient \hs, via a so-called metric operator.\footnote{Self-adjoint operators are usually called \emph{Hermitian} by physicists.}
  Although not self-adjoint,  such Hamiltonians have a real spectrum, usually discrete. Instead they are in general $\P\T$-symmetric, that is, invariant under the joint action of space reflection ($\P$) and complex conjugation ($\T$). Typical examples are  the $\P\T$-symmetric, 
  \index{$\P\T$-symmetry}
   but non-self-adjoint, Hamiltonians   $H = p^2 +ix^3$ and  $H = p^2 -x^4$. Surprisingly, both of them have a purely discrete spectrum, real and positive. A full analysis of $\P\T$-symmetric Hamiltonians may be found in the review paper of Bender \cite{bender}. The motivation comes  from a number of physical problems, mostly from condensed matter physics, but also from scattering theory (complex scaling), relativistic QM and quantum cosmology, or electromagnetic wave propagation in dielectric media.  

One may note also that the whole topic of Pseudo-Hermitian QM is covered in a series of annual international workshops, called ``Pseudo-Hermitian Hamiltonians in Quantum Physics", starting in 2003, the 12th edition having taken place in Istanbul in August 2013. In addition, several special issues of the Journal of Physics A have been devoted to it, the last one in November 2012 \cite{bender-specissue}. This special issue presents a panorama of theoretical/mathematical problems, and also a long list of physical applications such that  ``studies of classical shock-waves and tunnelling, supersymmetric models, spin chain models, models with ring structure, random matrix models, the Pauli equation, the nonlinear Schr\"{o}dinger equation, quasi-exactly solvable models, integrable models such as the Calogero model, Bose-Einstein condensates, thermodynamics, nonlinear oligomers, quantum catastrophes, the 
Landau-Zener problem and pseudo-Fermions".     Yet another, more recent, special issue   was published in the
Philosophical Transactions of the Royal Society  \cite{bender-specissue2}, with the aim of giving an up-to-date survey of the various attempts to use the techniques of $\P\T$ quantum mechanics for solving some of the outstanding problems in physics as well as for a number of concrete applications.

The $\P\T$-symmetric Hamiltonians are usually pseudo-Hermitian operators, a term introduced a long time ago by
Dieudonn\'e \cite{dieudonne} (under the name `quasi-Hermitian') for characterizing those bounded operators $A$ which satisfy a relation of the form
\be\label{eq-quasiH1}
GA=A \ha  G,
\en
  \index{operator!pseudo-Hermitian}
  where $G$ is a \emph{metric operator}, i.e., a  strictly positive self-adjoint operator. This operator $G$ then defines a new metric (hence the name) and a new Hilbert space (sometimes called physical) in which $A$ is symmetric and may possess a self-adjoint extension. For a systematic analysis of pseudo-Hermitian QM, we may refer to the review of Mostafazadeh \cite{mosta1} and, of course, the special issue \cite{bender-specissue}. We will come back to the terminology in Section \ref{sect_termin}.

Now, in most of the literature, the metric operators are assumed to be bounded   with bounded inverse. 
However, the example of  the Hamiltonian of the imaginary cubic oscillator, $H=p^2+ix^3$,   shows that bounded metric operators with unbounded inverse   \index{cubic oscillator}
do necessarily occur \cite {siegl}.
   Moreover, unbounded metric operators have also been  introduced in several recent works 
 \cite{bag,bag-fring,bag-zno,mosta2},
 and  an effort was made to put the whole machinery on a sound mathematical basis.
In particular, the Dieudonn\'e  relation implies that the operator $A$ is similar to its adjoint $A \ha $, in some sense, so that the notion of similarity plays a central role in the theory. One aim of the present chapter is to explore further the structure of unbounded metric operators, in particular, their incidence on similarity. Many of the results presented here are borrowed from our papers \cite{pip-metric,quasi-herm}.

To start with,  we examine, in Section \ref{sect_quasisim},   the notion of similarity between operators induced by a bounded metric operator with bounded inverse. Although this  notion is standard, it is too restrictive for applications, thus we are led to introduce several generalizations. The most useful one, called \emph{quasi-similarity}, applies when the metric operator is bounded and invertible, but has an unbounded inverse. 
The goal here is to study which spectral properties are preserved under such a quasi-similarity relation. This applies, for instance, to  non-self-adjoint operators with a discrete real spectrum.   
Later on, in Section \ref{sect_pipop}, we introduce an even weaker notion, called  \emph{semi-similarity}.

Next we notice, in Section \ref{sect-lattice}, that  an unbounded metric operator $G$ generates a lattice of seven \hs s, with lattice operations $\H_1 \wedge \H_2 := \H_1 \cap \H_2 \, ,   \H_1 \vee \H_2 := \H_1 + \H_2$  (see Fig. \ref{fig:diagram}). In addition, we consider  the infinite Hilbert scale generated by the powers of $G$.
This structure is then  extended in Section \ref{sect_LHS} to families of
metric operators, bounded or not. Such a family, if it contains unbounded operators, defines  a rigged \hs, and the latter in turn generates
 a canonical lattice of \hs s. This is a particular case of a  partial inner product space  (\pip),
a concept described at length in our monograph \cite{pip-book}.

Section \ref{sect_quasiH} is the heart of the chapter. It is devoted to the notion of \emph{quasi-Hermitian operators}, defined here in a slightly more general form than the original one of  Dieudonn\'e \cite{dieudonne}, in that both the operator $A$ and the metric operator $G$ in 
\eqref{eq-quasiH1} are allowed to be unbounded.  If  $G$  is bounded  with unbounded inverse,  it defines a new Hilbert space $\H(G)$ and the whole game consists in determining how operator properties are transferred from the original \hs\ $\H$ to $\H(G)$. This is exactly the situation encountered for Hamiltonians in  Hermitian QM, as explained above. Of particular interest is the case where $G$ is self-adjoint in the (physical) \hs\ $\H(G)$. We derive a number of conditions to that effect. As an aside we present a class of concrete examples, namely, operators defined from Riesz bases.

In  Section  \ref{sect_pipop}, we apply some of the previous results to operators on the scale of \hs s generated by the metric operator $G$. The outcome is that the \pip\ structure indeed  improves some of them. And finally, in   Section  \ref{sect_psH}, we present a construction, inspired from \cite{mosta2}, but significantly more general. Indeed, instead of requiring that the original pseudo-Hermitian Hamiltonian $H$ have a countable family of eigenvectors, we only need to assume that $H$ has a (large) set of analytic vectors.

Finally we summarize in the Appendix the essential facts about \pip s.

To conclude, let us fix our notations. The framework in a separable \hs\ $\H$, with inner product
$\ip{\cdot}\cdot$, linear in the first entry. Then, for any operator $A$ in   $\H$, we denote its domain by $D(A)$, its range by $R(A)$ and, if $A$ is positive,  its form domain by $Q(A):= D(A^{1/2})$.

\section{Some terminology}
\label{sect_termin}

We start with the central object of the chapter, namely, \emph{metric operators}.
  \index{operator!metric}
\bedefi By a metric operator in a \hs\ $\H$, we mean a  strictly positive self-adjoint operator $G$, that is, $G>0$ or $\ip{G\xi}{\xi}\geq 0$ for every $\xi \in D(G)$ 
and $\ip{G\xi}{\xi}= 0$ if and only if $\xi=0$.
\findefi
Of course,  $G$ is densely defined and invertible, but need not be bounded; its inverse $G^{-1}$ is also a metric operator,  bounded or not (in this case, in fact,  
0 belongs to the continuous spectrum of $G$).
For future use, we note that, given metric operators. $G, G_1, G_2$, one has  
\bei
\vspace*{-2mm}\item[(1)] If $G_1$ and $G_2$ are both bounded, then $G_1+G_2$ is a bounded metric operator;
\vspace*{-2mm}\item[(2)] $\lambda G $ is a metric operator if $\lambda>0$;
\vspace*{-2mm}\item[(3)]  $G^{1/2}$  and, more generally, $G^{\alpha} (\alpha \in \RN)$,  are metric operators.
\eni

 {As we noticed in the introduction,  in most of the literature on Pseudo-Hermitian quantum mechanics, the metric operators are assumed to be bounded   with bounded inverse, although there are exceptions. Following our previous work  \cite{pip-metric,quasi-herm}, we will envisage in this chapter all cases: $G$ and $G^{-1}$ both bounded, $G$ bounded with $G^{-1}$ unbounded  (such an operator $G$ is sometimes called \emph{quasi-invertible} \cite{hoover}), $G$ and $G^{-1}$ both unbounded.}

{Before proceeding, it is necessary to clarify the relationship between our metric operators and similar concepts commonly found in  Pseudo-Hermitian QM  \cite{quasi-herm}. }
To start with, Dieudonn\'e \cite{dieudonne} calls \emph{quasi-Hermitian} 
  \index{operator!quasi-Hermitian}
  a bounded operator $A$ on a \hs\ $\H$ for which there exists a  {bounded,}  positive,  self-adjoint operator $T\neq 0$ such that
\be\label{eq-quasiH}
TA=A \ha  T. 
\en
Thus $T$ is invertible, but its inverse $T^{-1}$ need not be bounded.  Notice that Dieudonn\'e calls `essentially trivial' the case where both $T$ and $T^{-1}$ are  bounded. In the sequel of this chapter, we will drop the boundedness assumption for both $T$ and $T^{-1}$.

Quasi-Hermitian operators contain, in particular,  spectral operators of scalar type and real spectrum, in the sense of Dunford \cite{dunford} (i.e., operators similar to a self-adjoint operator) and, \emph{a fortiori},  self-adjoint operators. Note the terminology is not uniform in the literature. Kantorovitz \cite{kantor}, for instance,  defines quasi-Hermitian operators  exactly as these spectral operators.
Now, if the operator $T^{-1}$ is bounded, then \eqref{eq-quasiH} implies that $A$ is similar to a self-adjoint operator, thus it is a spectral operator of scalar type and real spectrum. This is the case treated by Scholtz \emph{et al.}  \cite{scholtz} and Geyer \emph{et al.} \cite{geyer}, who introduced the concept in the physics literature.

A slightly more general notion is that of \emph{pseudo-Hermitian} operators, 
  \index{operator!pseudo-Hermitian}
  namely operators $A$ satisfying \eqref{eq-quasiH}, with $T$ and $T^{-1}$   bounded, but not necessarily positive {(this will unavoidably lead to indefinite metrics, see below).} This is the definition adopted also by Kretschmer-Szymanowski \cite{kretsch}, Mostafazadeh \cite{mosta1} or Albeverio \emph{et al.} \cite{alb-g-kuzhel}.
Later on,  Kretschmer-Szymanowski, Mostafazadeh \cite{mosta2} and Bagarello-Znojil \cite{bag-zno} adapted the definition to the case of an \emph{unbounded} operator $T$, claiming this is required for certain physical applications. Note that the last named authors have also coined the term \emph{cryptohermitian} for bounded quasi-Hermitian operators \cite{bag-zno}.

Another issue to clarify is the relation between pseudo-Hermitian operators and $J$-self-adjoint operators in a Krein space. 
  \index{Krein space}
  Assume  that 
$\H$ is a \hs\ with a \emph{fundamental symmetry} $J$, that is, a self-adjoint involution, $J=J^{*}, J^2 = I$. Defining the projections $P_{\pm} = \frac12(I\pm J)$,
we obtain the fundamental decomposition of $\H$:
\be
\H = \H_+ \oplus  \H_-, \quad \H_{\pm} := P_{\pm}\H.
\en
Then, the space $\H$  endowed with the indefinite inner product 
\be\label{indef-ip}
[\xi,\eta]_J := (J\xi, \eta)
\en
is a \emph{Krein space}. According to  \cite{bognar}, a Krein space is defined as a decomposable, nondegenerate  inner product space $\K = \K_+ \oplus  \K_-$, with inner product  $[\cdot,\cdot]$, where the 
subspace $\K_+$, resp. $ \K_-$, consists of vectors of positive, resp. negative norm $[\cdot,\cdot]^{1/2}$ and both subspaces $\K_{\pm}$ are complete in the
so-called intrinsic norm $|[\cdot,\cdot]|^{1/2}$. In that case, the $J$-inner product
\be
(\xi, \eta)_J := [J\xi,\eta]
\en
is positive definite and $\K$ is a \hs\ for the $J$-inner product.
 { Notice that exactly the same operators are bounded for the norms $|[\cdot,\cdot]|^{1/2}$ and 
$\norm{J}{\cdot} = (\cdot, \cdot)_J^{1/2}$  \cite{bognar}.}

Then,  a linear  densely defined operator $A$ in the Krein space $(\H, [\cdot,\cdot]_J )$ is called \emph{$J$-self-adjoint} if it satisfies the relation $A \ha  J = JA$. 
Thus   $J$-self-adjoint operators  are pseudo-Hermitian and constitute the appropriate class to study rigorously, as claimed in \cite{alb-g-kuzhel}.

Another notion commonly used it that of a $\mathcal {C}$-symmetry  \cite{bender,kuzhel,alb-g-kuzhel}. One says that a  $J$-self-adjoint operator $A$ in a Krein space has a \emph{$\mathcal{C}$-symmetry} if there exists a linear bounded operator $\mathcal{C}$ such that (i) $\mathcal{C}^2 = I$; (ii) $J \mathcal{C}>0$; and (iii) $A\mathcal{C} = \mathcal{C}A$. Thus the inner product
$$
   (\xi, \eta)_{\mathcal{C}} : = [\mathcal{C}\xi, \eta]_J = (J\mathcal{C}\xi, \eta)
$$
is positive definite, i.e., $J\mathcal{C}$ is a metric operator. Thus, if a Hamiltonian $H$ has the  $\mathcal {C}$-symmetry,   $(\H, (\cdot,\cdot)_\mathcal{C} )$ is a \hs, in which the dynamics  generated by $H$ should be described. Actually, there are two cases. If the operator $\mathcal {C}$ is bounded,   then it is   unique, up to equivalence. But if it is not unique, it must be unbounded (the definition may be adapted) \cite{kuzhel,bender-kuzhel}. So here too, one has to consider unbounded operators, as we shall do in the present chapter.

Finally, we note that non-self-adjoint operators (in Banach or \hs s) and their spectral properties are the object of a systematic analysis
by Davies \cite{davies}.

\section{Similar and quasi-similar operators}
\label{sect_quasisim}

In   this section we collect some basic definitions and facts about similarity of linear operators in Hilbert spaces and discuss
several generalizations of this notion. Throughout most of the section, $G$ will denote a \emph{bounded} metric operator. From now on, 
we will always suppose the domains of the given operators to be dense in $\H$.

\subsection{Similarity}\label{sect_sim}

In order to state precisely what we mean by similarity, we first define intertwining operators \cite{pip-metric}.
  \index{operator!intertwining!bounded}
\bedefi \label{def:qu-sim}
Let $\H, \K$ be Hilbert spaces, $D(A)$ and $D(B)$ dense subspaces
of $\H$ and $\K$,  respectively,   $A:D(A) \to \H$, $B: D(B) \to \K$ two linear operators.
A bounded operator $T:\H \to \K$ is called a \emph{ bounded intertwining operator}  for $A$ and $B$ if 
\begin{itemize}
\vspace{-2mm}\item[({\sf io$_1$})] $T:D(A)\to D(B)$;
\vspace{-2mm}\item[({\sf io$_2$})]$BT\xi = TA\xi, \; \forall\, \xi \in D(A)$.
\end{itemize}
\findefi
{ \berem \label{rem_adjoint}If $T$ is   {a bounded intertwining operator} for $A$ and $B$, then $T \ha  :\K \to \H$ is   a bounded intertwining operator  for $B \ha $ and $A \ha $.

\enrem 

\bedefi
Let  $A, B$ be two linear operators in the Hilbert spaces $\H$ and $\K$, respectively. Then, we say that

(i)  $A$ and $B$ are \emph{similar}, and write $A\sim B$,
  \index{operator!similar}
if there exists {a bounded intertwining operator} $T$ for $A$ and $B$ with bounded inverse $T^{-1}:\K\to \H$, which is intertwining for $B$ and $A$ .

(ii) $A$ and $B$ are \emph{unitarily equivalent} if $A\sim B$ and $T:\H \to \K$ is unitary, in which case we write $A\stackrel{u}{\sim} B$.

\findefi
We notice that  $\sim$ and $\stackrel{u}{\sim}$  ($\approx$) are equivalence relations.   Also, in both cases, one has $TD(A) = D(B)$.
\smallskip

The following properties of similar operators are easy  (see \cite{pip-metric} for a proof).

\begin{propaat} \label{prop_closedness}
 {Let $A$ and $ B$ be linear operators in $\H$ and $\K$, respectively, and $T$  {a bounded intertwining operator} for  $A$ and $ B$.
Then the following statements hold.}
\begin{itemize}
\vspace{-2mm}\item[(i)] $A\sim B$ if, and only if, $B \ha  \sim A \ha $.
\vspace{-2mm}\item[(ii)] $A$ is closed if, and only if, $B$ is closed.
\vspace{-2mm}\item[(iii)] $A^{-1}$ exists if, and only if,  $B^{-1}$ exists. Moreover, $B^{-1} \sim A^{-1}$.
\end{itemize}
\end{propaat}

Similarity of $A$ and $B$ is symmetric, preserves both the closedness of the operators and their spectra. But, in general, it does not preserve self-adjointness, as will result from Corollary \ref{cor-realsp} and Proposition \ref{prop-williams} below.

 {As we will see in Proposition \ref{prop_spectrum_sim} below, similarity} preserves also the resolvent set $\rho(\cdot)$ of operators and the parts in which the spectrum is traditionally decomposed: the point spectrum $\sigma_p (\cdot)$, the continuous spectrum $\sigma_c(\cdot)$ and the residual spectrum $\sigma_r(\cdot)$. Since we are dealing with closed, non self-adjoint operators, it is worth recalling the definitions of these various sets \cite{dunford-schwartz,reedsim1,schm}.

Given a closed operator $A$ in $\H$, consider $A-\lambda I : D(A) \to \H$ and the resolvent $R_A(\lambda) := (A-\lambda I)^{-1}$.
   \index{spectrum} 
  Then one defines:
\bei
\item The resolvent set 
$ \rho(A):= \{\lambda \in \CN : A-\lambda I $ is one-to-one and $(A-\lambda I)^{-1}$ 
\\ \mbox{is bounded$\}.$}
\item The spectrum $\sigma(A) := \CN \setminus \rho(A)$.
\item The point spectrum $\sigma_p (A):= \{\lambda \in \CN : A-\lambda I \mbox{ is not one-to-one}\}$, that is, the set of  eigenvalues of $A$.
\item The continuous spectrum $\sigma_c(A):=  \{\lambda \in \CN : A-\lambda I $ is one-to-one and has dense range, 
different from  $\H \}$,  hence $(A-\lambda I)^{-1}$ is densely defined, but unbounded.
\item The residual spectrum $\sigma_r(A):= \{\lambda \in \CN : A-\lambda I  $ is one-to-one, but its range is not dense,
hence $(A-\lambda I)^{-1}$ is not densely defined.
\eni
With these definitions, the three sets $\sigma_p (A), \sigma_c(A), \sigma_r(A)$ are disjoint
and  
\be\label{eq:spec}
\sigma (A) = \sigma_p (A) \cup \sigma_c(A)\cup \sigma_r(A).
\en
We note also that  $\sigma_r(A)= \overline{\sigma_p(A \ha)} =\{ \overline{\lambda}: \lambda \in \sigma_p(A \ha )\}$. Indeed, for any 
$ \lambda \in \sigma_r(A)$, there exists $\eta\neq 0$ such that
$$
0=\ip{(A-\lambda I)\xi}{\eta} = \ip{\xi}{(A \ha -\ov\lambda I)\eta} , \; \forall\, \xi\in D(A),
$$
which implies $\ov\lambda \in \sigma_p (A \ha )$. Also $\sigma_r(A)=\emptyset$ if $A$ is self-adjoint.
\berem
 Here we follow Dunford-Schwartz \cite{dunford-schwartz}, but other authors give a different definition of the continuous spectrum, implying that it is no longer disjoint from the point, for instance, Reed-Simon \cite{reedsim1} or Schm\"{u}dgen \cite{schm}. This alternative definition allows for eigenvalues embedded in the continuous spectrum, a situation common in many physical situations, such as the Helium atom, and a typical source of   resonance effects in scattering theory (see \cite[Sec. XII.6]{reedsim4}).
\enrem

 We proceed now to show the stability of the different parts of the spectrum under the similarity relation $\sim$, as announced above
\cite[Props. 3.7 and 3.9]{pip-metric}.

\begin{propaat}\label{prop_spectrum_sim} Let $A$, $B$ be closed operators such that $A\sim B$ with  {the bounded intertwining operator} $T$. Then,
 \begin{itemize} 
\vspace{-2mm} \item[(i)] $\rho(A)= \rho(B)$.
\vspace{-2mm} \item[(ii)] $\sigma_p(A)=\sigma_p(B)$. Moreover if $\xi \in D(A)$ is an eigenvector of $A$ corresponding to the eigenvalue $\lambda$,
 then $T\xi$ is an eigenvector of $B$ corresponding to the same eigenvalue. Conversely, if $\eta \in D(B)$ is an eigenvector of $B$ corresponding to the eigenvalue $\lambda$, then $T^{-1}\eta$ is an  eigenvector of $A$ corresponding to the same eigenvalue. Moreover, the multiplicity of $\lambda$ as eigenvalue of $A$ is the same as its multiplicity as eigenvalue of $B$.
\vspace{-2mm}\item[(iii)] $\sigma_c(A)= \sigma_c(B).$
\vspace{-2mm}\item[(iv)]$\sigma_r(A)= \sigma_r(B)$.
\end{itemize}
\end{propaat}
\index{spectrum}\index{eigenvalue}
 {\bf Proof. }(i) Let $\lambda \in \rho(A)$, so that $(A-\lambda I)^{-1}$ exists and it is bounded. The operator $X_\lambda :=T(A-\lambda I)^{-1}T^{-1}$  is bounded. Since $(A-\lambda I)^{-1}T^{-1}\eta \in D(A)$, for every $\eta \in \K$, we have
\begin{align*}
(B-\lambda I)X_\lambda \eta &= (B-\lambda I) T(A-\lambda I)^{-1}T^{-1}\eta \\
&= T(A-\lambda I) (A-\lambda I)^{-1}T^{-1}\eta = \eta , \quad \forall\, \eta \in \K.
\end{align*}
On the other hand, since $(B-\lambda I)T\xi = T(A-\lambda I)\xi$, for all $\xi \in D(A)$, taking $\xi =T^{-1}\eta$, $\eta \in D(B)$, we obtain
$(B-\lambda I)\eta = T(A-\lambda I)T^{-1}\eta$ and then $T^{-1}(B-\lambda I)\eta= (A-\lambda I)T^{-1}\eta)$.
Then, for every $\eta \in D(B)$ we get
\begin{align*}
 X_\lambda (B-\lambda I)\eta &= T(A-\lambda I)^{-1}T^{-1}(B-\lambda I)\eta\\
 &= T(A-\lambda I)^{-1}(A-\lambda I)T^{-1}\eta = \eta, \quad \forall\, \eta \in D(B).
\end{align*}
Hence $X_\lambda = (B-\lambda I)^{-1}$ and $\lambda \in \rho(B)$.
The statement follows by replacing $A$ with $B$, and $T$ with $T^{-1}$.

  (ii) is easy.
 
 (iii) Let $\lambda \in \sigma_c(B)$.  If $\eta \in \H$, then $\eta= T^{-1}\eta '$ for some $\eta ' \in \K$.
 Since $R(B-\lambda I)$ is dense in $\K$, there exists a sequence $\{\eta'_k \}\subset R(B-\lambda I)$ such that $\eta'_k \to \eta'$. Put $\eta'_k= (B-\lambda I)\xi' _k$,
 with $\xi'_k \in D(B)$. Since $TD(A)=D(B)$, for every $k \in \mb N$ there exists $\xi_k \in D(A)$ such that $ \xi'_k =T\xi_k$. Hence
$$ 
\eta'= \lim_{k\to \infty} (B-\lambda I)T\xi_k= \lim_{k\to \infty}T(A-\lambda I)\xi_k.
$$
This implies that
$$ 
\eta = T^{-1}\eta'= \lim_{k\to \infty} (A-\lambda I)\xi_k.
$$
Thus $R(A-\lambda I)$ is dense in $\H$.  {Since    $(B-\lambda I)^{-1}$ is unbounded, so is also  $(A-\lambda I)^{-1}$.}
 Hence $\sigma_c(A)\subseteq \sigma_c(B).$ Interchanging the roles of $A$ and $B$, one gets the reverse inclusion.

(iv) follows from \eqref{eq:spec} and (i)-(iii).
\qed
\medskip

\noi In the  proof of (i) above, the  assumption  `$T^{-1}$ bounded' guarantees that $X_\lambda$ (which is in any case a left inverse) is bounded,
whereas the  assumption `$TD(A)=D(B)$' allows one to prove that $X_\lambda$ is also a right inverse.
They both seem to be unavoidable.

Taking into account that, if $A$ is self-adjoint, its residual spectrum is empty, we obtain
\becor\label{cor-realsp}
Let $A$, $B$ be closed operators with $A$ self-adjoint. Assume that $A\sim B$. Then $B$ has real spectrum and $\sigma_r(B)=\emptyset$.
\encor
In other words, $B$ is then a spectral operator of scalar type with real spectrum, as discussed in Section \ref{sect_termin}.
This corollary  can be used to show the existence of nonsymmetric operators having real spectrum and empty residual spectrum.

Actually it is contained in the following result of Williams \cite{williams}.
\beprop\label{prop-williams}
If the operator $A$ satisfies the conditions $A =T^{-1} A \ha  T$, where $T^{-1}$ is bounded and $0 \not\in \ov{W(T)}$, then it is similar to a self-adjoint operator, hence has real spectrum. 
\enprop
\index{spectrum!real}
In this proposition, $\ov{W(T)}$ denotes the closure of the  numerical range of $T$, that is,  $W(T) := \{ \ip{T\xi}{\xi} : \norm{}{\xi} = 1\}$ \cite[Sec. 9.3]{davies}.  
The set $\ov{W(T)}$   is convex and contains $\sigma(T)$. 
The argument runs as follows. By the condition $0 \not\in \ov{W(T)}$,  one can separate $\ov{W(T)}$  from 0 
  {in such a way that $\ov{W(T)}$ belongs to a half-plane and 0 does not.} 
 Possibly replacing $T$ by 
$e^{i\theta}T$, this half-plane can be taken as $\mathrm{Re}\,z \geq \epsilon$, for some  $\epsilon>0$. Defining $B=\frac12(T+T \ha )$, one sees that $W(B) = \mathrm{Re}W(T)$ lies on the real axis to the right of $\epsilon$, thus 
$B$ is positive and boundedly invertible. Since $A =T^{-1} A \ha  T$, it follows that $L:= B^{-1/2} AB^{1/2}$  is self-adjoint and $A\sim L$. 
Note that one can assume, in particular,  $T>0$ or $T<0$.

In other words, for a  quasi-Hermitian operator satisfying $TA = A \ha  T$  to be similar to a self-adjoint operator, one needs both 
$T>0$ (or $T<0$) and $T^{-1}$   bounded. Indeed, Williams gives an example of an operator $A$ (the bilateral shift in $\ell^2$) where $T^{-1}$ is bounded but not positive and $\sigma(A)$ is not real. On the other hand, Dieudonn\'e \cite{dieudonne} gives an example of a quasi-Hermitian operator $A$ with $T^{-1}$ unbounded and $\sigma(A)$   not real.

\subsection{Quasi-similitarity and spectra}

The notion of similarity discussed in the previous section is too strong in many situations, thus we seek a weaker one. A natural step is to drop the boundedness of $T^{-1}$. 
  \index{operator!quasi-similar}
  \bedefi\label{def:quasi-sim}
We say that $A$  is \emph{quasi-similar} to $B$, and write $A\dashv B$, if there exists  {a bounded intertwining operator} $T$ for $A$ and $B$ which is  invertible, with inverse $T^{-1}$   densely defined (but not necessarily bounded).
\findefi
Note that, even if $T^{-1}$ is bounded, $A$ and $B$ need not be similar, unless $T^{-1}$ is also  an intertwining operator. 
 {Indeed, $T^{-1}$ does not necessarily map $D(B)$ into $D(A)$, unless of course if $T D(A)=D(B)$.}

As already remarked in \cite{quasi-herm}, there is a considerable confusion in the literature concerning the notion of quasi-similarity.  

  (1)  First, essentially all authors consider only (quasi-)similarity between two \\ \emph{bounded} operators.
 Next, a bounded invertible operator $T$ with (possibly) unbounded inverse $T^{-1}$ is called a \emph{quasi-affinity} by Sz.-Nagy and  Foia\c{s} \cite[Chap.II, Sec.3]{nagy} and  a  \emph{quasi-invertible} operator by other authors \cite{hoover}. Then, if  $A,B$ are two  bounded operators such that $TA=BT$, that is, $A\dashv B$,   $A$ is called a \emph{quasi-affine transform} of $B$. In this context, $A$ and $B$  are called \emph{quasi-similar}  if $A\dashv B$ \emph{and} $B\dashv A$  (so that quasi-similarity becomes also an equivalence relation).  

   (2) Tzafriri \cite{tzafriri} considers only bounded spectral operators,  
 in Dunford's sense \cite{dunford, dunford-schwartz}. For these, he introduces a different notion of quasi-similarity (but under the same name) based on the resolution of the identity. 

(3) Hoover \cite{hoover}  shows that if two  bounded spectral operators $A$ and $B$ are quasi-similar (i.e. $A\dashv B$ and $B\dashv A$), then $B$ is quasi-similar
to $A$ in the sense of Tzafriri (which he calls weakly similar).  On the other hand, Feldzamen \cite{feldzamen} considers yet another notion of generalized similarity, called \emph{semi-similarity}, but then Hoover shows that two  semi-similar bounded spectral operators $A$ and $B$ are in fact quasi-similar. 

 (4)  Quasi-similarity of unbounded operators  is considered by \^{O}ta and  Schm\"udgen  \cite{ota-schm}.  Namely, given two unbounded operators
$A$ and $B$ in \hs s $\H , \K$, respectively,  $A$ is said to be \emph{quasi-similar} to  $B$ if there exist two (quasi-invertible) intertwining operators (in the sense of Definition \ref{def:qu-sim}) $T_{AB}: D(A)\to D(B)$ and $T_{BA}: D(B)\to D(A)$. In other words, this notion is the straightforward generalization of the quasi-similarity of bounded operators defined by the previous authors. 
 \medskip

 In the sequel, we will stick to the asymmetrical notion of quasi-similarity given in Definition
\ref{def:quasi-sim}, namely, $A\dashv B$, because it appears naturally in the presence of a bounded metric operator with unbounded inverse, as shown in Theorem \ref{prop_292}   below. A concrete example is the Hamiltonian of the imaginary cubic oscillator \cite{siegl}.\footnote {There is a misprint in that paper, on page 2, l.-2. The correct statement is $\Theta [\mathrm{Dom}(H)] \subset  \mathrm{Dom}(H^\dag)$. which indeed satisfies the relation $H\dashv H^{\dag}$.}

Actually, there is a whole class of similar concrete examples, namely, $H= p^2 + x^2 + ix^3$ or, more generally, 
Schr\"{o}dinger Hamiltonians of the form
$$
H= p^2 + \sum_{m=1}^{2n}c_m x^m,
$$
where the constant $c_m$ has positive real and imaginary parts. These operators were already mentioned by Davies\cite{davies2}. 
Two further examples will be given below, in Examples \ref{ex1-3-23} and  \ref{ex1-3-24}.
 Another explicit example, that we will describe in detail in Section \ref{subsec-example} below, is the operator of second derivative on the positive half-line analyzed by Samsonov \cite{samsonov}.

  \index{operator!mutually quasi-similar}
Accordingly,  we will say that
two closed operators $A$ and $B$ are \emph{mutually quasi-similar} if they are quasi-similar in the sense of
\^{O}ta and  Schm\"udgen, that is, if we have both  $A\dashv B$ and  $B\dashv A$, which we denote by
$A\dashv \vdash B$. Clearly $\dashv \vdash $ is an equivalence relation. Moreover, $A\dashv \vdash B$ implies
$A\ha \dashv \vdash B\ha$. 

\bedefi If $A\sim B$ (resp., $A \dashv B$) and the intertwining operator is a metric operator $G$, we say that $A$ and $B$ are \emph{metrically} similar (resp., quasi-similar).
\findefi

If $A\sim B$ and $T$ is the corresponding intertwining operator, then $T=UG$, where $U$ is unitary and $G:=(T \ha  T)^{1/2}$ is a metric operator.
If we put $B'= U^{-1}BU$, then $B'$ and $A$ are metrically similar. Thus, up to unitary equivalence, one can always consider metric similarity instead of similarity.
\medskip

\begin{propaat} \label{prop_adjoint} If $A\dashv B$, with  {the bounded intertwining operator} $T$, then $B \ha \dashv A \ha $ with  {the bounded intertwining operator} $T \ha $.
\end{propaat}
 {This follows from Remark \ref{rem_adjoint} and from the fact that, since $T^{-1}$ exists, then ${(T \ha )}^{-1}$ exists too and ${(T \ha )}^{-1}={(T^{-1})} \ha $.}

{In the case of nonclosed operators, however, we must still weaken the notion of quasi-similarity, replacing conditions ({\sf io$_1$}) and ({\sf io$_2$}) by the following weak-type condition.}
 \index{operator!weakly quasi-similar}
  \bedefi
The operator $A$ is called \emph{weakly quasi-similar} to $B$, in which case we write $A\dashv_w B$, if $B$ is closable,
$T$ is invertible  with densely defined inverse $T^{-1}$    and  the following condition holds
\begin{itemize}
\item[({\sf ws})] $\ip{T\xi}{B \ha \eta}= \ip{TA\xi}{\eta}, \; \forall\, \xi \in D(A),\, \eta \in D(B \ha ).$
\end{itemize}
\findefi

{Of course, if the operator $B$ is closed, we recover the original Definition \ref{def:quasi-sim}.}
 \begin{propaat}
  $A\dashv_w B$  if and only if $T:D(A) \to D(B^{**})$ and $B^{**}T\xi= TA\xi$, for every $\xi \in D(A)$. In particular, if $B$ is closed, $A\dashv B$ if 
and only if $A\dashv_w B$.
\end{propaat}

\begin{propaat}\label{prop_closable} If $B$ is closable and $A\dashv_w B$, then $A$ is closable.
\end{propaat}
 {\bf Proof. }
Assume that $\{\xi_n\}$ is a sequence in $D(A)$ and $\xi_n \to 0$, $A\xi_n \to \eta$. Then, $T\xi_n \to 0$ and $TA\xi_n \to T\eta$.
But $TA\xi_n = B^{**}T\xi_n\to T\eta$. The closedness of $B^{**}$ then implies that $T\eta =0$ and, therefore, $\eta=0$.
\qed

The converse of the previous statement does not hold, in general, as shown by the following counterexample \cite{pip-metric}.

\beex \label{ex_1} In the Hilbert space $L^2 ({\mb R})$, consider   {the operator $Q$ of multiplication by $x$,}  defined on the dense domain
$$
 D(Q)= \left\{ f \in L^2 ({\mb R}): \int_{\mb R} x^2 |f(x)|^2 \ud x < \infty \right\}.
$$
Given $\varphi \in L^2 ({\mb R})$, with $\norm{}{\varphi}=1$,
let $P_\varphi:= \varphi \otimes \overline{ \varphi}$ denote the projection operator\footnote{ In physicists' Dirac notation, 
$P_\varphi= |\varphi\rangle  \langle\varphi|$.} onto the one-dimensional subspace generated by $\varphi$ 
and $A_\varphi$ the operator with domain $D(A_\varphi)= D(Q^2)$ defined by
$$
A_\varphi f = \ip{(I+Q^2)f}{\varphi}(I+Q^2)^{-1}\varphi, \quad { f \in D(A_\varphi)}.
$$
Then, it is easily seen that $P_\varphi \dashv A_\varphi$ with the  {bounded} intertwining operator $T:= $ \mbox{$(I+Q^2)^{-1}$}. Clearly $P_\varphi$ is everywhere defined and bounded, but the operator $A_\varphi$ is closable if, and only if, $\varphi \in D(Q^2)$.
  {This is seen as follows. Being densely defined, $A_\varphi$ is closable if and only if it has a densely defined adjoint. If $\varphi \in D(Q^2)$ we have, for every $g \in L^2 ({\mb R})$,
\begin{align*}
\ip{A_\varphi f}{g} = \ip{(I+Q^2)f}{\varphi} \ip{(I+Q^2)^{-1}\varphi}{g} 
 = \ip{f}{(I+Q^2)\varphi}\ip{(I+Q^2)^{-1}\varphi}{g}.
\end{align*}
Hence, $A_\varphi^*g= \ip{g}{(I+Q^2)^{-1}\varphi}(I+Q^2)\varphi = \ip{(I+Q^2)^{-1}g}{\varphi}(I+Q^2)\varphi $. This proves also that, in this case, $A_\varphi$ is bounded, since $g$ can be arbitrarily chosen in $\H$. On the other hand, 
if $g\in D(A_\varphi^*)$, 
$$
\ip{A_\varphi f}{g}=\ip{(I+Q^2)f}{\varphi}\ip{(I+Q^2)^{-1}\varphi}{g}= \ip{f}{A_\varphi^*g}
$$
Then the last equality shows that $\varphi \in D(Q^2)$ and $$A_\varphi^* g= \ip{(I+Q^2)^{-1}g}{\varphi}(I+Q^2)\varphi.$$}
\enex
\vspace*{-5mm}

Now we consider the relationship between the spectra of quasi-similar operators,   following  mostly \cite{pip-metric}.

\begin{propaat}\label{prop_sigmap} Let $A$ and $B$ be closed operators and assume that $A\dashv B$,  with  {the bounded intertwining operator} $T$. Then the following statements hold.
\begin{itemize}
\item[(i)] $\sigma_p(A)\subseteq \sigma_p(B)$ and for every $\lambda \in \sigma_p(A)$ one has $m_A(\lambda) \leq m_B(\lambda)$, where $m_A(\lambda)$, resp. $ m_B(\lambda)$, denotes  the multiplicity of $\lambda$ as eigenvalue of the operator $A$, resp. $B$.

\item[(ii)]  $\sigma_r(B) \subseteq\sigma_r(A)$.

\item[(iii)] If $TD(A)=D(B)$, then $\sigma_p(B)= \sigma_p(A)$. 
\item[(iv)] If $T^{-1}$ is bounded and $TD(A)$ is a core for B,  then $\sigma_p(B)\subseteq \sigma(A)$.
\end{itemize}
\end{propaat}
\index{spectrum!point}

 {\bf Proof. }

The statements (i) and (iii) can be proved as in Proposition \ref{prop_spectrum_sim}. We prove only the  statements (ii) and (iv).

(ii) By Proposition \ref{prop_adjoint}, $B \ha \dashv A \ha $, with  {the bounded intertwining operator} $T \ha $. Then, by (i), $\sigma_p(B \ha )\subseteq \sigma_p (A \ha )$. The statement follows the relation 
$\sigma_r(C)= \ov{\sigma_p(C \ha )}$, shown above.  

(iv): Let $\lambda \in \sigma_p(B)$. Then there exists $\eta \in D(B)\setminus\{0\}$ such that $B\eta= \lambda \eta$.
We may suppose that $\|\eta\|=1$.
Since  $TD(A)$ is a core for $B$, there exists a sequence $\{\xi_n\}\subset D(A)$ such that $T\xi_n \to \eta$ and $BT\xi_n \to B\eta$. Then,
\begin{align*}
\lim_{n \to \infty} T(A\xi_n -\lambda \xi_n)&= \lim_{n \to \infty} TA\xi_n -\lambda \lim_{n \to \infty}T\xi_n=\lim_{n \to \infty}BT\xi_n -\lambda \eta\\
&= B\eta -\lambda \eta =0.
\end{align*}
 {Since $T^{-1}$ is bounded, we get}
\begin{equation}\label{eqn_non} \lim_{n \to \infty}(A\xi_n -\lambda \xi_n)=0,\end{equation}
Assume that $\lambda \in \rho(A)$. Then $(A-\lambda I)^{-1} \in \BH$. We put, $\eta_{n}= (A-\lambda I)\xi_{n}$.
Then, by \eqref{eqn_non}, $\eta_{n}{\to} 0$. Hence, $\xi_{n}= (A-\lambda I)^{-1}\eta_{n}{\to} 0$. This in turn implies
that $T\xi_{n} {\to} 0$, which is impossible since $\|\eta\|=1$.
\qed

\begin{propaat}\label{lemma_one} Let $A$ and $B$ be closed operators. Assume that $A\dashv B$, with  {the bounded intertwining operator} $T$.
Then the following statements hold.
\begin{itemize}
\item[(a)] Let $\lambda \in \rho(A)$ and define
\begin{align*} D(X_\lambda) &=D(T^{-1}),\\
  X_\lambda\eta &=T(A-\lambda I)^{-1}T^{-1}\eta, \; \eta \in D(X_\lambda).
\end{align*}
Then,
\begin{itemize}\item[(a.1)] $(B-\lambda I) X_\lambda \eta = \eta, \; \forall\, \eta \in D(X_\lambda)$.
\vspace*{1mm}
\item[(a.2)] If $ (B-\lambda I)\eta \in D(T^{-1}), \; \forall\, \eta \in D(B)$, and $\lambda \not\in \sigma_p(B)$, then
$ X_\lambda (B-\lambda I)\eta = \eta ,\; \forall\, \eta \in D(B)$.
\end{itemize}
\vspace*{1mm}
\item[(b)] Let $\lambda \in \rho(B)$ and define
\begin{align*}  D(Y_\lambda) &=\{\xi \in \H:\, (B-\lambda I)^{-1}T\xi \in D(T^{-1})\},\\
  Y_\lambda\xi &=T^{-1}(B-\lambda I)^{-1}T\xi, \; \xi \in D(Y_\lambda).
\end{align*}
Then,
\begin{itemize}\item[(b.1)]
$Y_\lambda(A-\lambda I)\xi =\xi,\; \forall\, \xi \in D(A)$.
\vspace*{1mm}
\item[(b.2)] For every $\eta \in \H$ such that $Y_\lambda\eta \in D(A)$, $(A-\lambda I)Y_\lambda\eta=\eta.$
\end{itemize}
\end{itemize}
\end{propaat}
We skip the easy proof, which is given in \cite[Prop. 3.25]{pip-metric}.

\becor \label{cor_3.24} Let $A$, $B$ be as in Proposition \ref{lemma_one} and assume that $T^{-1}$ is everywhere defined and bounded. Then 
 {$\rho(A)\setminus \sigma_p(B) \subseteq \rho(B)$ and $ \rho(B)\setminus \sigma_r(A) \subseteq \rho(A)$.}
 
\encor
 {\bf Proof. }
The first inclusion  is an immediate application of (a) of the previous proposition and the second one is obtained by taking the adjoints. \makebox[3cm]{}
\qed

 {Actually, we may drop the assumption that $T^{-1}$ is bounded and still get a useful result.}
\becor \label{cor_two} Let $A,B$ be as in Proposition \ref{lemma_one}. Assume that $D(B)$ and  $ R(B)$ are subspaces of $D(T^{-1})$. Then
$\rho(A)\setminus \sigma_p(B)  \subseteq \rho(B)\cup  \sigma_c(B)$.
\encor
 {{\bf Proof. } Let $\lambda \in \rho(A)\setminus \sigma_p(B)$. By Proposition \ref{lemma_one}(a), the operator $(B-\lambda I)^{-1}$ has a densely defined inverse. If
$(B-\lambda I)^{-1}$ is bounded, then it has an everywhere defined bounded closure,
which coincides with $(B-\lambda I)^{-1}$, since the latter is closed, being the inverse of a closed operator. In this case, $\lambda \in \rho(B)$. 
 {On the other hand,} if $(B-\lambda I)^{-1}$
is unbounded, then $\lambda \in \sigma_c(B)$. Therefore, $\rho(A)\setminus \sigma_p(B)  \subseteq \rho(B)\cup  \sigma_c(B)$.
\qed

 Let us consider again the special case where $T^{-1}$ is also everywhere defined and bounded (but does not necessarily satisfy $TD(A)=D(B)$).
\beprop \label{prop_220}
Let $A$ and $B$ be closed operators. Assume that $A\dashv B$, with  {the bounded intertwining operator} $T$. Assume that $T^{-1}$ is everywhere defined and bounded and $TD(A)$ is a core for $B$. Then
$$
 \sigma_p (A)\subseteq \sigma_p (B) \subseteq \sigma (B) \subseteq \sigma(A).
$$
\enprop
 {\bf Proof. } We simply notice that, in this case, by (iv)  of Proposition \ref{prop_sigmap}, $\sigma_p(B)\subset \sigma(A)$. Hence,
$\rho(A)\setminus \sigma_p(B) =\rho(A) \subseteq \rho(B)$, by Corollary \ref{cor_3.24}.
\qed
 \berem The situation described in Proposition \ref{prop_220} is quite important for possible applications.
Even if the spectra of $A$ and $B$ may be different, it gives a certain number of informations on $\sigma(B)$
once $\sigma(A)$ is known.  For instance, if $A$ has a pure point spectrum, then $B$ is isospectral to $A$.
More generally, if $A$ is self-adjoint, then any operator $B$ which is quasi-similar to $A$ by
means of a bounded intertwining operator $T$ whose inverse is bounded too, has real spectrum.
\enrem  

 {We will illustrate the previous propositions by two examples, both taken from \cite{pip-metric}. In the first one,  $A\dashv B, A, B$ and $T$ are all bounded, and the two spectra, which are pure point, coincide. }

\beex \label{ex1-3-23}
Let us consider the operators $P_\vp$ and $A_\vp$ of Example \ref{ex_1} with $\vp \in D(Q^2)$. In this case $A_\vp$ is bounded and everywhere defined and, as noticed before, $P_\varphi \dashv A_\varphi$ with the intertwining operator $T:= $ \mbox{$(I+Q^2)^{-1}$} The spectrum of $A_\vp$ is easily computed to be $\sigma(A_\vp)=\{0, 1\}$. Thus it coincides with $\sigma(P_\vp)$, {in accordance with Proposition \ref{prop_sigmap} (iii)}. To see this, we begin by  looking for eigenvalues. The equation
\begin{equation}\label{eq_eigenvalues}
\ip{(I+Q^2)f}{\varphi}(I+Q^2)^{-1}\varphi -\lambda f=0
\end{equation}
has non zero solutions in two cases: if $\lambda=0$, then any element of $\{(I+Q^2)\vp\}^\perp$ is an eigenvector. If $\lambda \neq 0$, then a solution must be a multiple of $(I+Q^2)^{-1}\varphi $, i.e., $f= \kappa (I+Q^2)^{-1}\varphi$. Substituting in \eqref{eq_eigenvalues} one obtains $\lambda =1$ and the set of eigenvectors is the one-dimensional subspace generated by
$(I+Q^2)^{-1}\varphi$. On the other hand, if $\lambda \not\in \{0,1\}$, then, for every $g \in L^2({\mb R})$, the equation
$(A_\vp-\lambda I)f= g$ has the unique solution
$$
f= - \frac{1}{\lambda}g+ \frac{\ip{g}{(I+Q^2)\vp}}{\lambda(1-\lambda)}(I+Q^2)^{-1}\vp. 
$$
 {Thus, $(A_\vp-\lambda I)^{-1}$ is an everywhere defined  operator. Next, being the inverse of a closed (in fact, bounded) operator, it is closed. Therefore, by the closed graph theorem, it is bounded, $(A_\varphi-\lambda I)^{-1}\in {\mc B}(\H).$}
We  conclude that  {$\sigma_p(A_\vp)=\sigma_p(P_\vp)=\{0, 1\}$}.
\enex
 {In the second example, $T$ is bounded, but $A$ and $B$ are both unbounded. In that case, the two spectra coincide as a whole, but not their individual parts. In particular,  $A$ has a nonempty residual spectrum, whereas $B$ does not.}

\beex  \label{ex1-3-24}
Let $A$ be the operator in $L^2({\mb R})$ defined as follows:
\begin{align*}& (Af)(x)=f'(x)- \frac{2x}{1+x^2}f(x), \quad f\in D(A) = W^{1,2}({\mb R}).
\end{align*}
Then $A$ is a closed operator in $L^2({\mb R})$, being the sum of a closed operator and a bounded one. Let $B$ be the closed operator defined by
 \begin{align*}
& (Bf)(x)=f'(x), \quad f \in D(B) = W^{1,2}({\mb R}).
\end{align*}

Then $A\dashv B$ with the intertwining operator $T=(I+Q^2)^{-1}$. Indeed, it is easily seen that $T: W^{1,2}({\mb R})\to W^{1,2}({\mb R})$. Moreover, for every $f \in W^{1,2}({\mb R})$, one has
$$
(TAf)(x)= (1+x^2)^{-1} \left(f'(x)- \frac{2x}{1+x^2}f(x)\right) = \frac{f'(x)}{1+x^2} - \frac{2xf(x)}{(1+x^2)^2}
$$
and
$$
(BTf)(x)=\frac{d}{dx}\left(\frac{f(x)}{1+x^2}\right)=  \frac{f'(x)}{1+x^2} - \frac{2xf(x)}{(1+x^2)^2}.
$$
 Thus, indeed, $TD(A) \subseteq D(B)$ and $TAf = BTf, \; \forall\, f \in D(A)$.
It is easily seen that $\sigma_p (A)=\emptyset$. As for $B$, one has, as it is well known, $\sigma(B)=\sigma_c(B)= i{\mb R}$. On the other hand, $0\in \sigma_r (A)$, since, if $h(x)=(1+x^2)^{-1}$, then $\ip{Af}{h}=0$, for every $f \in W^{1,2}({\mb R})$, so that the range $R(A)$ is not dense. Actually one has $\sigma_r(A)= \{0\}$, as one can easily check by computing $\sigma_p(A \ha )$. Thus, by Corollary \ref{cor_3.24}, $\sigma(A)=\sigma(B)$, but the quasi-similarity does not preserve the relevant parts of the spectra. 
\enex

\subsection{Quasi-similarity with an unbounded intertwining operator}
\label{subsect_33}

 As shown in \cite{quasi-herm}, it is   easy   to generalize the preceding analysis to the case of an unbounded intertwining operator.
First we adapt the definition.
\bedefi \label{def-intertwin2}
Let $A,B$ be two  densely defined  linear operators on the Hilbert spaces  $\H, \K$, respectively.
 {A closed (densely defined) operator $T: D(T) \subseteq \H \to \K$ is called an \emph{intertwining operator}  for $A$ and $B$ if}
 \index{operator!intertwining!unbounded}
 \begin{itemize}
\vspace{-2mm}\item[({\sf io$_0$})] $D(TA) = D(A)\subset D(T)$;
\vspace{-2mm}\item[({\sf io$_1$})] $T:D(A)\to D(B)$;
\vspace{-2mm}\item[({\sf io$_2$})]$BT\xi = TA\xi, \; \forall\, \xi \in D(A)$.
\end{itemize}
\findefi
The first part of condition ({\sf io$_0$}) means that $\xi\in D(A)$ implies $A\xi\in D(T)$.
Of course, this definition reduces to the usual one, Definition \ref{def:qu-sim}, if $T$ is bounded, since then condition ({\sf io$_0$}) is satisfied automatically.

 \index{operator!quasi-similar}
 In terms of this definition, we say again
 that $A$   is \emph{quasi-similar} to $B$ , and write $A\dashv B$, if there exists a (possibly unbounded) intertwining operator $T$ for $A$ and $B$ which is  invertible, with inverse $T^{-1}$   densely defined  {(that is, $T$ is quasi-invertible, in the terminology of \cite{hoover}).
This definition implies easily that $A$ is quasi-similar to $B$ if, and only if, $A \subseteq T^{-1}BT$, where $T$ is a closed densely defined operator which is injective and has dense range.}  
  {Indeed, $T^{-1}$ exists by assumption and by ({\sf io$_2$}), if $\xi \in D(A)$, $BT\xi$ is an element of the range of $T$; thus, we can apply $T^{-1}$ to both sides and get $T^{-1}BT\xi=A\xi$. This is equivalent to say that $A \subseteq T^{-1}BT$. }

 {Notice that, contrary to Remark \ref{rem_adjoint}, if $T$ is an   unbounded intertwining operator  for $A$ and $B$,   its adjoint $T \ha  : D(T \ha) \subseteq \K \to \H$ need not  be an intertwining operator for $B \ha $ and $A \ha $, because Condition ({\sf io$_0$}) may fail for $B \ha $, unless  $T$ is bounded.
As a matter of fact, quasi-similarity with an unbounded intertwining operator  may occur only under \emph{singular}, even pathological, circumstances, as shown by the next proposition.}.

\begin{propaat} Let $A\dashv B$ with intertwining operator $T$. If  the resolvent set $\rho(A)$ is not empty, then $T$ is necessarily bounded.

\end{propaat}
 {\bf Proof. } From $A \subseteq T^{-1}BT$ it follows that $A-\lambda I \subseteq T^{-1}(B-\lambda I)T$, for every $\lambda \in {\mb C}$. If $\lambda \in \rho(A)$, then, for every $\eta \in \H$, there exists $\xi \in D(A)$ such that $(A-\lambda I)\xi =\eta$. Thus, $\xi \in D(T^{-1}(B-\lambda I)T)$ and
$T^{-1}(B-\lambda I)T\xi=\eta$. This clearly implies that $\eta \in D(T)$. Hence $D(T)=\H$ and  {thus $T$ is bounded, since it is closed by definition.}
\qed

What remains to do now is to see how much of Propositions  \ref{prop_sigmap} - \ref{prop_220} remains true when the intertwining operator $T$ is unbounded. The first result parallels part of Proposition \ref{prop_sigmap}.

\begin{propaat}\label{prop_sigmap-unbdd} Let $A$ and $B$ be closed operators and assume that $A\dashv B$,  with the   (possibly) unbounded intertwining operator $T$. Then the following statements hold.
\begin{itemize}
\item[(i)] 
$\sigma_p(A)\subseteq \sigma_p(B)$. If $\xi \in D(A)$ is an eigenvector of $A$ corresponding to the eigenvalue $\lambda$,
 then $T\xi$ is an eigenvector of $B$ corresponding to the same eigenvalue.
Thus, for every $\lambda \in \sigma_p(A)$, one has  {for the multiplicities $m_A(\lambda) \leq m_B(\lambda)$.}
\item[(ii)] If $TD(A) = D(B)$ and $T^{-1}$ is bounded, then $\sigma_p(A)= \sigma_p(B)$.
\item[(iii)] If $T^{-1}$ is bounded and $TD(A)$ is a core for B,  then $\sigma_p(B)\subseteq \sigma(A)$.
\end{itemize}
\end{propaat}
 {\bf Proof. }
 (i) Let $\lambda \in \sigma_p(A)$, i.e. $ \psi\in D(A)$ and $ A \psi = \lambda \psi$. Then, by ({\sf io$_0$}), $A \psi\in D(T)$ and
$ TA \psi = \lambda T\psi$. The rest is obvious.

(ii) If $\eta\in D(B)$, there exists $\xi\in D(A)$ such that $\eta = T\xi$ and $T^{-1}\eta =  \xi$. Then
$B\eta=  \lambda\eta$  implies $A\xi=  \lambda\xi$. 

(iii) The proof of Proposition \ref{prop_sigmap}(iv) remains valid. The last argument states that
 $\xi_{n}\to 0$ and $T\xi_{n} \to \eta$. Since $T$ is closed, this again implies
that $T\xi_{n} \to 0$,   which is impossible since $\|\eta\|=1$.
\makebox[2cm]{}\qed
\medskip

Next, Proposition \ref{lemma_one} goes through, with the domain $D(Y_\lambda)$ replaced by
$$
 D(Y_\lambda)  =\{\xi \in D(T):\, (B-\lambda I)^{-1}T\xi \in D(T^{-1})\}.
 $$
 As a consequence,  {the first half of Corollary \ref{cor_3.24} goes through.}
\becor \label{cor_3.24-unbdd} Let $A$, $B$ be as in Proposition \ref{lemma_one} and assume that $T^{-1}$ is everywhere defined and bounded. Then 
$$
\rho(A)\setminus \sigma_p(B) \subseteq \rho(B).
$$
\encor
  Once again, the second statement, about  $\sigma_r(A)$, does not hold in general, since it  relies on the adjoints, for which we have no information.
 In the same way, Corollary  \ref{cor_two} still  holds, but Proposition \ref{prop_220} does not.

 Things improve if we assume the operators $A$ and $B$ to be mutually quasi-similar, \mbox{$A\dashv \vdash B$.}
First, we can improve Proposition \ref{prop_sigmap-unbdd}(i). 
\begin{propaat}\label{prop_mutsigmap}
  Let $A$ and $B$ be closed operators and assume that $A\dashv \vdash B$,
with possibly unbounded intertwining operators  $T_{AB}: D(A)\to D(B)$ and $T_{BA}: D(B)\to D(A)$. Then:
 $\sigma_p(A)=\sigma_p(B)$.
If $\xi \in D(A)$ is an eigenvector of $A$ corresponding to the eigenvalue $\lambda$,
 then $T_{AB}\xi$ is an eigenvector of $B$ corresponding to the same eigenvalue.
 If $\eta \in D(B)$ is an eigenvector of $B$ corresponding to the eigenvalue $\mu$,
 then $T_{BA}\xi$ is an eigenvector of $A$ corresponding to the same eigenvalue. In both cases, the multiplicities are the same.
 \end{propaat}
\index{spectrum!point}\index{eigenvalue} In order to obtain   identity of the spectra, we have to assume that both intertwining operators have a bounded inverse. Indeed, applying    Corollary \ref{cor_3.24-unbdd} above, we get immediately: 
\begin{propaat}\label{prop_mutspectra}
 { Let $A$ and $B$ be closed operators such that $A\dashv \vdash B$. Assume that both intertwining operators $T_{AB},T_{BA}$ have a bounded inverse. Then one has, in addition to the statements of Proposition \ref{prop_sigmap}, $\rho(A) = \rho(B)$, hence  $\sigma (A)= \sigma (B)$.
}\end{propaat}
 {\bf Proof. } 
{  Let $T_{AB}^{-1}$ be everywhere defined and bounded. Then, by Corollary \ref{cor_3.24-unbdd} and Proposition \ref{prop_mutsigmap}, we have
$$
\rho(A)\setminus \sigma_p(B) = \rho(A)\setminus \sigma_p(A) = \rho(A)\subseteq \rho(B).
$$
Exchanging $A$ and $B$, we get  $\rho(B)\subseteq \rho(A)$, which proves (i).
Then the statement follows immediately. }
\qed

Under these conditions, it follows that $\sigma_c (A) \cup \sigma_r (A) = \sigma_c (B) \cup \sigma_r (B)$, but 
 {we cannot   compare separately the two remaining parts of the spectra of $A$ and $B$, for the same reason as before.}

 {These results show that  mutual quasi-similarity is a strong property. As another testimony of that fact, 
it is worth quoting a result from  \^{O}ta and  Schm\"udgen  \cite{ota-schm}. }
 \beprop
   Let $A$ and $B$ be closed operators. Then:
\begin{itemize}
\vspace{-2mm}\item[(i)] Let $A$ and $B$ be  normal (in particular, self-adjoint) and $A\dashv \vdash B$. Then
they are unitarily equivalent, $A\stackrel{u}{\sim} B$.
\vspace{-2mm}\item[(ii)] Let $A$ be symmetric and $B$ self-adjoint, with $A\dashv \vdash B$. Then $A$ is self-adjoint
and $A\stackrel{u}{\sim} B$.
\vspace{-2mm}\item[(iii)] Let $A$ be symmetric  and $A\dashv \vdash A\ha$. Then $A$ is self-adjoint.
\eni
 \enprop

\section{The lattice  generated by a single metric operator}
\label{sect-lattice}

 {Let us consider first the general case,  where both $G$ and $G^{-1}$ may be unbounded. We consider the domain $D(G^{1/2})$ and we equip it with the following norm
\be\label{norm-RG}
\norm{R_G}{\xi}^2 =  \norm{}{(I+G)^{1/2}\xi}^2,   
\; \xi \in D(G^{1/2}).
\en
Since this norm is  equivalent to the graph norm, 
\be\label{norm-graph}
\norm{\rm gr}{\xi}^2 := \norm{}{\xi}^2 +\norm{} {G^{1/2}\xi}^2,
\en
this yields a \hs, denoted $\H(R_G)$, dense in $\H$. Next,  we equip that space with the norm
$\norm{G}{\xi}^2 := \norm{}{G^{1/2}\xi}^2$ and denote by $\H(G)$ the completion of $\H(R_G)$ in that norm and corresponding inner product $\ip{\cdot}{\cdot}_G :=\ip{G^{1/2}\cdot}{G^{1/2}\cdot}$.  Hence, we have  $\H(R_G) = \H \cap \H(G)$, with the so-called projective norm \cite[Sec.I.2.1]{pip-book}, which here is simply the graph norm \eqref{norm-graph}.
Then we define  $R_G:= I+G$, which justifies the notation $\H(R_G)$, by comparison  of \eqref{norm-RG} with the norm $\norm{G}{\cdot}^2$  of $\HG$.}

Now we perform the construction described in \cite[Sec. 5.5]{pip-book}, and largely inspired by interpolation theory \cite{berghlof}.
 {First we notice that} the conjugate dual  $\H(R_G)^\times$ of  $\H(R_G)$ is $\H(R_G^{-1})$ and one gets the triplet
\be  \label{eqtr1}
\H(R_G) \;\subset\; \H   \;\subset\;   \H(R_G^{-1}).
\en
Proceeding in the same way with the inverse operator $G^{-1}$, we obtain another \hs, $\H(G^{-1})$, and another triplet
\be \label{eqtr2}
\H(R_{G^{-1}})  \;\subset\;  \H   \;\subset\;  \H(R_{G^{-1}}^{-1}).
\en
Then, taking conjugate duals,  it is easy to see that one has
\begin{align}
\H(R_G)^\times &= \H(R_G^{-1}) = \H + \H(G^{-1}),  \label{cup1}\\ 
\H(R_{G^{-1})})^\times &= \H(R_{G^{-1}}^{-1}) = \H + \H(G).   \label{cup2}
\end{align}
 In these relations, the r.h.s. is meant to carry the inductive norm (and topology) \cite[Sec.I.2.1]{pip-book}, so that both sides are in fact unitary equivalent, hence identified.

 By the definition of the spaces $\H(R_{G^{\pm 1}})$  and the relations \eqref{cup1}-\eqref{cup2},
 it is clear that all the seven spaces involved 
 constitute a lattice with respect to the lattice operations
 \begin{align*}
 \H_1 \wedge \H_2& := \H_1 \cap \H_2 \, , \\
 \H_1 \vee \H_2& := \H_1 + \H_2 \, .
  \end{align*}
Completing that lattice by the extreme spaces $\H(R_G)\cap\H(R_{G^{-1})}) = \H(G)\cap\H(G^{-1})$ and $\H(R_G^{-1}) + \H(R_{G^{-1}}^{-1}) =
\H(G) + \H(G^{-1}) $ (these equalities follow from interpolation), we obtain the diagram shown on Fig.  \ref{fig:diagram}, which completes the corresponding one from \cite{pip-metric}. Here also   every embedding is continuous and has dense range.

\begin{figure}[t]
\centering \setlength{\unitlength}{0.38cm}
\begin{picture}(8,8)

\put(3.5,4){
\begin{picture}(8,8) \thicklines
\footnotesize
 \put(-3.4,-0.9){\vector(3,1){2.2}}
\put(-3.6,2.3){\vector(3,1){2}}
 \put(-3.4,-2.1){\vector(3,-1){2.2}}
\put(-3.4,1.2){\vector(3,-1){2.2}}
\put(1.3,0.5){\vector(3,1){2.2}}
\put(1.3,-2.9){\vector(3,1){2.2}}
 \put(1.3,-0.4){\vector(3,-1){2.2}}
\put(1.45,2.9){\vector(3,-1){2}}
\put(0,3.4){\makebox(0,0){$ \H(G^{-1})$}}
\put(-0.1,0){\makebox(0,0){ $\H$}}
\put(0,-3.4){\makebox(0,0){ $\H(G)$}}
\put(-11.5,0){\makebox(0,0){ $\H(G) \cap  \H(G^{-1})$}}
\put(12,0){\makebox(0,0){ $\H(G) +  \H(G^{-1})$}}
\put(-9.5,0.7){\vector(3,1){2}}
\put(7.2,-1.3){\vector(3,1){2}}
\put(7.2,1.3){\vector(3,-1){2}}
\put(-9.5,-0.5){\vector(3,-1){2}}
\put(-5.3,1.7){\makebox(0,0){ $\H(R_{G^{-1}})$}}
\put(-5.2,-1.7){\makebox(0,0){$\H(R_{G})$}}
\put(5.3,1.7){\makebox(0,0){ $ \H(R_G^{-1})$}}
\put(5.3,-1.7){\makebox(0,0){$ \H(R_{G^{-1}}^{-1})$}}

\end{picture}
}
\end{picture}
\caption{\label{fig:diagram}The lattice of \hs s generated by a metric operator.}

\end{figure}

\index{lattice of Hilbert spaces (LHS)}
\index{operator!metric}

Next, on the space $\H(R_{G})$, equipped with the norm $\norm{G}{\cdot}^2$,
the operator $G^{1/2}$ is isometric onto $\H$, hence it extends to a unitary operator from  $\H(G)$ onto $\H$. Analogously, $G^{-1/2}$ is a
unitary operator from  $ \H(G^{-1})$ onto $\H$. In the same way, the operator $R_{G}^{1/2}$ is unitary from $\H(R_{G})$ onto $\H$, and from  $\H$ onto
  { $ \H(R_G^{-1})$.\footnote{ 
  The space $\H(R_G^{-1})$ is (three times) erroneously denoted $\H(R_{G^{-1}})$ in \cite[p.4]{pip-metric}; see Corrigendum.}
 Hence $R_{G}$ is the Riesz unitary operator mapping $\H(R_{G})$ onto its conjugate dual  $ \H(R_G^{-1})$, and similarly 
 $R_{G}^{-1}$ from $ \H(R_G^{-1})$ onto
 $\H(R_{G})$, that is, in the triplet \eqref{eqtr1}. Analogous relations hold for $G^{-1}$, i.e. in the triplet \eqref{eqtr2}.}

  Since all spaces $\H(A)$ are indexed by the corresponding operator $A$,
  we can as well apply the lattice operations on the operators themselves.
  This would give the diagram shown in Fig. \ref{fig:diagram2}.

   \begin{figure}[t]
\centering \setlength{\unitlength}{0.4cm}
\begin{picture}(8,8)

\put(4.2,4){
\begin{picture}(8,8) \thicklines
\small
 \put(-3.4,-0.9){\vector(3,1){2.2}}
\put(-3.4,2.2){\vector(3,1){2}}
 \put(-3.4,-1.9){\vector(3,-1){2.2}}
\put(-3.4,1.2){\vector(3,-1){2.2}}
\put(1.3,0.4){\vector(3,1){2.2}}
\put(1.3,-2.8){\vector(3,1){2.2}}
 \put(1.3,-0.4){\vector(3,-1){2.2}}
\put(1.45,2.6){\vector(3,-1){2.15}}
\put(-8.9,0.6){\vector(3,1){2}}
\put(6.9,-1.1){\vector(3,1){2}}
\put(-8.9,-0.4){\vector(3,-1){2}}
\put(6.9,1.2){\vector(3,-1){2}}
\put(0,3.2){\makebox(0,0){ $ G^{-1}$}}
\put(-0.1,0){\makebox(0,0){ $I$}}
\put(0,-3.2){\makebox(0,0){ $G$}}

\put(-5.3,1.7){\makebox(0,0){ $I\wedge G^{-1}$}}
\put(-5.2,-1.5){\makebox(0,0){$I\wedge G$}}
\put(-10.5,0){\makebox(0,0){ $G\wedge G^{-1}$}}
\put(10.7,0){\makebox(0,0){ $G\vee G^{-1}$}}
\put(5.3,1.7){\makebox(0,0){ $I \vee G^{-1}$}}
\put(5.3,-1.5){\makebox(0,0){$I\vee G$}}

\end{picture}
}
\end{picture}
\caption{\label{fig:diagram2}The lattice generated by a metric operator.}

\end{figure}
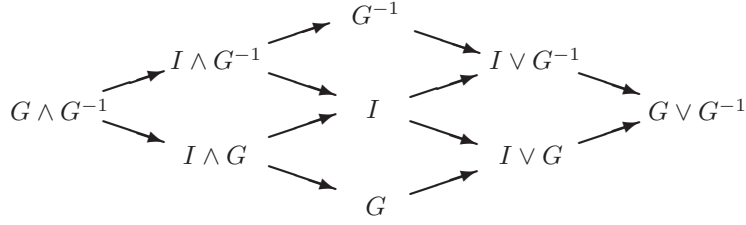

The link between the  {lattices  of Fig. \ref{fig:diagram}    and  Fig. \ref{fig:diagram2}}  is given in terms of an order relation:
 $G_1\preceq G_2$ if and only if $\H(G_1) \subset \H(G_2)$, where the embedding is continuous and has dense range. 
In particular, if $G$ is bounded and $G^{-1}$  unbounded, the relation \eqref{eq:triplet} becomes
$$
G^{-1} \preceq I \preceq G.
$$
In Section \ref{sect_LHS}, we will extend these considerations to families of metric operators.

Before proceeding, let us give two (easy) examples, in which $G$ and $G^{-1}$ are multiplication operators in
$\H = L^2(\RN,\ud x)$,  both unbounded, so that the three middle spaces are mutually noncomparable.
\begin{enumerate}
\item The first example comes from  \cite[Sec. 5.5.1]{pip-book}, namely, $G = x^2$, so that $R_G = 1 + x^2$.
{Then all spaces appearing in Fig. \ref{fig:diagram} are weighted $L^2$ spaces, for instance
 $\H(G)=L^ 2(\RN, x^2 \ud x)$, $ \H(R_{G})=L^ 2(\RN,(1 + x^2)\ud x)$, etc. The complete lattice is given in  \cite[Fig.2]{pip-book}.
As expected, all the norms   are equivalent to the corresponding projective norms, resp.
 inductive norms (see the proof of a similar statement for sequences in \cite[Sec. 4.3.1]{pip-book}).}

    \begin{figure}[t]
\centering \setlength{\unitlength}{0.4cm}
\begin{picture}(8,8)

\put(4,4){
\begin{picture}(8,8) \thicklines
\small
 \put(-3.4,-0.9){\vector(3,1){2}}
\put(-3.4,2.4){\vector(3,1){2}}
 \put(-3.4,-1.9){\vector(3,-1){2}}
\put(-3.4,1.2){\vector(3,-1){2}}
\put(1.3,0.4){\vector(3,1){2}}
\put(1.3,-2.8){\vector(3,1){2}}
 \put(1.3,-0.4){\vector(3,-1){2}}
\put(1.45,2.6){\vector(3,-1){2}}
\put(-9.7,0.6){\vector(3,1){2}}
\put(7.2,-1.1){\vector(3,1){2}}
\put(-9.7,-0.4){\vector(3,-1){2}}
\put(7.2,0.9){\vector(3,-1){2}}
\put(0,3.2){\makebox(0,0){ $L^2(-a)$}}
\put(-0.1,0){\makebox(0,0){ $L^2(0)$}}
\put(0,-3.2){\makebox(0,0){ $L^2(a)$}}

\put(-5.6,1.7){\makebox(0,0){ $L^2(0\wedge-a)$}}
\put(-5.5,-1.5){\makebox(0,0){$L^2(0\wedge a)$}}
\put(-12,0){\makebox(0,0){ $L^2(a\wedge-a)$}}
\put(11.5,0){\makebox(0,0){ $L^2(a\vee -a)$}}
\put(5.5,1.7){\makebox(0,0){ $L^2(0\vee -a)$}}
\put(5.5,-1.5){\makebox(0,0){$L^2(0\vee a)$}}

\end{picture}
}
\end{picture}
\caption{\label{fig:diagram3}The lattice generated by a metric operator. Note that $L^2(0)\equiv L^2$.}

\end{figure}
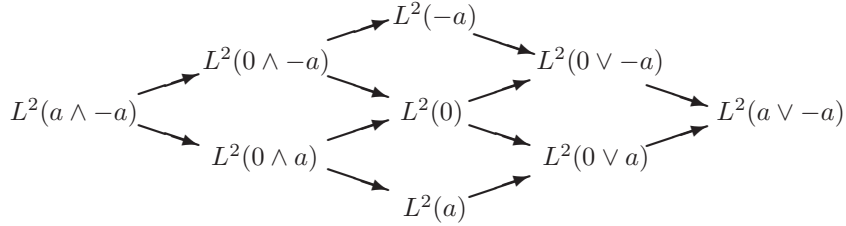
\index{lattice of Hilbert spaces (LHS)}
\index{operator!metric}

\item 
{For the second example, inspired by \cite{ali-bag-gaz} and \cite{kretsch}, one chooses $G = e^{ax}, G^{-1} = e^{-ax}$ and proceeds in the same way. The resulting lattice plays a significant role in quantum scattering theory, as discussed in \cite[Sec. 4.6.3]{pip-book}, so that it is worthwhile to go into some details. Keeping the notation of that reference, we define the \hs
\be
L^2(a) := \{ f  : \int_{-\infty}^{+\infty}     e^{ax}  \,|f(x)|^2 \; \ud x < \infty \} = L^2(r_a), \; \mbox{ with }r_a(t) = e^{-ax}.
\en
Then consider the lattice generated by the family 
$ L^2(a) ,  L^2(0) \!=\!L^2 $ and \linebreak
$L^2(-a)$.
The infimum is $L^2(a) \wedge L^2(b) = L^2(a) \cap L^2(b) = L^2(a \wedge b) $, with  
$r_{a\wedge b}(x) = \min(r_{a}(x),r_{b}(x)) $,
and the supremum  $L^2(a) \vee L^2(b) = L^2(a) + L^2(b) =  L^2(a \vee b)$, 
$r_{a \vee b}(x) = \max(r_{a}(x),r_{b}(x))$. 
As usual, these norms are equivalent to the projective, resp. inductive, norms.
For instance, the following two norms are equivalent
$$
\norm{L^2(r_{a\wedge -a})}{f}^2 = \int_{-\infty}^{+\infty}e^{a|x|} |f(x)|^2  \, \ud x \; \asymp\; 
\int_{-\infty}^{+\infty}(e^{ax}+e^{-ax})|f(x)|^2  \, \ud x .
$$
The resulting lattice is shown in  Fig.  \ref{fig:diagram3}}.
\end{enumerate}

{As a matter of fact, in both examples, the discrete lattice of  nine spaces may be converted into a continuous one by interpolation.
In the first case, the spaces of the central column are $\{L^ 2(\RN, x^\alpha \ud x), -2 \leq \alpha \leq 2\}$.  Similarly, the   second example yields  $\{L^ 2(a), -1 \leq a \leq 1\}$. And actually the same holds true in the general, abstract case, where one gets instead the family
$\{\H(G^\alpha), -1 \leq \alpha \leq 1\}$.

\index{operator!metric!bounded}

\subsection{Bounded metric operators}
\label{subsec-bddmetric}

 {Now, if $G$ is bounded, the triplet \eqref{eqtr1} collapses, in the sense that all three spaces coincide as vector spaces, with equivalent norms. Similarly, one gets
$\H(R_{G^{-1}}) = \H(G^{-1})$ and
$\H(R_{G^{-1}}^{-1}) = \H(G)$. So we are left with the triplet
\be
  \H(G^{-1}) \;\subset\; \H \;\subset\;  \H(G).
\label{eq:triplet}
\en
Then  $G^{1/2}$ is a unitary operator from $\HG$ onto $\H$ and from $\H$ onto $\H(G^{-1})$, whereas $G^{-1/2}$ is a unitary operator  
$\H(G^{-1})$ onto $\H$ and from $\H$ onto $\HG$.}

 {If $G^{-1}$ is also bounded, then the spaces $\H(G^{-1})$ and $\H(G)$ coincide with $\H$  as vector spaces and  their norms are equivalent to (but different from) the norm of $\H$.}

\berem \label{rem_extensions} If $B$ is a bounded operator from $\HG$ into itself, for every $\eta \in \H\subseteq \HG$, we have
$$
 \norm{G}{B\xi} \leq  \gamma \norm{G}{\xi}  \leq  \gamma'  \norm{}{\xi},
 $$ 
 for some $\gamma, \gamma'>0$.

This means that $B_0:=B\uph_\H$ is bounded from $(\H,\norm{}{\cdot})$ into $(\HG ,\norm{G}{\cdot})$. Then there exists a bounded operator $B_0^\dag: \HG\to \H $ such that
$$
 \ip{B_0\xi}{\eta}_G = \ip{\xi}{B_0^\dag \eta}, \quad \forall\, \xi \in \H, \, \eta\in \HG.
 $$
This in particular applies to $G$ and $G^{1/2}$ which can be viewed as restrictions to $\H$ of their natural extensions $\widetilde{G}$ and $\widetilde{G^{1/2}}$ to $\HG$. In this case, one can easily prove that $ G^\dag= G^{3/2}\widetilde{G^{1/2}}$, $\widetilde{G^{-1/2}}= (G^{1/2})^\dag$. In particular, if $G^{-1}$ is also bounded, $G^\dag=G^2$ and $ (G^{1/2})^\dag= {G^{-1/2}}$.
\enrem

\subsection{Unbounded metric operators}
\label{subsec-unbddmetric}

\index{operator!metric!unbounded}
Actually one can go further, following a construction made in \cite{scales}.  {Let $G$ be \emph{unbounded}, with $G > 1$ . }Then the norm 
$\norm{G}{\cdot}$ is equivalent to the   norm  $\norm{R_G}{\cdot}$ on $D(G^{1/2})$, so that  
$\H(G) = \H(R_{G})$ as vector spaces and thus also $\H(G^{-1}) = \H(R_{G}^{-1})$. On the other hand,  $G^{-1}$ is bounded. 
Hence we get the triplet
\be\label{eq:tri>1}
\H(G) \; \subset\; \H \; \subset\; \H(G^{-1}).
\en
 {In the general case, we have  $R_{G} = 1+G > 1$ } and it is also a metric operator. Thus we have now
\be\label{eq:tri<1}
\H(R_{G}) \; \subset\; \H \; \subset\; \H(R_{G}^{-1}).
\en
In both cases
one recognizes   that the triplet \eqref{eq:tri>1}, resp. \eqref{eq:tri<1}, is the central part of the discrete scale of Hilbert spaces built on the powers of $G^{1/2}$, resp. $R_{G}^{1/2}$.
This means, in the first case, $V_{\G}:= \{\H_{n}, n \in \ZN \}$,
where $\H_{n} = D(G^{n/2}),  n\in \NN$, with a norm equivalent to the graph norm, and $ \H_{-n} =\H_{n}^\times$:
\be\label{eq:scale}
 \ldots\subset\; \H_{2}\; \subset\;\H_{1}\; \subset\; \H \; \subset\; \H_{-1} \subset\; \H_{-2} \subset\; \ldots
\en
Thus $\H_{1} =  \H(G )$ and $\H_{-1} =  \H(G^{-1})$.
In the second case, one simply replaces $G^{1/2}$ by $R_{G}^{1/2}$  { and performs the same construction}.

As in the original construction, this raises the question of identifying the end spaces of the scale, namely,
\be \label{eq:endscale}
 { \H_{\infty}(G):=\bigcap_{n\in \ZN} \H_n, \qquad \H_{-\infty}(G):=\bigcup_{n\in \ZN} \H_{n}.}
\end{equation}
In fact, one can go one more step. Namely, following \cite[Sec. 5.1.2]{pip-book}, we can use quadratic interpolation theory and build a continuous scale of Hilbert spaces
$\H_{\alpha}$, \linebreak  $0\leq \alpha\leq 1$, between  $\H_{1}$  and $\H $, where $\H_{\alpha}=  D(G^{\alpha/2})$,  {with the graph norm  $\|\xi\|_{\alpha}^2 = \|\xi\|^2 + \|G^{\alpha/2}\xi\|^2$ or, equivalently, the norm
$\norm{}{(I+G)^{\alpha/2}\xi}^2$.
Indeed every $G^\alpha, \alpha\geq 0$, is   an unbounded metric operator.}

Next we define $\H_{-\alpha} =\H_{\alpha}^\times$ and iterate the construction to the full continuous scale $V_{\widetilde \G}:= \{\H_{\alpha}, \alpha \in \RN \}$.
Then, of course, one can replace $\ZN$ by $\RN$ in the definition \eqref{eq:endscale} of the end spaces of the scale.

Let us give three (trivial) examples. Take first $\H = L^2(\RN, \ud x)$ and define
$G_x$ as the operator of multiplication  by $(1+x^2)^{1/2}$, which is an unbounded metric operator  {(and the square root of the operator    $\H(R_{G})$ of the first example above.} In the same way, define 
$G_p := (1-\ud^2\!/\!\ud x^2)^{1/2} = \F G_x \F^{-1}$, where $\F$ is the Fourier transform. Similarly, in $L^2(\RN^3)$, we can take $G_p := (1-\Delta)^{1/2}$.
For these examples, the end spaces  of the scale \eqref{eq:scale} are easy to identify:
$\H_{\infty}(G_x)$ consists of square integrable,  fast decreasing functions, whereas
the scale built on $G_p$ is precisely the scale of Sobolev spaces, $\H_n = W^{n/2,2}$ \cite[Sec. 3.2]{davies}.
  {The notation, of course, refers to the operators of position $x$ and momentum $p$ in quantum mechanics.}

 {For the third example, take $\H = L^2(\RN, \ud x)$ and $G_{\rm osc} := (1-\ud^2\!/\!\ud x^2 + x^2)^{1/2}$, where one recognizes the Hamiltonian of a one-dimensional quantum harmonic oscillator. Then $\H_{\infty}(G_{\rm osc})$ is simply the Schwartz space $\SS$ of smooth fast decreasing functions and 
$\H_{\infty}(G_{\rm osc}) = \SS^\times$, the space of tempered distributions. In addition, $\H_{\infty}(G_x)\cap \H_{\infty}(G_p) = H_{\infty}(G_{\rm osc})$.} 

More generally, given any unbounded self-adjoint operator $A$ in $\H, \,G_A:= $ \linebreak $(1+A^2)^{1/2}$ is an unbounded metric operator, larger than 1, and the construction of the corresponding scale is straightforward.

\section{Quasi-Hermitian operators}
\label{sect_quasiH}

Intuitively, a quasi-Hermitian operator $A$ is an operator which is Hermitian when the space is endowed with a new inner product.
We will make this precise in the sequel, generalizing the original definition of Dieudonn\'e \cite{dieudonne}.

\index{operator!quasi-Hermitian}
 \bedefi \label{quasihermitian}
A closed operator $A$, with dense domain $D(A)$ is called \emph{quasi-Hermitian} if there exists a metric operator $G$, with dense domain $D(G)$ such that $D(A)\subset D(G)$ and 
\be \label{eq_quasihermitian}
\ip{A\xi}{G\eta}= \ip{G\xi}{A\eta}, \quad \xi, \eta \in  D(A)\en
\findefi
Of course, if the condition $D(A)\subset D(G)$ is not satisfied,   the relation  \eqref{eq_quasihermitian} may hold for every $ \xi, \eta \in D(G)\cap D(A)$, but
 {the definition is not reliable,} since it may happen that $D(G)\cap D(A)=\{0\}$.   {Thus, to make sense,} this more general definition would require additional conditions on $G$, that will depend whether $G$ and $G^{-1}$ are bounded or not.  }{For that reason, we will keep Definition \ref{quasihermitian} in the sequel.}

\subsection{Changing the norm: Two-\hs\ formalism}
\label{subsec-changing}

\index{Two-Hilbert space formalism}

Take first $G$   bounded and $G^{-1}$ possibly unbounded. According to the analysis of Section \ref{sect-lattice}, we are facing the triplet \eqref{eq:triplet}, namely,
$$
 \H(G^{-1}) \;\subset\; \H \;\subset\;  \H(G),
 $$
where $\H(G)$ is a    \hs,  {the completion of $\H$ in the norm $\norm{G}{\cdot}$. Thus we have now two different 
\hs s and the question is how} operator properties are transferred from $\H$  to $\H(G)$.
In particular,   {two different  adjoints} may be defined and we have to compare them before analyzing quasi-Hermitian operators.
 {Notice that we are recovering here the standard situation in pseudo-Hermitian quantum mechanics \cite{bender,bender-specissue}, even the three--\hs\ formulation developed in \cite{znojil}.}
Here we follow again \cite{quasi-herm}. 

 {We begin by the following easy result \cite{pip-metric}.}
\begin{propaat}\label{prop_212} (i) Given the bounded metric operator $G$, let $A$ be  a linear operator in $\H$ with $D(A)$ dense in $\H$. Then $D(A)$ is dense in $\H(G)$. 

(ii) Let $A$ be  a linear operator in $\H$. If $A$ is closed in $\H(G)$, then it closed in $\H$.
\end{propaat}
\smallskip

 Notice that, in both statements of Proposition \ref{prop_212}, one can replace the space $\H(G)$ by $\H(G^\alpha)$, for any $\alpha>0$.
 We emphasize that the converse of (ii) does not hold in general: if $A$ closed in $\H$, it need not be closed, not even closable,  in $\H(G)$. See Corollary \ref{cor25}  below.

 {Next, for the reader's convenience, we prove the following well-known result \cite[Theor. 4.19]{weidmann}.}
\begin{lem} \label{lem_25} Let us consider a closed operator $S$ in $\H$ with dense domain $D(S)$. 
 If $B$ is bounded, one has $ (BS) \ha =S \ha B\ha$.
\end{lem}
\bdim
It is standard that $ (BS) \ha \supseteq S \ha B\ha$. 
Let $\eta\in D((BS) \ha )$. Then there exists $\eta \ha \in \H$ such that
$$
 \ip{BS\xi}{\eta} =\ip{\xi}{\eta \ha } , \quad \forall\, \xi \in D(S)=D(BS).
 $$
Thus
$$
\ip{S\xi}{B\ha \eta}  =\ip{\xi}{\eta \ha } , \quad \forall \,\xi \in D(S).
$$
Hence $B\ha \eta \in D(S \ha )$ and $S \ha B\ha \eta= (BS) \ha \eta$. Thus $(BS) \ha \subseteq S \ha B\ha$.
\edim
\smallskip

Let again   $S$  be a closed densely defined operator in $\H$.  By Proposition \ref{prop_212}, $D(S)$ is dense in $\HG$.
So $S$ has a well-defined adjoint in $\HG$. We denote it by $S^\#$,  {while we denote by $S \ha $   the usual adjoint in $\H$. 
We compute $S^\#$, recalling that $\widetilde{G}$ is the natural extension of $G$ to $\H(G)$.}

\beprop \label{prop_242} Let $G$ be a bounded metric operator in $\H$ and $S$ a closed, densely defined operator in $\H$. Then:
\smallskip

 (i)  $\widetilde{G}D(S^\#) \subseteq D(S \ha )$ and $S \ha \widetilde{G}\eta= \widetilde{G}S^\#\eta$, for every $\eta\in D(S^\#)$, where $\widetilde{G}$ denotes the natural extension of $G$ to $\HG$.

(ii) If $G^{-1}$ is also bounded, then   $D(S^\#)=G^{-1}D(S \ha )$ and $S^\#\eta= G^{-1}S \ha G\eta$, for every $\eta\in D(S^\#)$.
\enprop
\bdim  (i) Let $\eta \in D(S^\#)$. Then there exists $\eta^\#\in \HG$ such that
$$
 \ip{S\xi}{\eta}_G= \ip{\xi}{\eta^\#}_G, \quad \forall \xi\in D(S)
$$
and $S^\#\eta=\eta^\#$.
Since
$$
 |\ip{S\xi}{\eta}_G|= |\ip{\xi}{\eta^\#}_G|  \leq  \|\xi\|_G \, \|\eta^\#\|_G  \leq   \gamma \|\xi\| \, \|\eta^\#\|_G, \quad \forall \,\xi\in D(S),
 $$
there exists $\eta \ha \in \H$ such that
$$
 \ip{S\xi}{\widetilde{G}\eta}=\ip{\xi}{\eta \ha }.
 $$ 
 This implies that $\widetilde{G}\eta\in D(S \ha ).$ We recall that $\widetilde{G}\HG\subseteq \H$ (see Remark \ref{rem_extensions}).
It is easily seen that $\eta \ha = \widetilde{G}\eta^\#$.
Hence
 $$
 \ip{S\xi}{\widetilde{G}\eta}=  \ip{\xi}{S \ha \widetilde{G}\eta }=\ip{\xi}{\widetilde{G}S^\#\eta}, \quad \forall\, \xi \in D(S).
 $$
This, in turn, implies that $S \ha \widetilde{G}\eta= \widetilde{G}S^\#\eta$, for every $\eta\in D(S^\#)$.

(ii) If $G$ and $G^{-1}$ are both bounded, $\H(G^{-1}) = \H = \H(G)$ as vector spaces, but with different norms. 
From (i), it follows that $GD(S^\#) \subseteq D(S\ha)$ and $S \ha {G}\eta= {G}S^\#\eta$, for every $\eta\in D(S^\#)$.

Let  now $\zeta\in D(S\ha)$. Then $\zeta=G\eta$ and $S\ha G \eta = G \eta\ha$, for some $\eta, \eta\ha\in \H$.
Then we have, for any $\xi\in D(S)$,
$$
\ip{S\xi}{\eta}_G = \ip{S\xi}{G\eta} = \ip{S\xi}{\zeta}= \ip{\xi}{S\ha \zeta} =  \ip{\xi}{G\eta\ha} =  \ip{\xi}{\eta\ha}_G.
$$
Hence $\eta\in D(S^\#)$ and $S^\#\eta = \eta\ha = G^{-1}S\ha \zeta=  G^{-1}S\ha G \eta$, i.e., $GS^\#\eta = S\ha G \eta$.
\edim
\medskip

 {The second part of the proof of (ii) does not hold if $G^{-1}$ is unbounded. In this case, we have only the following partial result.
\beprop
Let $G$ be a bounded metric operator, with $G^{-1}$ (possibly) \linebreak unbounded.
Then, for every $\zeta\in D(S\ha) \cap D(G^{-1/2})$ such that $S\ha \zeta \in D(G^{-1/2})$,
there exists $\eta\in  D(S^\#)$ such that $\widetilde{G} \eta = \zeta$ and $GS^\# \eta = S\ha \widetilde{G} \eta$.
\enprop}
\bdim
The proof is similar to the previous one,
noting that $D(G^{-1}) = \H(G^{-1})$ and $\zeta, S\ha \zeta \in \H(G^{-1})$. Hence there exist $\eta, \eta\ha \in \H$ such that 
$$
\widetilde{G^{-1/2}}\zeta=\widetilde{G^{1/2}}\eta \quad \mbox{and} \quad  \widetilde{G^{-1/2}}\zeta \ha = \widetilde{G^{1/2}}\eta \ha .
$$
 The rest is as before.
\edim
\belem \label{lemma_onehalf2}Let $G$   be   bounded in $\H$ and let $S$ be closed and densely defined.
Define $D(K)=G^{1/2}D(S)$  and  $K\xi={G^{1/2}}SG^{-1/2}\xi$, $\xi \in D(K)$. Then $K$ is densely defined, with adjoint   $K \ha ={G^{-1/2}}S \ha G^{1/2}.$
\enlem
\bdim It is easy to see that $D(K)$ is dense in $\H$ and that ${G^{-1/2}}S \ha G^{1/2}\subset K \ha $. We prove the converse inclusion.
Let $\eta \in D(K \ha )$. Then there exists $\eta \ha  \in \H$ such that
$$ 
\ip{{G^{1/2}}SG^{-1/2}\xi}{ \eta}= \ip{\xi}{\eta \ha }, \quad \forall\, \xi \in G^{1/2}D(S).
$$
This implies that
$$
 \ip{S\zeta}{{G^{1/2}} \eta}= \ip{G^{1/2}\zeta}{\eta \ha }=\ip{\zeta}{G^{1/2}\eta \ha }, \quad \forall\, \zeta \in D(S).
 $$
Hence, ${G^{1/2}} \eta \in D(S \ha )$ and
$$
 \ip{G^{-1/2}\xi}{{S \ha G^{1/2}} \eta}= \ip{\xi}{\eta \ha }, \quad \forall\, \xi\in  G^{1/2}D(S).
 $$
This in turn implies that $S \ha G^{1/2} \eta\in D(G^{-1/2})$ and $K \ha \eta={G^{-1/2}}S \ha G^{1/2}\eta$.
\edim

In Proposition \ref{prop_242} (ii), we have obtained the expression of $S^\#$ in the case where $G$ and $G^{-1}$ are both bounded in $\H$, namely, $S^\#=G^{-1}S \ha G $. If $G^{-1}$ is unbounded,  {we can  only determine  the restriction  of $S^\#$ to $\H$.}

\begin{propaat}\label{prop_214}Given  the closed  operator  $S$, put
 {\begin{align}\label{eq-domain}
 D(S^\star)&:= \{\eta \in \H: G\eta \in D(S \ha ), \, S \ha  G\eta \in D(G^{-1}) \}
\\
S^\star \eta &:= G^{-1}S \ha  G\eta, \quad \forall\, \eta \in D(S^\star). \nn 
\end{align}
}
Then $S^\star$ is the restriction to $\H$ of the adjoint $S^\#$ of $S$ in $\H(G)$.
\end{propaat}
 {\bf Proof. } Let $\xi \in D(S)$ and $\eta \in D(S^\star)$. Then,
\begin{align*}
\ip{S\xi}{\eta}_G &= \ip{GS\xi}{\eta} =\ip{S\xi}{G\eta}\\
&= \ip{\xi}{S \ha G\eta}=\ip{G^{-1}G\xi}{S\ha  G\eta}\\
 &=\ip{G\xi}{G^{-1}S \ha  G\eta}=\ip{\xi}{G^{-1}S \ha  G\eta}_G.
 \\[-14mm]
&
\end{align*}
\qed

\becor \label{cor25} 
If the domain $ D(S^\star)$ is dense, then $S^\#$ is densely defined and $S$ is closable in $\H(G)$.
\encor
However, we still don't know  whether $S$ is  closed in $\H(G)$, that is, whether one has \linebreak $(S^\#)^\# = S$.
\medskip

\medskip

 Let now $A,B$ be closed  operators in $\H$. Assume that they are   metrically quasi-similar  and let $G$ be the  {bounded} metric intertwining operator
for $A,B$.
Then, by ({\sf ws}), we have
$$
 \ip{G\xi}{B \ha \eta}=\ip{GA\xi}{\eta}, \quad \forall\, \xi \in D(A), \, \eta \in D(B \ha ).
 $$
This equality can be rewritten as
$$ 
\ip{\xi}{B \ha \eta}_G=\ip{A\xi}{\eta}_G, \quad \forall \,\xi \in D(A), \, \eta \in D(B \ha ).
$$
This means that $B \ha $ is a restriction of $A^\#$, the adjoint of $A$ in $\H(G)$.

Since $D(B \ha )$ is dense in $\H$, then, by Proposition \ref{prop_212} (i), $A^\#$ is densely defined in $\H(G)$ and $(A^\#)^\# \supseteq A$.
Then we can consider the operator $A^\star$ defined in Proposition \ref{prop_214}. This operator is an extension of $B \ha $ in $\H$.

Clearly $A^\star$ satisfies the equality
$$ \ip{\xi}{A^\star\eta}_G=\ip{A\xi}{\eta}_G, \quad \forall\, \xi \in D(A), \, \eta \in D(A^\star)$$
and by Proposition \ref{prop_214}, $A^\star \eta = G^{-1}A \ha G\eta, \quad \forall\, \eta \in D(A^\star)$. Since $A^\star\supset B \ha $, then $(A^\star) \ha  \subset B$. Put $B_0:= (A^\star) \ha $. 
 {The previous discussion means that $A\dashv B_0$, and that $B_0$ is minimal among the closed operators $B$ such that $A\dashv B$,
 for fixed $A$ and $G$.}
From these facts it follows easily that
$GD(A)$ is a core for $B_0$. Indeed, it is easily checked that $GD(A)$ is dense in $\H$. Let $B_1$ denote closure of the restriction of $B_0$ to $GD(A)$.
 {Then $B_1\subseteq B_0$ and it is easily seen that  $A\dashv B_1$. Hence $B_1=B_0$.}

 Thus we have proved \cite[Lemma 3.23]{pip-metric}:
\belem Let $A,B$ be closed and  $A\dashv B$ with a bounded metric intertwining operator $G$. Then
$A^\star$ is densely defined, $B_0:= (A^\star) \ha $  is minimal among the closed operators $B$ satisfying, for fixed $A$ and $G$, the conditions
\begin{align*} & G:D(A)\to D(B),
\\[1mm]
& BG\xi = GA\xi, \; \forall\, \xi \in D(A),
\end{align*}
i.e., $B_0$ is minimal among the closed operators $B$ satisfying $A\dashv B$.
Moreover,  $GD(A)$ is a core for $B_0$.
\enlem

\subsection{Bounded quasi-Hermitian operators}

\index{operator!quasi-Hermitian!bounded}
Let $A$ be a bounded operator in $\H$. Assume that $A$ is quasi-Hermitian and that the metric operator $G$ in \eqref{eq_quasihermitian} is bounded with  bounded inverse. Then
\be
\ip{GA\xi}{\eta}=\ip{A\xi}{G\eta} = \ip{G\xi}{A\eta}= \ip{\xi}{GA\eta}, \quad \forall\, \xi, \eta \in \H.
\en
 {Thus $GA$ is self-adjoint in $\H$.}
\beprop Let $A$ be bounded.
The following statements are equivalent.
\begin{itemize}
\vspace*{-2mm}\item[(i)] $A$ is quasi-Hermitian.
\vspace*{-2mm}\item[(ii)] There exists a bounded metric operator $G$, with bounded inverse, such that $GA\; (=A \ha G)$ is self-adjoint.
\vspace*{-2mm}\item[(iii)]$A$ is metrically similar to a self-adjoint operator $K$.
\end{itemize}
\enprop
\bdim \underline{(i)$\Rightarrow$(ii)} is easy.
\\[1mm]
\underline{(ii)$\Rightarrow$(iii}): 
We put $K=G^{1/2}AG^{-1/2}$. Since $A \ha G$ is self-adjoint we get
\begin{align*}
K \ha &=G^{-1/2}A \ha G^{1/2}=G^{-1/2}(A \ha G)G^{-1/2}=G^{-1/2}(GA)G^{-1/2}\\
&=G^{1/2}AG^{-1/2}
\end{align*}
Hence, $K$ is self-adjoint and $A=G^{-1/2}KG^{1/2}$, i.e., $A\sim K$.
\\[1mm]
\underline{(iii)$\Rightarrow$(i)}: 
Assume $A=G^{-1/2}KG^{1/2}$ with $K=K \ha $. Then, for every $\xi, \eta \in \H$
\begin{align*}
\ip{GA\xi}{\eta}&=\ip{A\xi}{G\eta}=\ip{G^{-1/2}KG^{1/2}\xi}{G\eta}
\\&=\ip{G^{1/2}\xi}{KG^{1/2}\eta}=\ip{G^{-1/2}G\xi}{KG^{-1/2}G\eta}
\\&=\ip{G\xi}{G^{-1/2}KG^{1/2}\eta}=\ip{G\xi}{A\eta}.
\\[-14mm]
& \makebox[1cm]{} 
\end{align*}
\edim
\medskip

\noi  {As a consequence of this proposition, bounded quasi-Hermitian operators coincide with bounded    spectral operators  of scalar type and real spectrum,  mentioned in Section \ref{sect_termin}  \cite{dunford}.}

\subsection{Unbounded quasi-Hermitian operators}

 Let again $G$   be bounded, but now we take $A$ unbounded and quasi-Hermitian in the sense of Definition \ref{quasihermitian}.  {The first result is immediate.}
\index{operator!quasi-Hermitian!unbounded}

\beprop \label{prop510}
If $G$ is    bounded, then  $A$ is quasi-Hermitian if, and only if, $GA$ is symmetric in $\H$.
\enprop
\bdim If $A$ is quasi-Hermitian,   \eqref{eq_quasihermitian} implies immediately 
\be
\ip{GA\xi}{\eta}=\ip{A\xi}{G\eta} = \ip{G\xi}{A\eta}= \ip{\xi}{GA\eta}, \quad \forall\, \xi, \eta \in D(A).
\en
Hence $GA$ is symmetric.

On the other hand, if $GA$ is symmetric,
\be
\ip{A\xi}{G\eta} =\ip{GA\xi}{\eta}= \ip{\xi}{GA\eta} = \ip{G\xi}{A\eta} , \quad \forall\, \xi, \eta \in D(A).
\en
Thus, $A$ is quasi-Hermitian.
\edim
\medskip

  {Next we investigate the self-adjointness of $A$ as an operator in $\HG$. This is equally easy.}

\beprop \label{prop_28}Let  $G$ be bounded. If $A$ is self-adjoint in $\HG$, then $GA$ is symmetric in $\H$
 {and $A$ is quasi-Hermitian.}
If  $G^{-1}$ is also bounded, then  $A$ is self-adjoint in $\HG$ if, and only if, $GA$ is self-adjoint  in $\H$.
\enprop
\bdim Let $A= A^\#$. Then, by Lemma \ref{lem_25}, $(GA) \ha =A \ha G, \, \forall\, \xi\in D(A)$, 
$GD(A) \subseteq D(A \ha )$ and $A \ha G\xi =GA\xi$.
Hence $GA$ is symmetric  {and, thus, $A$ is quasi-Hermitian by Proposition \ref{prop510}.}

If  $G^{-1}$ is bounded, one has
$$
 A=A^\#= G^{-1}A \ha G \;\Longleftrightarrow\; GA=A \ha G=(GA) \ha \; \Longleftrightarrow\; GA \mbox{ self-adjoint}.  
\vspace*{-8mm}
$$
 \edim

\vspace*{4mm}
Now we turn the problem around. Namely, given the closed densely defined operator $A$, possibly unbounded, we seek whether there is a metric operator $G$ that makes $A$ quasi-Hermitian and self-adjoint in $\HG$. The first result is rather strong.

\beprop \label{prop_29}  {Let $A$ be  closed and densely defined}. Then the following statements are equivalent:
\begin{itemize}
\vspace*{-2mm}\item[(i)]There exists a bounded metric operator $G$, with bounded inverse, such that $A$ is self-adjoint in  $\H(G)$.
\vspace*{-2mm}\item[(ii)]There exists a bounded metric operator $G$, with bounded inverse, such that $GA=A \ha G$, i.e.,  $A$ is similar to its adjoint $A \ha $, with intertwining operator $G$.  
\vspace*{-2mm}\item[(iii)]There exists a bounded metric operator $G$, with bounded inverse, such that $G^{1/2} A G^{-1/2}$ is self-adjoint.  
\vspace*{-2mm}  \item[(iv)] $A$ is a spectral operator of scalar type with real spectrum.
\end{itemize}
\enprop
\index{operator!spectral of scalar type}\index{spectrum!real}
\bdim 
 {The implication (i) $\Rightarrow $(ii)  is clear. Next we  prove} that (ii) $\Rightarrow$ (iii) and (iii) $\Rightarrow $(i), which implies the equivalence of the first three statements.
\\[1mm]
\underline{(ii) $\Rightarrow$ (iii)}: {Let  $K:=G^{1/2}AG^{-1/2}$. This operator is self-adjoint.}  Indeed, by Lemma \ref{lemma_onehalf2}, we have
\begin{align*}
 K \ha &= G^{-1/2}A \ha G^{1/2} =G^{-1/2}A \ha GG^{-1/2}= G^{-1/2}GAG^{-1/2}=G^{1/2}AG^{-1/2}
 \\
 &=K.
\end{align*}
\underline{(iii) $\Rightarrow$ (i)}: Let $G^{1/2} A G^{-1/2}$ be self-adjoint. Then by Lemma \ref{lemma_onehalf2}, we get
$$ 
GA= G^{1/2}(G^{1/2}AG^{-1/2})G^{1/2}= G^{1/2}(G^{-1/2}A \ha G^{1/2}) G^{1/2}= A \ha G= (GA) \ha .
$$
The statement follows from Proposition \ref{prop_28}.
\vspace*{1mm}

 {Now we turn to the fourth statement.}
\\[1mm]
  {\underline{(iii) $\Rightarrow$ (iv)}:  {Put again $K = G^{1/2} A G^{-1/2}$, which is self-adjoint by assumption. }Hence $K=\int_\RN \lambda \ud E(\lambda)$, 
where $\{E(\lambda)\}$ is a self-adjoint spectral family. From this it follows that
$$
A=\int_\RN \lambda \ud X(\lambda), \quad \mbox{where} \quad X(\lambda)= G^{-1/2}E(\lambda)G^{1/2}.
$$
That is, $A$ is a spectral operator of   {scalar type with real spectrum, since $\sigma(A) = \sigma(K) \subseteq \RN$.}
\\[1mm]
  {\underline{(iv) $\Rightarrow$ (iii)}:  If $A$ is a spectral operator of scalar type with $\sigma(A)  \subseteq \RN$, then
$A=\int_\RN \lambda \ud X(\lambda)$, where $\{X(\lambda)\}$  is a countably additive resolution of the identity  {(not necessarily self-adjoint)}  \cite{inoue-trap}.
By a result of Mackey \cite[Theorem 55]{mackey}, there exists a bounded operator $T$ with bounded inverse and a self-adjoint resolution of the identity 
$\{E(\lambda)\}$
such that  $X(\lambda) = T^{-1} E(\lambda) T$. Put $G = | T |^2$. By the polar decomposition, $T = U G^{1/2}$ with $U$ unitary. Hence,
$X(\lambda) = G^{-1/2} U^{-1} E(\lambda) U G^{1/2}$. Put $F(\lambda) = U^{-1} E(\lambda) U$. Then $\{F(\lambda)\}$ is a self-adjoint resolution of the identity. Thus $K:=\int_\RN \lambda \ud F(\lambda)$ is self-adjoint. Clearly $K = G^{1/2} A G^{-1/2}$, as announced.\footnote{ 
Using Dunford's result \cite[Sec.XV.6]{dunford-schwartz}, we may conclude directly that $A = T^{-1} S T$, with $S$ self-adjoint. The rest follows by putting again $T = U G^{1/2}$.}}
\edim

 Condition (i) of Proposition \ref{prop_29} suggests the following definition.
\bedefi \label{def:qsa}Let $A$ be closed and densely defined. We say that $A$ is \emph{quasi-self-adjoint}   if there exists a bounded metric operator  $G$,  such that $A$ is self-adjoint in  $\H(G)$.
\findefi
\index{operator!quasi-self-adjoint}

 {In particular, if any of the conditions of Proposition \ref{prop_29} is satisfied, then $A$ is quasi-self-adjoint.
Notice, however, that Definition \ref{def:qsa} is slightly more general than the set-up of the proposition, since we do not require $G^{-1}$ to be bounded.}

 {Proposition \ref{prop_29} characterizes quasi-self-adjointness in terms of similarity of $A$ and $A^*$, if the intertwining metric operator is bounded with bounded inverse.
Instead of requiring that $A$ be similar to $A \ha $,   we may ask  that they be only quasi-similar.   The price to pay is that now $G^{-1}$ is no longer bounded and, therefore, the equivalences stated in Proposition \ref{prop_29} are no longer true. Instead we have the following weaker result \cite{quasi-herm}.}

\beprop \label{prop_292}  {Let $A$ be  closed and densely defined}.  Consider the statements
\begin{itemize}

\vspace*{-2mm}\item[(i)]There exists  a bounded metric operator $G$ such that $GD(A)= D(A \ha ),  \,A \ha G\xi $\\
$=GA\xi$, for every $\xi \in D(A)$, in particular, $A$ is quasi-similar to its adjoint $A \ha $, with intertwining operator $G$.

\vspace*{-2mm}\item[(ii)]There exists a bounded metric operator $G$, such that $G^{1/2} A G^{-1/2}$ is self-adjoint.
\vspace*{-2mm}\item[(iii)]There exists a bounded metric operator $G$  such that $A$ is self-adjoint in $\HG$; { i.e., $A$ is quasi-selfadjoint}.
\vspace*{-2mm}\item[(iv)]{ There exists a bounded metric operator $G$  such that $GD(A)= D(G^{-1}A \ha )$, $A \ha G\xi =GA\xi$, for every $\xi \in D(A)$, in particular, $A$ is quasi-similar to its adjoint $A \ha $, with intertwining operator $G$}.
\end{itemize}
Then, the following implications hold :
$$
 (i) \Rightarrow (ii) \Rightarrow (iii) \Rightarrow (iv).$$
{  If the range $R(A\ha)$ of $A\ha$ is contained in $D(G^{-1})$, then the four conditions (i)-(iv) are equivalent}.
\enprop
\bdim \underline{(i) $\Rightarrow$ (ii)}:
 {We put $K:=G^{1/2}AG^{-1/2}$ and show it is   self-adjoint. }
As in Lemma \ref{lemma_onehalf2}, take $ \xi \in D(K), \, \eta \in D(K \ha )$.
Then, taking into account that $\xi \in D(G^{-1/2})$ and that, since $G^{1/2}\eta \in D(A \ha )$ and $D(A \ha )=GD(A)$,
$G^{1/2}\eta =G\zeta$ for some $\zeta \in D(A)$, we have
\begin{align*}
 \ip{K\xi}{\eta}
 &= \ip{\xi}{G^{-1/2}A \ha  G^{1/2}\eta}v= \ip{G^{-1/2}\xi}{A \ha  G^{1/2}\eta}\\
&=\ip{G^{-1/2}\xi}{A \ha  G\zeta}= \ip{G^{-1/2}\xi}{GA\zeta}\\
&=\ip{G^{-1/2}\xi}{GAG^{-1/2}\eta}=\ip{\xi}
={G^{1/2}AG^{-1/2}\eta}=   \ip{\xi}{K\eta}. 
\end{align*}
{Hence $K=K \ha $ is self-adjoint.}\
\\[1mm]
\underline{(ii)$\Rightarrow$ (iii)}: First, we prove that $A$ is symmetric in $\H(G)$,  i.e., $A\subseteq A^\#$.
Indeed, if $\xi, \eta \in D(A)$, we have, by putting $\zeta=G^{1/2}\xi$ and $\varsigma = G^{1/2}\eta$,
$$ 
\ip{A\xi}{\eta}_G= \ip{GA\xi}{\eta}=\ip{G^{1/2} A G^{-1/2}\zeta}{\varsigma}=\ip{\zeta}{G^{1/2} A G^{-1/2}\varsigma}=\ip{\xi}{A\eta}_G.
$$
Let now $\eta\in D(A^\#)\subseteq \H(G)$. Then, there exists  {an element $\eta^*\in \H(G)$, }such that
$$ 
\ip{A\xi}{\eta}_G =\ip{\xi}{\eta^*}, \quad \forall\, \xi \in D(A);
$$
or, equivalently,
$$
 \ip{A\xi}{\widetilde{G}\eta}_G =\ip{\xi}{\widetilde{G}\eta^*}, \quad \forall \,\xi \in D(A).
 $$
Since, as noticed before, $\widetilde{G} \H(G)=D(G^{-1/2}= G^{1/2}(\H)$ and $\widetilde{G}^{1/2}\H(G)=\H$, we get the equality $\widetilde{G}=G^{1/2}\widetilde{G}^{1/2}$. Then,
$$
\ip{G^{1/2}A\xi}{\widetilde{G}^{1/2}\eta}=\ip{G^{1/2}\xi}{\widetilde{G}^{1/2}\eta^*}, \quad \forall \,\xi \in D(A).
$$
Let $\zeta:=G^{1/2}\xi$, we get
$$
\ip{G^{1/2}AG^{-1/2}\zeta}{\widetilde{G}^{1/2}\eta}=\ip{\zeta}{\widetilde{G}^{1/2}\eta^*}, \quad \forall \,\zeta \in G^{1/2}D(A).
$$
This implies that $\widetilde{G}^{1/2}\eta \in D((G^{1/2} A G^{-1/2})^*)= D(G^{1/2} A G^{-1/2})=G^{1/2}D(A).$ Thus, in particular, $\widetilde{G}^{1/2}\eta\in D(G^{-1/2}=\widetilde{G} \H(G)$. Hence $\widetilde{G}^{1/2}\eta= G^{1/2}\widetilde{G}^{1/2}\varphi$ for some $\varphi \in \H(G)$. The injectivity of $\widetilde{G}^{1/2}$, then implies that $\eta = \widetilde{G}^{1/2}\varphi\in \H$. Therefore, $G^{1/2}\eta=\widetilde{G}^{1/2}\eta\in D(AG^{-1/2})$. This, in turn, implies that $\eta \in D(A)$. In conclusion, $A$ is self-adjoint in $\H(G)$.
\\[1mm]
{ \underline{(iii)$\Rightarrow$ (iv)}: Assume that $A$ is self-adjoint in $\H(G)$, i.e., $A=A^\#$. Then, by Proposition \ref{prop_214}, it follows that
\begin{equation}\label{eq-domA} D(A)=\{\eta \in \H: G\eta \in D(A \ha ), \, A \ha  G\eta \in D(G^{-1}) \}.\end{equation}
Now,
$ \zeta \in GD(A)$ if and only if $G^{-1}\zeta \in D(A)$. By \eqref{eq-domA}, this is equivalent to say that $\zeta \in D(A\ha)$ and $A\ha \zeta \in D(G^{-1})$. The latter two conditions define the domain of $D(G^{-1}A\ha)$. Hence, $GD(A)=D(G^{-1}A\ha)$.
Furthermore, if $\xi \in D(A)$, Proposition \ref{prop_214} implies also that $A\xi =G^{-1}A\ha G \xi$. Then, since $A\ha G \xi\in D(G^{-1})$, by applying $G$ to both sides we conclude that $GA\xi=A\ha G\xi$.

 Finally, if $ R(A\ha)\subset D(G^{-1})$, then $D(G^{-1}A\ha)=D(A\ha)$ and (iv)$\Rightarrow$(i) is obvious.
}
\edim
 
 \berem   Condition (ii) of Proposition \ref{prop_29} is equivalent to the self-adjoint\-ness of the operator $GA$ ($G$ is there bounded, with bounded inverse). So one could expect that the self-adjointness of $GA$ plays a role also when studying, as in Proposition \ref{prop_292}, the quasi-self-adjointness of $A$ in a more general context. However, it seems not to be so. One can easily prove that the condition (i) in Proposition \ref{prop_292} implies the self-adjointness of $GA$. But the self-adjointness of $GA$ seems not to be sufficient for the quasi self-adjointness of $A$.
\enrem

 Let us now assume that one of the equivalent conditions (ii), (iii) or (iv) of   {Proposition \ref{prop_292} } holds for a certain bounded metric operator $G$ and define $H:= G^{1/2}AG^{-1/2}$ on $D(H)=G^{1/2}D(A)$. Then, $H$ is self-adjoint and $HG^{1/2}\xi =G^{1/2} A\xi$ for every $\xi \in D(A)$. Clearly $G^{1/2}$ intertwines $A$ and $H$ and $A\dashv H$. We notice, on the other hand, that $G^{-1/2}$  intertwines $H$ and $A$ in the sense of Definition \ref{def:qu-sim}.

 Let now $\{E(\lambda)\}$ denote the spectral family of $H$. Let $\xi \in \H$ and consider the conjugate linear functional $\Omega_{\lambda, \xi}$ defined on $D(G^{-1/2})$ by
$$
\Omega_{\lambda, \xi} (\eta)= \ip{E(\lambda)G^{1/2}\xi}{G^{-1/2}\eta}, \quad \eta \in D(G^{-1/2}).
$$
We consider here again  $D(G^{-1/2})$ as a Hilbert space, denoted by $\H(G^{-1})$, with norm $\|\cdot\|_{G^{-1}}=\|G^{-1/2}\cdot\|$ (see Section \ref{sect_termin}).
Then,
$$
|\Omega_{\lambda, \xi} (\eta)|= |\ip{E(\lambda)G^{1/2}\xi}{G^{-1/2}\eta}|\leq \|G^{1/2}\xi\|\|G^{-1/2}\eta\| =\|G^{1/2}\xi\|\|\eta\|_{G^{-1}}.
$$
Hence $\Omega_{\lambda, \xi}$ can be represented as follows
$$
\Omega_{\lambda, \xi} (\eta)= \ip{E(\lambda)G^{1/2}\xi}{G^{-1/2}\eta}=\ip{\Phi}{\eta}, \quad \eta \in D(G^{-1/2}),
$$
for a unique $\Phi\in \H(G^{-1})^\times$, the conjugate dual of $\H(G^{-1})$. It is a standard fact that $\H(G^{-1})^\times$ can be identified with $\H(G)$. We define $X(\lambda)\xi =\Phi$. Then $X(\lambda)$ is linear and maps $\H$ into $\H (G)$ continuously. One can easily prove that
$$
X(\lambda)\xi = \widetilde{G^{-1/2}}E(\lambda)G^{1/2}\xi, \quad \xi \in \H, 
$$ 
and it obviously satisfies
$$
 \ip{X(\lambda)\xi}{\eta}=\ip{E(\lambda)G^{1/2}\xi}{G^{-1/2}\eta}, \quad \forall \,\xi \in \H, \eta \in D(G^{-1/2}).
$$
 \beprop The family $\{X(\lambda)\}$ enjoys the following properties.
\begin{itemize}
\vspace*{-2mm}\item[(i)] $\dis\lim_{\lambda \to -\infty}\ip{X(\lambda)\xi}{\eta}=0; \;\lim_{\lambda \to \infty}\ip{X(\lambda)\xi}{\eta}=\ip{\xi}{\eta}, \quad \forall \,\xi \in \H, \eta \in D(G^{-1/2}).$
\vspace*{-2mm}\item[(ii)] $\dis\lim_{\lambda \downarrow \mu}\ip{X(\lambda)\xi}{\eta}= \ip{X(\mu)\xi}{\eta} , \quad \forall \,\xi \in \H, \eta \in D(G^{-1/2}).$
\vspace*{-2mm}\item[(iii)] The function $f_{\xi, \eta}:\lambda \mapsto \ip{X(\lambda)\xi}{\eta}$ is of bounded variation, for every  \\
$\xi \in \H, \eta \in D(G^{-1/2})$, and its total variation $V(f_{\xi, \eta})$ does not exceed \\ \mbox{$\|G^{1/2}\xi\|\|G^{-1/2}\eta\|$.}
\vspace*{-2mm}\item[(iv)] The following equality holds:
$$
 \ip{A\xi}{\eta}=\int_{\mb R}\lambda \ud\ip{X(\lambda)\xi}{\eta}, \quad \forall \,\xi \in D(A), \eta \in D(G^{-1/2}).$$
\end{itemize}
\enprop

\bdim 
 The proof of these statements reduces to simple applications of the spectral theorem for a self-adjoint operator, similar to those given in \cite{burnap} in an analogous situation. We simply check (iv). One has, in fact, for $\xi \in D(A)$ and $\eta \in D(G^{-1/2})$,
\begin{align*}
\int_{\mb R}\lambda \ud\ip{X(\lambda)\xi}{\eta} &= \int_{\mb R}\lambda \ud\ip{E(\lambda)G^{1/2}\xi}{G^{-1/2}\eta} \\
&=\ip{HG^{1/2}\xi}{G^{-1/2}\eta}\\
& =\ip{(G^{1/2}AG^{-1/2})G^{1/2}\xi}{G^{-1/2}\eta}\\ 
&= \ip{A\xi}{\eta}\\[-14mm]
&\end{align*} 
\qed

\medskip

 Hence, $A$ is a {spectral operator of scalar type} in a generalized sense. We notice that the representation in (iv) does not imply that  
$\sigma(A)=\sigma(H)$.
\index{operator!spectral of scalar type}

\subsection{Quasi-Hermitian operators with unbounded metric operators}
\label{subsec:unbddmetrop}

Assume now that   $G$ is also unbounded.  {Given a  closed and densely defined operator $A$,
  we still say that $A$ is \emph{quasi-Hermitian}}  if it verifies Definition \ref{quasihermitian}.
We say that $A$ is \emph{strictly quasi-Hermitian} if, in addition, $AD(A) \subset D(G)$ or, equivalently, $D(GA) = D(A)$.

In that case,  $\eta\in D(A), A\eta\in D(G)$ implies $G\eta \in D(A \ha )$, so that we may write
$$
  \ip{\xi}{A \ha G\eta} = \ip{A\xi}{G\eta} = \ip{G\xi}{A\eta} = \ip{\xi}{GA\eta}, \quad \forall\, \xi, \eta \in D(A).
$$
Therefore 
\be\label{eq:strqH}
A \ha G\eta = GA\eta, \quad \forall\, \eta  \in D(A).
\en
 {The relation \eqref{eq:strqH}}  means $A$ is quasi-Hermitian in the sense of Dieudonn\'e, that is, it satisfies the relation 
 $A \ha G = GA$ on the dense domain $D(A)$.
 
 Now, as a consequence of \eqref{eq_quasihermitian},
 the condition $D(GA) = D(A)$ is in fact equivalent to $G:D(A) \to D(A \ha )$. Thus, comparing the discussion above with the definition of (generalized) quasi-similarity given in Section \ref{subsect_33}, we see that $A$ is  strictly quasi-Hermitian if, and only if, $A$ is quasi-similar to  $A \ha $,  $A\dashv A \ha $.
  We will come back to this point in Section \ref{sect_psH}.

\medskip
{ Although these results have some interest, they do not solve the main problem, namely, given the quasi-Hermitian operator $A$, how does one construct an appropriate metric operator $G$? We suspect there is no general answer to the question: it has to be analyzed for each specific operator $A$.}
 
{  A partial answer may be given if one uses the formalism of \pip s, as we will see in Section \ref {subsect_71} below.}

Another open question is the following. Given two closed operators $A,B$, under which conditions are they (quasi-)similar to each other? According to the discussion so far, these conditions will be of a spectral nature, such as equality of the spectra or of some parts of the spectra.

\subsection{Example: Operators defined from Riesz bases}
\label{subsec-Riesz}

 \index{bases!Riesz}
\index{operator!defined from Riesz basis}
  For the case where $G$ and $G^{-1}$ are both bounded, an interesting class of examples has been given recently in  \cite{bag-in-tra}, namely, 
quasi-Hermitian operators defined in terms of a Riesz basis \cite{christ}. We recall that $\F_\phi=\{\phi_n,\,n\geq0\}$ is a Riesz basis if there exist an orthonormal basis  $\{e_n,\, n\geq 0\}$ and a bounded operator $T$,
invertible and with bounded inverse $T^{-1}$ such that $\phi_n=Te_n$ for all $n\geq0$. Actually, a Riesz basis  is the same thing as an exact frame, that is, a frame  $\{\phi_n\}$ that ceases to be a frame if one removes any vector from it \cite{christ,young}. Moreover, a family $\{\phi_n,\,n\geq0\}$ is a
Riesz basis if and only if it is a bounded unconditional basis, that is, $0 < \inf \norm{}{\phi_n} \leq \sup \norm{}{\phi_n} < \infty$ and the series
$f=\sum_n c_n \phi_n$ converges unconditionally for any $f\in \H$\cite{young}.  

Let $A$ be a  closed  operator with a purely discrete simple spectrum, that is,  the spectrum $\sigma(A)$ consists only of isolated eigenvalues with multiplicity one, but not necessarily real. Assume that the corresponding  eigenvectors form a Riesz basis $\F_\phi=\{\phi_n,\,n\geq0\}$ for $\H$.  This is a rather strong assumption,  interesting from a mathematical point of view, but not so  frequent  in concrete physical models  \cite{bagnew}.

Writing again $\phi_n=Te_n$, define $\psi_n :=(T^{-1})\ha e_n$ and  $\F_\psi=
\{\psi_n,\,n\geq0\}$. It is clear that $\F_\psi$ is a Riesz basis, too. Moreover, it is biorthogonal to $\F_\phi$: $\ip{\phi_n}{\psi_m}=\delta_{n,m}$, and
$$
\xi=\sum_{n=0}^\infty\ip{\xi}{\phi_n}\psi_n=\sum_{n=0}^\infty \ip{\xi}{\psi_n}\phi_n,\quad
\forall\, \xi \in \H.
$$
Furthermore, the operators $S_\phi$ and $S_\psi$ defined by
\be\label{eq-Sphipsi}
S_\phi \xi=\sum_{n=0}^\infty \ip{\xi}{\phi_n}\phi_n,\qquad S_\psi \xi=\sum_{n=0}^\infty \ip{\xi}{\psi_n}\psi_n,
\en
are bounded, everywhere defined  in $\H$, positive and self-adjoint. Hence they are both metric operators,  inverse of each other: $S_\phi=(S_\psi)^{-1}$ and $S_\phi=TT\ha$.
In addition, they are both intertwining operators between $A$ and $A\ha$:
$$
 S_\phi A\ha = A S_\phi , \mbox{ and }  S_\psi A = A\ha S_\psi ,
 $$
so that indeed $A$ is quasi-Hermitian. Hence, by  Proposition \ref{prop510}, $S_\psi A$ is a symmetric operator.  
Moreover, by Proposition \ref{prop_29},  $S_\psi A$ is   self-adjoint if and only if   $A$ is  similar to a self-adjoint operator, 
$S_\psi^{1/2} A S_\psi^{-1/2}$. 

According to  \cite{bag-in-tra}, where all the proofs may be found, the Riesz basis $\F_\phi$ generates a whole class of quasi-Hermitian operators, that we describe now.
Given   a  sequence $\balpha=(\alpha_n)$  of complex numbers, we define two operators 
 $$
 A_{\phi, \psi}^{\balpha}= \sum_{n=0}^\infty \alpha_n \phi_n \otimes \overline{\psi}_n, \mbox{ and } 
A_{\psi, \phi}^{\balpha}=\sum_{n=0}^\infty \alpha_n \psi_n \otimes \overline{\phi}_n
$$
as follows:
$$ \left\{\begin{array}{l} D(A_{\phi, \psi}^{\balpha})=\left\{\xi\in \H: \sum_{n=0}^\infty \alpha_n \ip{\xi}{\psi_n}\phi_n \mbox{ exists in }\H \right\}
\\[1mm]
A_{\phi, \psi}^{\balpha} \xi= \sum_{n=0}^\infty \alpha_n \ip{\xi}{\psi_n}\phi_n, \; \xi \in D(A_{\phi, \psi}^{\balpha})\, ,
\end{array}\right.
$$
$$ \left\{\begin{array}{l} D(A_{\psi, \phi}^{\balpha})=\left\{\xi\in \H: \sum_{n=0}^\infty \alpha_n \ip{\xi}{\phi_n}\psi_n \mbox{ exists in }\H \right\} 
\\[1mm]
A_{\psi, \phi}^{\balpha} \xi= \sum_{n=0}^\infty \alpha_n \ip{\xi}{\phi_n}\psi_n, \; \xi \in D(A_{\phi, \psi}^{\balpha})\, .
\end{array}\right.  
$$
Then we have immediately
\begin{align*} 
\vspace*{-1mm}
&\D_\phi:= \mbox{span}\{\phi_n\} \subset D(A_{\phi, \psi}^{\balpha} )\;\mbox{and} \;
  \D_\psi:= \mbox{span}\{\psi_n\} \subset D(A_{\psi, \phi}^{\balpha})\, , 
  \\ 
& A_{\phi, \psi}^{\balpha} \phi_k =\alpha_k \phi_k, \; k=0,1, \ldots,\;\mbox{and} \;
  A_{\psi, \phi}^{\balpha} \psi_k =\alpha_k \psi_k, \; k=0,1, \ldots\, ,  
\end{align*}
Hence, $A_{\phi, \psi}^{\balpha}$ and $A_{\psi, \phi}^{\balpha}$ are densely defined.
\begin{propaat}\label{prop_2.1}
The following statements hold.
\begin{itemize}
\item[(i)]$D(A_{\phi, \psi}^{\balpha})=\left\{\xi \in \H :\sum_{n=0}^\infty |\alpha_n|^2 |\ip{\xi}{\psi_n}|^2 <\infty \right\}$, \\[1mm]
$D(A_{\psi, \phi}^{\balpha})=\left\{\xi\in \H : \sum_{n=0}^\infty |\alpha_n|^2 |\ip{\xi}{\phi_n}|^2 <\infty \right\}$.

\vspace*{-1mm}\item[(ii)] $A_{\phi, \psi}^{\balpha}$ and $A_{\psi, \phi}^{\balpha}$ are closed.

\vspace*{-1mm}\item[(iii)] $(A_{\phi, \psi}^{\balpha})\ha=A_{\psi, \phi}^{\balphabar}$, where $\balphabar = (\overline{\alpha}_n)$.

\vspace*{-1mm}\item[(iv)] $A_{\phi, \psi}^{\balpha}$ is bounded if and only if $A_{\psi, \phi}^{\balpha}$ is bounded and if and only if $\balpha$ is a bounded sequence.
In particular $A_{\phi, \psi}^{\bf 1}=A_{\psi, \phi}^{\bf 1}=I$, where ${\bf 1}$ is the sequence constantly equal to $1$.
\end{itemize}
\end{propaat}

It turns out that $ S_\phi ,  S_\psi $ are still intertwining operators for the new operators. Moreover, one has
\begin{propaat}\label{prop_2.3} The following equalities hold:
\vspace*{-5mm}

\begin{align} 
 S_\psi A_{\phi, \psi}^{\balpha}&= A_{\psi, \phi}^{\balpha} S_\psi =S_\psi^{\balpha} :=\sum_{n=0}^\infty \alpha_n \phi_n \otimes \overline{\phi}_n,  \label{2add1}
 \\
 S_\phi A_{\psi, \phi}^{\balpha}&= A_{\phi, \psi}^{\balpha} S_\phi=S_\phi^{\balpha}:=\sum_{n=0}^\infty \alpha_n \psi_n \otimes \overline{\psi}_n. \label{2add2}
\end{align}
\end{propaat}
It follows again that both $ A_{\phi, \psi}^{\balpha}r$  and $A_{\psi, \phi}^{\balpha}$ are quasi-Hermitian operators. Thus we may proceed as before, in Section \ref{subsec-changing}, and consider the triplet \eqref{eq:triplet}, that reads now as follows, since $(S_\psi)^{-1}= S_\phi$:
\be\label{eq-triple-phi}
 \H(S_\phi) \;\subset\; \H \;\subset\;  \H(S_\psi).
\en
The three spaces coincide as vector spaces, with equivalent, but different norms. Thus on $\H(S_\psi)$ we consider the  inner product $\ip{\cdot}{\cdot}_{S_\psi}$ given by 
$\ip{\xi}{\eta}_{S_\psi}= \ip{S_\psi \xi}{ \eta}, $ $\xi,\eta \in \H$. Of course, 
$A_{\phi, \psi}^{\balpha}$ is symmetric with respect to this new inner product, i.e.,
$$
\ip{A_{\phi, \psi}^{\balpha} \xi}{\eta}_{S_\psi}= \ip{\xi}{A_{\phi, \psi}^{\balpha} \eta}_{S_\psi}, 
 \mbox{  for }  \xi, \eta \in D(A_{\phi, \psi}^{\balpha}).
$$
This is, in a certain sense, not surprising, since the set of eigenvectors of $A_{\phi, \psi}^{\balpha}, \F_\phi$, is an   orthonormal basis in $\H$, when endowed with the  inner product $\ip{\cdot}{\cdot}_{S_\psi}$; indeed, $\left<\phi_n,\phi_m\right>_{S_\psi}=\delta_{n,m}$.

However, we are looking for self-adjoint operators. We define the operators ${\sf a}_{\phi, \psi}^{\balpha}$ and ${\sf a}_{\psi, \phi}^{\balpha}$ as follows:
\be\label{2add3} 
 {\sf a}_{\phi, \psi}^{\balpha} =S_\psi^{1/2} A_{\phi, \psi}^{\balpha} S_\phi^{1/2},\quad
{\sf a}_{\psi, \phi}^{\balpha} = S_\phi^{1/2} A_{\psi, \phi}^{\balpha} S_\psi^{1/2}.
 \en
Then, by Proposition \ref{prop510}, $A_{\phi, \psi}^{\balpha}$ is   self-adjoint in $\H(S_\psi)$ whenever   the operator ${\sf a}_{\phi, \psi}^{\balpha}$
 is   self-adjoint in $\H$. A criterion to that effect is given by the following result.
\begin{propaat}\label{prop 2.4} The following statements hold:
\begin{itemize}

\item[(i)] $D({\sf a}_{\phi, \psi}^{\balpha})=\{ S_\psi^{1/2}\xi :  \xi \in D(A_{\phi, \psi}^{\balpha})\}$, \mbox{and} 
$D({\sf a}_{\psi, \phi}^{\balpha})=\{ S_\phi^{1/2}\xi :  \xi \in D(A_{\psi, \phi}^{\balpha})\}$  
\\[1mm]
 and they are dense in $\H$.

\vspace*{-1mm}\item[(ii)] $({\sf a}_{\phi, \psi}^{\balpha})\ha = {\sf a}_{\psi, \phi}^{\balphabar}$.

\vspace*{-1mm}\item[(iii)] If $\{\alpha_n\}\subset {\RN}$, then ${\sf a}_{\phi, \psi}^{\balpha}$ is self-adjoint.
\end{itemize}

\end{propaat}
This is the kind of result one hopes for  Hamiltonians in Pseudo-Hermitian QM. 

Two remarks in conclusion.
\bee
\item Instead of the operators $S_\psi^{\balpha},S_\phi^{\balpha}$ defined in \eqref{2add1}-\eqref{2add2}, one can consider   operators 
$S_\psi^{\bbeta},S_\phi^{\bbeta}$, with another sequence
$\beta=(\beta_n)$  of complex numbers and study their interplay with $A_{\phi, \psi}^{\balpha}$ and 
$A_{\psi, \phi}^{\balpha}$. The results are very similar to those given in  Proposition \ref{prop_2.1}.

\item  Since the whole machinery is symmetric in $\phi, \psi$, one can interchange $S_\phi$ and $S_\psi$, thus considering, instead of \eqref{eq-triple-phi}, the triplet
\be\label{eq-triple-psi}
 \H(S_\psi) \;\subset\; \H \;\subset\;  \H(S_\phi).
\en
Then on $\H(S_\phi)$ we consider the  inner product $\ip{\cdot}{\cdot}_{S_\phi}$ given by 
$\ip{\xi}{\eta}_{S_\phi}= \ip{S_\phi \xi}{ \eta}, \xi,\eta \in \H$. Of course, 
$A_{\psi, \phi}^{\balpha}$ is symmetric with respect to this new inner product, and the whole development may be repeated.
\ene

\section{The LHS generated by metric operators }
\label{sect_LHS}

\index{operator!metric}
\index{lattice of Hilbert spaces (LHS)}
\index{operator!metric!LHS generated by }
Let $\M(\H)$ denote the family of all metric operators and $\M_b(\H)$ that of all  {the bounded metric operators.}
As said in Section \ref{sect_termin},
 there is a natural order in $\M(\H) : G_1\preceq G_2$ if and only if  $\H(G_1)\subset \H(G_2)$, where the embedding is continuous and has dense range.
   If $G_1$ and $G_2$ are both bounded, a sufficient condition for $G_1\preceq G_2$ is that
 there exists $\gamma >0$  such that $G_2\leq \gamma G_1$.
Then  one has 
$$G_2^{-1} \preceq G_1^{-1} \;\Longleftrightarrow \; G_1\preceq G_2 \; \mbox { if } \;G_1, G_2 \in \M_b(\H) 
$$
and 
$$
G^{-1} \preceq I \preceq G, \quad \forall \, G \in \M_b(\H).
$$

\berem
The family $\M(\H)$ is not necessarily directed upward with respect to $\preceq$. For instance, if $X, Y \in \M_b(\H)$, then $X,Y$ have the null operator 0 as a lower bound; 0
is not the greatest lower bound \cite[Ex. 2.8.18]{kadison}, but we cannot say that a positive lower bound exists. If a positive lower bound $Z$ exists, then $Z\in \M_b(\H)$ and by definition $X, Y \preceq Z$.
\enrem

As we will see now, the spaces $\{\H(X) : X \in \M(\H) \}$ constitute  a lattice of Hilbert spaces (LHS) $V_{\J}$ in the sense of \cite[Definition 2.4.8]{pip-book}. 
  {For the convenience of the reader, we have summarized in the Appendix the necessary notions about LHSs and, more generally, partial inner product spaces (\pip s), and operators on them.}

Let first $\O \subset \M(\H)$ be a family of metric operators 
and assume that
$$
\D:= \bigcap_{G\in \O} D(G^{1/2}) 
$$ 
is a dense subspace of $\H$. 
Of course, the condition is nontrivial only  if $\O$  contains unbounded elements, for instance, unbounded inverses of bounded operators.
We may always suppose that $I \in \O$.

  As shown in \cite{interpolation,pip-metric}  and  in \cite[Section 5.5.2]{pip-book},   the family $\O$ generates a canonical lattice of Hilbert spaces (LHS).
 The lattice operations are defined by means  of the operators
\begin{align*} &X\wedge Y:= X \dotplus Y,\\ &X\vee Y:= (X^{-1} \dotplus Y^{-1})^{-1},
\end{align*}
where $\dotplus$ stands for the form sum and $X, Y \in \O$.
We recall that the form sum $T_{1 } \dotplus T_{2}$ of two positive operators is the  positive self-adjoint operator associated to the quadratic form $\t =\t_{1} + \t_{2}$,
where  $\t_{1}, \t_{2}$ are the quadratic forms of  $T_{1 },T_{2}$, respectively \cite[\S VI.2.5]{kato}.

 We notice that $X\wedge Y$  is  a metric operator,  but it need not belong to $ \O$.
First, it is self-adjoint and bounded from below by a positive quantity.
In addition, $(X\wedge Y)\xi=0$ implies $\xi=0, \, \forall\, \xi\in
Q(X \dotplus Y)= Q(X) \cap Q(Y)$, which is dense. Indeed, $\ip{(X+Y)\xi}{\xi} = \ip{X\xi}{\xi} +\ip{Y\xi}{\xi}=0$ implies
$\ip{X\xi}{\xi}=\ip{Y\xi}{\xi}=0$, since both $X$ and $Y$ are positive. This in turn implies $\xi=0$. Thus $X\wedge Y$ is  a metric operator, but it need not belong to $ \O$.
The same argument applies to the operator $X\vee Y$ .

In particular, if we take for  $\O$ the set $\M(\H)$ of all metric operators, we see that
it is stable under the lattice operations, i.e., it is a lattice by itself (but the corresponding domain $\D$ may fail to be dense).
This is not true in the general case envisaged in \cite[Section 5.5.2]{pip-book}.

For the corresponding Hilbert spaces, one has
\be\label{eq:lattice}
\begin{array}{rl} &\H(X\wedge Y):= \H(X) \cap \H(Y)\, ,
\\[1mm]
&\H(X\vee Y):= \H(X) + \H(Y)\, ,
\end{array}
\en
equipped, respectively, with the projective and the inductive norm, namely,
\be
\begin{array}{rl}
\| \xi\|_{X\wedge Y}^2 &\!\!\!= \;\| \xi\|_{X}^2  + \| \xi \|_{Y}^2 \, ,
\\[2mm]
\| \xi\|_{X\vee Y}^2 &\!\!\!=  \;
\inf_{\xi=\eta+\zeta}\left(\| \eta \|_{X}^2  + \| \zeta \|_{Y}^2\right), \; \eta\in \H(X), \zeta\in \H(Y)\, .
\end{array}
\en

{Define the set $\R= \R(\O):= \{G^{\pm 1/2}: G \in \O\}$ and the corresponding domain $\D_\R:= \bigcap_{X\in \R} D(X)$.
Let now $\Sigma$ denote the minimal set of self-adjoint operators containing $\O$, stable under inversion and form sums, with the property 
that $\D_\R$ is dense
 in every $H_Z$, $Z \in \Sigma$ (i.e., $\Sigma$ is an admissible cone of self-adjoint operators, in the sense of  \cite[Def. 5.5.4]{pip-book}). 
 Then, by \cite[Theorem 5.5.6]{pip-book}, $\O$ generates a lattice of \hs s $\J :=\J_\Sigma = \{\H(X), \, X \in \Sigma\}$ and a 
 \pip\ $V_\Sigma$ with central Hilbert space $\H= \H(I)$ and total space $V=\sum_{G\in \Sigma}\H(G)$. The ``smallest'' space is $V^\#=\D_\R$.}
 The compatibility and the partial inner product read, respectively, as
\begin{align*}
& \xi \# \eta\;\Longleftrightarrow\;  \exists \, G \in \Sigma  \; \mbox{ such that} \;\xi \in \H(G ),\, \eta \in \H(G^{-1}),
\\
&\hspace*{2cm} \ip{\xi}{\eta}_\J = \ip {G^{1/2} \xi} { G^{-1/2} \eta}_\H.
\end{align*}
{From now on, we shall denote the partial inner product simply as 
$\ip{\xi}{\eta} := \ip{\xi}{\eta}_\J$, since it coincides with the inner product of $\H$ whenever $\xi,\eta \in \H$.

 {For instance, if $\O = \{I,G\}$, the set $\Sigma$ consists of the nine operators of Fig. \ref{fig:diagram2}.  On the other hand, every power of $G$  is a metric operator.
Thus, if we take $\O = \{G ^{\alpha}, \alpha\in\ZN \; \textrm{or}\;\RN \}$ and some $G ^{\alpha_0}$ is bounded, then the set 
$\O$ is totally ordered and we obtain the scales $V_\G$ and $V_{\widetilde \G}$, 
 which are the \pip s generated by the construction above.}

We denote by ${\rm Op}(V_\Sigma)$ the space of operators in $V_\Sigma$, described at length in Section \ref{subsec:oper}. 
 {Given $(X,Y)\in {\sf j}(A)$, we denote by  $A\subn{YX} : \H(X) \to \H(Y)$   the $(X,Y)$-representative of $A$,
i.e., the restriction of $A$ to $\H(X)$.}
Then $A$ is identified with the collection of its representatives:
 $$ 
A\simeq \{A\subn{YX}: (X,Y)\in {\sf j}(A)\}.
$$
In particular,  $E\subn{YX}: \H(X) \to \H(Y) $ is the representative of the identity operator (embedding) when $\H(X) \subset\H(Y)$. 

In addition, we define
 ${\sf s}(A)=\{X \in \Sigma :  (X,X)\in {\sf j}(A) \}$, so that ${\sf s}(A^\times)=\{X^{-1} : X \in{\sf s}(A)\} $.
From the definitions \eqref{eq:lattice}, it is clear that  the set ${\sf s}(A)$ is invariant under the lattice operations $\cap$ and $+$.
Coming back to the scale \eqref{eq:scale} or  its continuous extension $V_{\widetilde \G}:= \{\H_{\alpha}, \alpha \in \RN \}$,
associated to the fixed metric operator $G$,
we may identify  $\alpha \in \RN$ with $\H_{\alpha}= \H(G^{\alpha})$ and consider the subset
${\sf s}_{G}(A)=\{\alpha\in \RN : (\alpha,\alpha)\in  {\sf j}(A)  \} \subset \RN$.
Then ${\sf s}_{G}(A^\times) = \{-\alpha:  \alpha\in {\sf s}_{G}(A) \}$.

\section{Similarity for  \pip\ operators}
\label{sect_pipop}

\subsection{General  \pip\ operators}
\label{subsect_genpipop}

\index{{\sc pip} space (partial inner product space)}
\index{operator!on {\sc pip}  space}
\index{operator!similar}
 Let us assume now that $G\in {\sf s}(A)$, that is, $(G,G) \in {\sf j}(A)$, for some    $G \in \M(\H)$, bounded or not.  Then $A\subn{GG} $ is a bounded operator from $\H(G)$ into itself, i.e., there exists $c> 0$ such that
$$
\|G^{1/2}A\subn{GG} \xi\| \leq c\|G^{1/2}\xi\|, \quad \forall \,\xi \in \H(G).
$$
This means that
$$
\|G^{1/2}A\subn{GG}G^{-1/2} \eta\| \leq c\|\eta\|, \quad \forall\, \eta \in \H.
$$
Hence, ${\sf B}:=G^{1/2}A\subn{GG}G^{-1/2}$ is a bounded operator on $\H$.
Then the operator $A\subn{GG}\in {\mc B}(\H(G))$ is \emph{quasi-similar} to ${\sf B}\in \BH$, that is, $A\subn{GG}\dashv {\sf B} $,    
with respect to the (possibly unbounded) intertwining operator $G^{1/2}$, in the sense of Definition \ref{def-intertwin2}.

More   generally, by an argument similar to that used above for the couple $(G,G)$, one can prove the following
\begin{propaat} \label{prop62}
Consider $A\in  {\rm Op}(V_\Sigma)$. Then,
 $(X,Y)\in {\sf j}(A)$ if, and only  if,  \\
 $Y^{1/2}AX^{-1/2}$ is a bounded operator in $\H$.
\end{propaat}

\berem \label{rem72} Since ${\sf B}=G^{1/2}A\subn{GG}G^{-1/2}$ is  a bounded operator on $\H$,
its restriction ${\sf B}_{0}$   to $V^\#$ is continuous from $V^\#$ into $V$, hence ${\sf B}_{0}$ determines a unique operator
 $B \in {\rm Op}(V_\Sigma)$ \cite[Proposition 3.1.2]{pip-book} such that $B \xi = {\sf B}_{0}\xi$, for every $\xi \in V^\#$ and $B\subn{II}= {\sf B}$.
The previous statement then reads as follows: 
$G \in {\sf s}(A)$, for some  $G \in \M(\H)$, implies $I \in {\sf s}(B)$ (here $I$ is the identity operator, corresponding to $\H(I) = \H$).
\enrem

 Take first $G \in \M_b(\H)$, with unbounded inverse,  so that    $\H(G^{-1}) \subset\H \subset \H(G) $. 
 Let $A\in  {\rm Op}(V_\Sigma)$ and assume that $G\in {\sf s}(A)$.  Then, following the standard approach (see Section \ref{subsubsec-sym}),   we can consider the restriction {\sf A} of  $A\subn{GG}$ to $\H$,
on the domain $D({\sf A}) = \{ \xi \in \H: \, A\subn{GG}\xi=A\xi \in \H\}$.
However, in general $D({\sf A})$ need not be  dense in $\H$.
A sufficient condition for the density of $D({\sf A})$ can be given in terms of the  adjoint operator $A^\times$,  defined in \eqref{eq:adjoint}.
 Indeed, by the assumption, $G^{-1}\in  {\sf s}(A^\times)$.
 Then the space
$ D({\sf A}^\sharp):=\{\xi \in \H: \, A^\times\xi \in \H\}$ is dense in $\H$ since it contains $\H(G^{-1})$.
We define ${\sf A}^\sharp\xi =A^\times \xi$ for $\xi \in D({\sf A}^\sharp)$. Then we have

\beprop Let $G \in {\sf s}(A)$, with $G \in \M_b(\H)$. Then the domain $D({\sf A})$ is dense in $\H$
 if and only if  the operator ${\sf A}^{\sharp{\textstyle \ast}}$  
 is a restriction of $A$.
  \enprop
 {\bf Proof. }  Let us assume first that ${\sf A}^{\sharp{\textstyle \ast}}$ is a restriction of $A$. Then, since the domain  $D({\sf A})$ is maximal in $\H$,  we get that ${\sf A}^{\sharp{\textstyle \ast}}={\sf A}$. Hence ${\sf A}$ is densely defined in $\H$.
{Conversely, for $\xi\in D({\sf A}^{\sharp{\textstyle \ast}}) \subset \H \subset \H(G)$ and $ \eta\in \H(G^{-1})$, we have
\begin{align*}
\ip{{\sf A}^{\sharp{\textstyle \ast}}\xi}{\eta} &= \ip{\xi}{{\sf A}^{\sharp}\eta} = \ip{\xi}{A^\times \eta} 
=  \ip{\xi}{A^\times\subn{G^{-1}G^{-1}} \eta}\\
 &= \ip{A\subn{GG}\xi}{\eta} = \ip{A\xi}{\eta},
\end{align*}
the last equality being valid on the dense domain  $D({\sf A})$.}
\qed

Assume that the domain $D({\sf A})=\{ \xi \in \H:\, A\xi \in \H\}$ is dense in $\H$. 
 {Since $G^{1/2}$ is bounded, we have  $G^{1/2}: D({\sf A}) \to D({\sf B}) = \H$,  where, as above, ${\sf B} = G^{1/2}A\subn{GG}G^{-1/2}$ } and
$$
 {\sf B} \,G^{1/2} \eta = G^{1/2}{\sf A}\, \eta, \quad \forall \,\eta \in D({\sf A}).
$$
This means that ${\sf A}\dashv {\sf B}$,  {with $G^{1/2}$ as intertwining operator.}

On the other hand,
$$
{\sf B} \,G^{1/2} \eta = G^{1/2}{A}\, \eta, \quad \forall \,\eta \in \H(G).
$$
So, if $G \in  {\sf s}(A)$, then $A$ is similar to a bounded operator in $\H$. But $G^{1/2}$ is a unitary operator from $\H(G)$ onto $\H$; hence $A$ and ${\sf B}$ are
 unitarily equivalent, while for the restriction  ${\sf A}$ of $A$ to $\H$ only quasi-similarity may hold.

 Next, take $G $   unbounded, with $G^{-1}$ bounded, so that  $\H(G)\subset \H\subset \H(G^{-1})$.
Then    $A: \H(G)\to \H(G)$ is a densely defined operator in $\H$.
As before,   ${\sf B}:=G^{1/2}A\subn{GG}G^{-1/2}$ is bounded and everywhere defined on $\H$. 
Hence  $G^{-1/2} : D({\sf B}) = \H \to D(A\subn{GG})= \H(G)$ and 
$ G^{-1/2}{\sf B} \xi =A\subn{GG} G^{-1/2}\xi, \;\forall\, \xi \in \H$, i.e., ${\sf B} \dashv A\subn{GG}$ with respect to the bounded intertwining operator 
$G^{-1/2}$. On the other hand, we had already $A\subn{GG}\dashv  {\sf B} $, with respect to $G^{1/2}$,
so that, in this case, $A\subn{GG}\dashv \vdash {\sf B}$. 
In addition, since $G^{\pm1/2}$ are unitary between $\H$ and $\H(G)$, it follows that $A\subn{GG}$ and ${\sf B}$ are unitarily equivalent.

 {If $G $  and $G^{-1}$ are both  unbounded, then $A: \H(R_G)= \H \cap \H(G) \to \H(G)$ is a densely defined operator in $\H$, and the argument goes through.}

\subsection{The case of symmetric \pip\ operators}
\label{subsect_71}

\index{operator!on {\sc pip}  space!symmetric}
In many applications, it is essential to show that a given symmetric operator $A$  in a \hs\ $\H$ is self-adjoint. This is, for instance,  the crucial  question in quantum mechanics,  for the Hamiltonian  of the system under consideration.     {The same question can be asked for other observables of the system.}
More generally, we may ask whether $A$ is similar in a some sense to a self-adjoint operator.
 {In that case, we might  start} from   a pseudo-Hermitian operator or a quasi-Hermitian operator $A$    on $\H$, for instance, a $\P\T$-symmetric Hamiltonian. 
If $A$ is a symmetric, densely defined, operator in the \hs\  $\H$,  
 it makes sense to ask for the existence of self-adjoint \emph{extensions} of $A$  (if $A$ itself is not self-adjoint).  The standard technique is to  use   quadratic forms (the Friedrichs extension) or  von Neumann's theory of self-adjoint extensions \cite{gitman,schm}. This approach is taken, for instance, by Albeverio \emph{et al.} \cite{alb-g-kuzhel},  in the language of $J$-self-adjoint operators in a Krein space  (see Section \ref{sect_termin}).
 
 However, there is another possibility. Namely, given a  operator $A$ in a space $\K \supset \H$, symmetric in some sense, it is natural to ask directly whether $A$ has \emph{restrictions} 
 that are self-adjoint in $\H$.  The answer is given essentially by the KLMN theorem
 that  we recall here.\footnote{KLMN stands for Kato, Lax, Lions,  Milgram, Nelson.} The framework is a scale of three Hilbert spaces, 
$\H_{1} \subset \H_{0} \subset \H_{\ov{1}}\,  , $
where  the embeddings are continuous and have dense range, and $\H_{\ov{1}}$ is the conjugate dual of $\H_{1}$. Such are, for instance, the triplets \eqref{eq:triplet} or \eqref{eq:tri>1}
or the central triplet of the scale \eqref{eq:scale}.
Then the theorem reads as follows.
\index{theorem!KLMN}
\begin{theo}[KLMN theorem]\label{theo:KLMN}
Let $A_{o}$ be a continuous map from  $\H_{1}$ into  $\H_{\ov{1}}$, such that $A_{o}-\lambda I$ is a bijection for some $\lambda\in \RN$ and that
$\ip{f}{A_{o}g}=\ip{A_{o}f} {g}, $ $\forall\,f, g \in\H_{1} $. Then there is a unique self-adjoint operator $A$ in $\H_{0}$
with domain $D(A)=\{ g\in\H_{1}: Ag\in\H_{0} \}\subset\H_{1} $ and $\ip{f}{Ag}= \ip{f}{A_{o}g}, \, \forall\, g\in D(A) $ and $ f \in\H_{1} $.
\entheo

This celebrated theorem (which has already a \pip\ flavor in its \hs\ formulation) can be extended to a \pip\ context \cite[Theor. 3.3.27-3.3.28]{pip-book}.
Thus  we  formulate the question in a \pip\ context
and we assume   that $V$ is a nondegenerate, positive definite \ipip\ with  central \hs\ $V_{o}= V_{\ov{o}}= \H$,  for instance,  {the  LHS $V_{J}$. 
However, since every operator $A \in \mathrm{Op}(V_{J}) $ satisfies the condition $A\taa = A $, there is no room for extensions, only the KLMN approach may be used.  
Thus, in order to obtain a self-adjoint representative in $\H$, by restriction from a larger space, we have to start from a \emph{symmetric operator}  in the \pip\ sense, i.e., an operator $A\in \mathrm{Op}(V_{J})$ that satisfies  $A=A^\times$}
(see Section \ref{subsubsec-sym}). This is the class that can give rise to self-adjoint \emph{restrictions} to $\H$, thanks to the generalized KLMN Theorem \ref{theo:genKLMN2}.
 
Let $A =A^\times\in \mathrm{Op}(V_{J}) $ be a symmetric operator {on  the LHS $V_{J}$.}
Then  $X\in {\sf s}(A)$ if, and only if $X^{-1}\in {\sf s}(A)$,
and this implies $I\in {\sf s}(A)$, by \cite[Cor. 3.3.24]{pip-book} (see Remark \ref{rem72} and Section \ref{subsubsec-sym}).  In addition, since  ${\sf s}(A)$ is invariant under the lattice operations $\cap$ and $+$, it becomes a genuine (involutive) sublattice of $ \J$.
Thus, if there is a metric operator $G \in \M(\H)$, with $G\in {\sf s}(A)$, it follows that $A$ has a bounded representative 
$A\subn{II}$ in $\H$.
Moreover, $A$ fixes all three middle spaces in Fig. \ref{fig:diagram} and, therefore, all nine spaces of the lattice.

This applies, in particular, to all three  spaces in the triplet \eqref{eq:triplet} if $G$ is bounded, or in the triplet   \eqref{eq:tri>1} or   \eqref{eq:tri<1} if $G$ is unbounded.  
Moreover, by the interpolation property (iii) of \cite[Sec. 5.1.2]{pip-book},
$A$  leaves invariant every space $\H_{\alpha}, \alpha\in \ZN \mbox{ or } \RN$, in the scales $V_{\G}$ and $V_{\widetilde \G}$. In other words,
$A$ is a totally regular operator in these \pip s (see Section \ref{subsubsec:regops}), hence, ${\sf s}_{G}(A) = \ZN$ or $\RN$, respectively.

Thus we may state:
\beprop \label{prop61}
 {Every symmetric operator $A\in  {\rm Op}(V_\Sigma)$ such that $G \in {\sf s}(A)$, with $G \in \M(\H)$, 
has a bounded  representative $A\subn{II}$ in $\H$.}
\enprop
There is a sort of converse to the previous statement. Given a closed  unbounded  operator $B$ in $\H$, one may consider the self-adjoint operator
$G:= 1+(B\ha  B)^{1/2}$ and the scale $V_{\G}$  built on the powers of $G^{1/2}$ (this is essentially the only \pip\ one can build intrinsically from $B$ alone). Then $T=G^{-1/2}$ is a bounded metric operator. Hence,  according to 
Proposition \ref{prop510}, $B$ is quasi-Hermitian with respect to $T$ if and only if $A_0= TB$ is symmetric in $\H$. Next, since $D(A_0)$ is dense in $\H$, 
$A_0 $ defines a unique symmetric operator $A=A^\times$ in the scale $V_{\G}$. 

Clearly, the assumption that $G\in {\sf s}(A)$ is too strong for applications, since it implies that $A$ has a \emph{bounded} restriction to $\H$.
Assume instead that $(G^{-1},G)\in {\sf j}(A)$
for some $G \in \M_b(\H)$ with an unbounded inverse. Then, according to  \eqref{eq:triplet}, $\H(G^{-1}) \subset \H \subset \H(G)$  and we can apply the KLMN theorem in its \pip\ version  \cite[Theorem 3.3.27]{pip-book}.
\index{theorem!KLMN!{\sc pip} space version}
\beprop\label{prop-KLMN1}
Given a symmetric operator $A=A^\times$, assume there is a metric operator $G \in \M_b(\H)$ with an unbounded inverse, for which there exists  a $ \lambda \in \RN$ such that $A  - \lambda I$ has
a boundedly invertible representative $(A  - \lambda I)\subn{G G^{-1}}: \H(G^{-1}) \to \H(G).$
Then  $A\subn{G G^{-1}}$ has a unique restriction to a self-adjoint operator  {\sf A} in the
 Hilbert space $\H$, with dense domain $D({\sf A})=\{ \xi \in \H:\, A\xi \in \H\}$.
 In addition,  $\lambda \in \rho({\sf A} )$.
\enprop
 In a nutshell, the argument runs as follows. Let $R\subn{G^{-1}G} = ((A  - \lambda I)\subn{G G^{-1}})^{-1} : \H(G) \to \H(G^{-1})$ be the 
bounded inverse of the   invertible representative \linebreak
$(A  - \lambda I)\subn{G G^{-1}}$.
 Define $R\subn{II}= E\subn{IG^{-1}}R_{G^{-1}G}E\subn{GI}$, which is a restriction of $R\subn{G^{-1}G}$.
(we recall that  $\H(I):= \H$). 
Then, by the assumption,
$R\subn{II}$ is bounded and, by \cite[Lemma 3.3.26]{pip-book},
it has a self-adjoint inverse ${\sf A}-\lambda  I$, which is a restriction of $(A  - \lambda I)\subn{G G^{-1}}$. The rest is obvious. 

\beprop\label{prop-KLMN11}
Given a symmetric operator $A=A^\times$, assume there is a metric operator $G$ such that $A\subn{G G^{-1}}$ has a  restriction to a self-adjoint operator  {\sf A} in the Hilbert space $\H$. Then, if  the natural embedding $ \H(G^{-1}) \to \H(G) $ is compact, the operator  {\sf A} has a purely point spectrum of finite multiplicity, thus $\sigma({\sf A}) = \sigma_p({\sf A}),  \,m_{\sf A}(\lambda_j) <\infty$ for every $\lambda_j\in  \sigma_p({\sf A})$ and $\sigma_c({\sf A})= \emptyset$.
\index{spectrum!point}\enprop
{\bf Proof. }
According to the KLMN Theorem  \ref{theo:genKLMN2},
the resolvent $({\sf A}  - \lambda I)^{-1}$ is  compact if and only if the natural embedding $ \H(G^{-1}) \to \H(G) $ is compact. In that case,  {\sf A} is a self-adjoint operator with compact resolvent, which implies the statements.
\qed
\medskip

Even if we don't have  $(G^{-1},G)\in {\sf j}(A)$ for some $G \in \M_b(\H)$ with an unbounded inverse, we can still obtain a self-adjoint restriction to $\H$, 
 by exploiting Theorem 3.3.28 in \cite{pip-book},  that is, Theorem \ref{theo:genKLMN2}, by restricting the \pip\ to one of the scales $V_{\G}$ and $V_{\widetilde \G}$ built  on
 the powers of $G^{-1/2}$.
  \beprop\label{prop-KLMN2}
Let $V_\G = \{\H_{n}, n\in \ZN\}$ be the Hilbert scale built on the powers of the  operator $G^{-1/2}$, where
$G \in \M_b(\H)$ with $G^{-1}$ unbounded. Given $A=A^\times \in {\rm Op}(V_\G)$,
assume there is a $ \lambda \in \RN$ such that $A  - \lambda I$ has
a boundedly invertible representative $(A  - \lambda I)_{nm}: \H_m \to \H_n$, with $\H_{m} \subset \H_n$. Then the conclusions of Proposition \ref{prop-KLMN1} hold true.
\enprop
According to the proof of \cite[Theorem 3.3.28]{pip-book}, the assumption implies that the operator $R_{mn}= (A_{mn}  - \lambda I_{mn})^{-1}: \H_n \to \H_m$
has a self-adjoint representative $R_{oo}$ in $\H$, which is injective and has dense range. Therefore, its inverse
${\sf A}- \lambda I= R_{oo}^{-1}$, thus also {\sf A} itself, is defined on a dense domain and is self-adjoint.

In the discussion above, we assumed that  $G$ is bounded and $G^{-1}$ unbounded, so that the natural environment is the scale built on the powers of $G^{-1/2}$.
If $G$ is unbounded and $G^{-1}$ bounded, one can perform  the same construction using the powers of  $G^{1/2}$.
If  $G$ is   and $G^{-1}$ are both  unbounded,  we can proceed by taking  {the powers of $(R_G)^{1/2}$ or $(R_{G^{-1}})^{1/2}$, }thus getting the scales around the triplets \eqref{eqtr1} or \eqref{eqtr2}.
Thus globally, we may  state
\index{theorem!KLMN!{\sc pip} space version}
\betheo\label{theo-KLMN3}
Let $V_\G= \{\H_{n}, n\in \ZN\}$ be the Hilbert scale built on the powers of the  operator $G^{\pm1/2}$ or
  { $(R_{G^{\pm 1}})^{-1/2}$,}  depending on the (un)boundedness of $G^{\pm 1} \in \M(\H)$   and let $A=A^\times$ be  a symmetric operator in $V_\G$.

(i)  Assume there is a $ \lambda \in \RN$ such that  $A  - \lambda I$ has
a boundedly invertible representative $(A  - \lambda I)_{nm}: \H_m \to \H_n$, with $\H_{m} \subset \H_n$. 
Then $A_{nm}$  has a unique restriction to a self-adjoint operator  {\sf A} in the
 Hilbert space $\H$, with dense domain $D({\sf A})=\{ \xi \in \H:\, A\xi \in \H\}$.
 In addition,  $\lambda \in \rho({\sf A} )$.
 
(ii) If  the natural embedding $\H_m \to \H_n$  is compact, the operator  {\sf A} has a purely point spectrum of finite multiplicity, thus $\sigma({\sf A}) = \sigma_p({\sf A}),  \,m_{\sf A}(\lambda_j) <\infty$ for every $\lambda_j\in  \sigma_p({\sf A})$ and $\sigma_c({\sf A})= \emptyset$.
\entheo
\index{spectrum!point}
This condition on the natural embeddings is familiar in the theory of topological vector spaces. For instance, the end space of the scale $V_\G$, namely,
$\H_{\infty}(G^{1/2})=\bigcap_{n\in \ZN} \H_n$ (see \eqref{eq:endscale}), is nuclear if, for every $m\in \ZN$, there exists $n\in \ZN$ such that the
natural embedding $\H_m \to \H_n$  is a Hilbert-Schmidt operator (analogous results hold when the embedding is compact).
\medskip

  {At this stage, we do have a self-adjoint restriction ${\sf A}$  of $A$ in $\H$, but we don't know if
 there is any (quasi-)similarity relation between  $A\subn{G G^{-1}}$ or ${\sf A}$ and another operator.}  
 
  {On the contrary,   assume that  $G$ is bounded and $G^{-1}$ unbounded, and that $A$ 
maps $\H(G^{-1})$ continuously into $\H(G)$. Then, following the discussion preceding Proposition \ref{prop62}, there exists $c> 0$ such that
$$
\|G^{1/2}A\subn{GG^{-1}} \xi\| \leq c\|G^{-1/2}\xi\|, \quad \forall \,\xi \in \H(G^{-1}).
$$
This means that
$$
\|G^{1/2}A\subn{GG^{-1}}G^{1/2} \eta\| \leq c\|\eta\|, \quad \forall\, \eta \in \H.
$$
Hence, ${\sf B}:=G^{1/2}A\subn{GG^{-1}}G^{1/2}$ is a bounded operator on $\H$.
However,  the operator $A\subn{GG^{-1}}$ is \emph{not}  quasi-similar to ${\sf B}\in \BH$. Indeed, 
condition ({\sf io$_1$}) imposes that $T = G^{-1/2}$, hence unbounded, but the conditions  ({\sf io$_0$}) and 
({\sf io$_2$}) cannot be satisfied.}

\subsection{Semi-similarity}
\label{subsect_72}

So far we have considered only the case of one metric operator $G$ in relation to $A$. Assume now we take two different metric operators 
$G_{1}, G_{2}\in \M(\H)$.
What can be said concerning $A$ if it maps $\H(G_{1})$ into $ \H(G_{2})$?
 
\index{operator!semi-similar}
One possibility is to introduce, following \cite{pip-metric}, a notion slightly more general than quasi-similarity, called \emph{semi-similarity}.
\bedefi
Let $\H, \K_{1}$ and $\K_{2}$ be three Hilbert spaces,  $A$ a closed, densely defined operator from $\K_{1}$ to  $\K_{2}$, $B$ a closed, densely defined operator on $\H$. Then $A$ is said to be
\emph{semi-similar} to $B$, which we denote by $A\dashvv B$, if there exist two bounded operators  $T:\K_{1}\to \H$ and $S:\K_{2}\to \H$ such that (see Fig. \ref{fig:semisim}):
\begin{itemize}
\vspace*{-2mm}\item[(i)] $T:D(A)\to D(B)$;
\vspace*{-2mm}\item[(ii)] $BT\xi = SA\xi, \; \forall\, \xi \in D(A)$.
\end{itemize}
The pair $(T,S)$ is called an \emph{intertwining couple}.
\findefi
Of course, if  $\K_{1}=\K_{2}$ and $S=T$, we recover the notion of quasi-similarity and $A\dashv B$   {(with a bounded intertwining operator).}

\begin{figure}[t]
\centering \setlength{\unitlength}{0.5cm}
\begin{picture}(5,5 )

\put(4,2.4){\begin{picture}(5,5) \thicklines

 \put(-1.6,2){\vector(-2,-1){3.2}}
 \put(-1.6,-2){\vector(-2,1){3.2}}
 \put(-1,1.8){\vector(0,-1){3.2}}

 \put(-6,0){\makebox(0,0){ $\H $}}
\put(-1,2.5){\makebox(0,0){ $\K_{1} $}}
\put(-1,-2.4){\makebox(0,0){ $\K_{2}$}}

 \put(-0.4,0){\makebox(0,0){$A$}}
\put(-3.6,1.8){\makebox(0,0){ $T$}}
\put(-3.6,-2){\makebox(0,0){ $S$}}
\end{picture}}
\end{picture}
\caption{\label{fig:semisim}The semi-similarity scheme {(from  \cite{pip-metric})}.}
\end{figure}
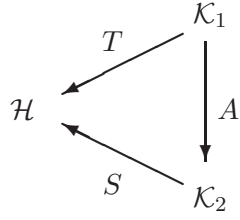

Now we come back to the case envisaged above: $A:\H(G_{1}) \to \H(G_{2})$ continuously, for the two metric operators $G_{1}, G_{2}\in \M_b(\H)$, but $A$ is \emph{not} supposed to be symmetric.
Under this assumption, we essentially recover the previous situation.  Since $A\subn{G_{2}G_{1}}$ is a bounded operator from $\H(G_{1})$ into $\H(G_{2})$,  there exists $c>0$ such that
$$
\|G_{2}^{1/2}A\subn{G_{2}G_{1}} \xi\| \leq   c\|G_{1}^{1/2}\xi\|, \quad \forall\, \xi \in \H(G_{1)}.
$$
This means that
$$
\|G_{2}^{1/2}A\subn{G_{2}G_{1}}G_{1}^{-1/2} \eta\| \leq  c \|\eta\|, \quad \forall\, \eta \in \H.
$$
Hence, ${\sf B}:=G_{2}^{1/2}A\subn{G_{2}G_{1}}G_{1}^{-1/2} $ is bounded in $\H$.
Then the operator $A\subn{G_{2}G_{1}}$ is \emph{semi-similar} to ${\sf B}\in \BH$, that is, $A\subn{G_{2}G_{1}}\dashvv {\sf B} $,
with respect to the intertwining couple  $T=G_{1}^{1/2}, S=G_{2}^{1/2} $.

  {Next we take   a   symmetric operator $A=A^\times\in   {\rm Op}(V_\I)$, where $V_\I$ is any LHS (or \pip) containing $\H(G_{1}) $ and $\H(G_{2})$. }
Then  $A:\H(G_{1}) \to \H(G_{2})$ continuously implies  $A:\H(G_{2}^{-1}) \to \H(G_{1}^{-1})$ continuously as well.
Assume that $G_{1} \preceq G_{2}$, that is,  $\H(G_{1}) \subset \H(G_{2}) $. This yields the following situation:
\be\label{eq:quintuplet}
\H(G_{2}^{-1}) \;\subset\;  \H(G_{1}^{-1}) \;\subset\;  \H \;\subset\;  \H(G_{1})\;\subset\;  \H(G_{2}).
\en
Therefore, since $\H(G_{1}^{-1}) \hookrightarrow \H(G_{2})$, the operator $A$ maps $\H(G_{2}^{-1})$ continuously into $\H(G_{2})$, that is,
we are back to the situation of Proposition \ref{prop-KLMN1} and we can state:

\beprop\label{prop-KLMN4}
Given a symmetric operator $A=A^\times\in  {\rm Op}(V_\I)$, assume there exists two metric operators $G_{1},G_{2}\in \M_b(\H)$ such that
$G_{1}\preceq G_{2}$ and  $(G_{1},G_{2})\in {\sf j}(A)$.
Assume   there exists  a $ \lambda \in \RN$ such that $A  - \lambda I$ has  an invertible representative
$(A  - \lambda I)_{G_{2} G_{2}^{-1}}: \H(G_{2}^{-1}) \to \H(G_{2}).$
Then there exists a unique restriction of $A\subn{G_{2} G_{2}^{-1}}$ to a self-adjoint operator  {\sf A} in the   Hilbert space $\H$. 
The number $\lambda$  does not belong to the spectrum of  {\sf A}.
  The dense domain of  {\sf A}  is  given by $D({\sf A})=\{ \xi \in \H:\, A\xi \in \H\}$.
The resolvent $({\sf A}  - \lambda I)^{-1}$ is  compact (trace class, etc.) if and only if the natural embedding $ \H(G_1) \to \H(G_2) $ is compact (trace class, etc.).
\enprop
The analysis may be extended to the three other cases, assuming again that $A:\H(G_{1}) \to \H(G_{2})$:
\begin{enumerate}
\item    {$G_1$ unbounded, $G_2$ bounded: then
$$
\H(G_{1}) \; \subset\; \H  \; \subset\; \H(G_{2})
$$
and $A$ maps the small space into the large one. Then the KLMN theorem applies.}
\item    { $G_1$ and $G_2$ both  unbounded, with $\H(G_{1}) \subset \H(G_{2})$;  then we are back to the first situation, with a bounded representative $(A  - \lambda I)_{G_{1}^{-1} G_{1}}: \H(G_{1}) \to \H(G_{1}^{-1})$
and the KLMN theorem applies.}
\item   {$G_1$ bounded, $G_2$ unbounded: then
$$
\H(G_{2}) \; \subset\; \H \; \subset\; \H(G_{1}) \quad \mbox{and}
\quad \H(G_{1}^{-1}) \; \subset\; \H  \; \subset\; \H(G_{2}^{-1}),  
$$
so that, in both cases,  $A$ maps the large  space into the small  one. Then the  KLMN theorem does \emph{not} apply.}\footnote{
 The statement of Case 3. given in \cite[Sec.5.6]{pip-metric} is wrong. See Corrigendum.}
\end{enumerate}
  {In conclusion, if $A=A^\times$ is  symmetric and $(G_{1},G_{2}) \in {\sf j}(A)$, with 
$G_{1} \!\preceq\! G_{2}$, the KLMN theorem applies and yields a self-adjoint restriction in $\H$, in two cases: either $G_{1}$ is unbounded, or $G_1$ and $G_2$ are both  bounded.}

\section{The case of pseudo-Hermitian Hamiltonians}
\label{sect_psH}

\index{operator!metric}
\index{pseudo-Hermitian Hamiltonian}
Metric operators appear routinely in the so-called pseudo-Hermitian quantum mechanics \cite{bender}, but in general only bounded ones are considered. In some recent work \cite{bag-zno,kretsch,mosta2}, however, unbounded metric operators have been discussed. The question is, how do these operators fit in the present formalism?

Following the argument of \cite{mosta2}, the starting point   is a reference \hs\ $\H$ and a quasi-Hermitian operator \footnote{
The author of  \cite{mosta2} calls this a $G$-pseudo-Hermitian operator, but in fact it is simply a quasi-Hermitian operator, in the original sense of Dieudonn\'e \cite{dieudonne}, but unbounded.}
$H$ on $\H$, which means  there exists an unbounded metric operator $G$  satisfying the relation
\be\label{eq:psH}
H \ha  G = G H.
\en
This operator $H$ is the putative non-self-adjoint (but $\P\T$-symmetric) Hamiltonian of a quantum system.

In the relation \eqref{eq:psH}, the two operators are assumed to have the same  domain, $D(H \ha  G)=D(GH)$,
which is supposed to be dense. This condition is not necessary, however, if we assume that $H$ is quasi-Hermitian in the sense of Definition \ref{quasihermitian}.
This means that $D(H) \subset D(G)$ and 
\be\label{eq_quasihermitian4}
\ip{H\xi}{G\eta}= \ip{G\xi}{H\eta},  \;\; \forall\, \xi, \eta \in \D(H).
\en 

 If $G$ is bounded, we get $H\dashv H\ha$ and then  Proposition \ref{prop_292} implies that $G^{1/2} H G^{-1/2}$ is self-adjoint, 
 {i.e., $H$ is quasi-self-adjoint}.

On the other hand, If $G$ is unbounded and if we assume that $H$ is strictly quasi-Hermitian, we still have $H\dashv H\ha$, but we cannot conclude.
However, if in addition
$G^{-1}$ is bounded, we get $G^{-1}H \ha  G\eta = H\eta, \; \forall \,\eta\in \D(H)$, which is a more restrictive form of similarity.}

\medskip

Finally, let us assume   that the quasi-Hermitian operator $H$  possesses a (large) set of vectors,  
$\D_G^\omega(H)$, which are  analytic in the norm  $\norm{G}{\cdot}$ and are contained in $D(G)$  \cite{barut-racz,nelson}.
This means that every vector  $\phi\in\D_G^\omega(H)$ satisfies the relation
$$
\sum_{n=0}^{\infty}\frac{ \norm{G}{H^n \phi}}{n!} \, t^n < \infty, \mbox{ for some }  t\in \RN.
$$
This implies that every $\phi\in\D_G^\omega(H)$ satisfies the relation  $H^n \phi \in D(G^{1/2}), \, \forall\, n = 0,1,\ldots$.
Thus one has
\be\label{eq:analytic}
\D_G^\omega(H) \subset   D(H) \subset D(G)  \subset D(G^{1/2}) \subset \H.
\en
Then the construction given in \cite[Sec.6]{pip-metric} can be performed

Endow  $\D_G^\omega(H)$ with the norm $\norm{G}{\cdot}$ and take  the completion $\H_G$, which is a closed subspace of $\H(G)$, as defined in Section \ref{sect_termin}. An immediate calculation then yields
$$
\ip{\phi}{H\psi}_G = \ip{H\phi}{\psi}_G, \; \forall\, \phi,\psi\in \D_G^\omega(H),
$$
that is, $H$ is a densely defined symmetric operator in $\H_G$. Since it has a dense set of analytic vectors, it is essentially self-adjoint, by Nelson's theorem \cite{barut-racz,nelson},\cite[Theor.7.16]{schm}, hence its closure $\ov H$ is a self-adjoint operator in $\H_G$.  The pair $(\H_G, \ov H)$ is then interpreted as the physical quantum system.

{Next, by definition, $W_\D:= G^{1/2}\uph \D_G^\omega(H)$ is isometric from  $\D_G^\omega(H)$ into $\H$, hence it extends to an isometry 
$W =\ov{W_\D}: \H_G \to \H$.  {The range of $W$}  is a closed subspace of $\H$,  denoted  $\H_{\rm phys}$, and the operator $  W$ is unitary  from $\H_G$ onto $\H_{\rm phys}$. Therefore, the operator $h=  W\, \ov H\,W^{-1}$ is self-adjoint in $\H_{\rm phys}$. This operator $h$ is interpreted as the genuine Hamiltonian of the system, acting in the physical \hs\  $\H_{\rm phys}$.}

{ Things simplify if  {$\D_G^\omega(H)$ is dense} in $\H$. 
Then $ W(\D_G^\omega(H))$ is also dense, $\H_G= \H(G)$, 
 $\H_{\rm phys}= \H$ and $W = G^{1/2}$ is unitary from $\H(G)$ onto $ \H$.
Also, if $G$ is bounded, it is sufficient to  assume that the vectors in $\D_G^\omega(H)$ are analytic with respect to the original norm of $\H$.
But then $H$ has a dense set of analytic vectors, so that it is essentially self-adjoint  in $\H$, by Nelson's theorem again.
Note that we are back to the situation of Proposition \ref{prop_29}, since then $\H(G) = \H$, with equivalent, but different norms.

Now, every eigenvector of an operator is automatically analytic, hence this construction generalizes that of \cite{mosta2}.
This applies, for instance, to the example given there, namely, the $\P\T$-symmetric operator
$H=\frac12 (p-i\alpha)^2 + \frac12 \omega^2 x^2$  in $\H= L^2(\RN)$, for any $\alpha\in \RN$, which has an orthonormal basis of eigenvectors.

\subsection{An example}
\label{subsec-example}

A beautiful example of the situation just analyzed has been given recently by Samsonov \cite{samsonov}, namely, the second derivative on the positive half-line
(this example stems from Schwartz \cite{schwartz}):
\be\label{eq-H}
H = -\frac{\ud ^2}{\ud  x^2}, \quad x \geq 0,
\en
with domain
\be\label{eq-DH}
D(H) = \{\xi \in L^2(0,\infty) : \xi ''  \in L^2(0,\infty), \xi'(0) + (d+i b)\xi(0) = 0\}.
\en
For $d<0$, this operator has a purely continuous spectrum. Its adjoint $H\ha$ is given again by \eqref{eq-H}, on the domain
$D(H\ha)$ defined as in \eqref{eq-DH}, with $b$ replaced by $-b$.

Next introduce the unbounded operator
\be
G = -\frac{\ud ^2}{\ud  x^2} -2 ib \frac{\ud }{\ud  x} + d^2 +b^2,
\en
on the domain $D(G) = D(H)$. Then a direct calculation shows that $G$ is self-adjoint, strictly positive and invertible, i.e., it is a metric operator. Since its spectrum is 
$\sigma(G) = \sigma_c(G) = [d^2, \infty)$, it follows that  $G^{-1}$ is bounded.

Since both $H$ and $G$ are second order differential operators, an element of the domain $D(GH)$ should have a square integrable fourth derivative. Hence one defines
\be\label{eq:tildeD}
\widetilde{D}(H) = \{\xi \in D(H) : \xi^{\rm (iv)} \in L^2(0,\infty)\} \subset D(H).
\en
and $ \widetilde{H} = H \uph \widetilde{D}(H)$.
Then the analysis of \cite{samsonov} yields the following results:
\bee
\vspace*{-2mm}\item[(i)] $H$ is quasi-Hermitian in the sense of Definition \ref{quasihermitian}, that is, it satisfies the relation \eqref{eq_quasihermitian4}
on $ D(G)=  D(H)$.

\vspace*{-2mm}\item[(ii)] $G$ maps $\widetilde{D}(H) $ into $D(H\ha)$.

\vspace*{-2mm}\item[(iii)] $H$ is quasi-Hermitian in the sense of Dieudonn\'e, that is, $GH = H\ha G$ on  the  dense domain $\widetilde{D}(H)$
(we have to restrict ourselves to $\widetilde{D}(H)$ because of the requirement on the fourth derivative).

\vspace*{-2mm}\item[(iv)] The operator $h = G^{1/2} H G^{-1/2} = G^{-1/2} H\ha G^{1/2}$ is self-adjoint on the domain 
$D(h) = \{\eta=  G^{1/2}\xi, \, \xi \in\eta \}$.
\ene
In conclusion, by (i) and (ii), we   get $\widetilde{H}\dashv H\ha$. 

In fact, one can use as metric operator $T:=G^{-1}$, which is bounded, with unbounded inverse $T^{-1}=G$, a more standard situation. Writing $H = H_\lambda$, with 
$\lambda = (d+i b)$, we have that $H\ha = H_{\ov{\lambda}}$ and the relation $GH = H\ha G$ becomes 
\be\label{eq:dieu}
GH_\lambda =H_{\ov{\lambda}} G
\en
 which yields a symmetry between  $H_\lambda$ and $H_{\ov{\lambda}}$. Multiplying both sides of \eqref{eq:dieu} from the left and from the right by $T^{-1}=G$,
 we get 
 $$
 H_\lambda T \eta= T H_{\ov{\lambda}}\,\eta, \; \forall \, \eta\in \widetilde{D}(H_{\ov{\lambda}}),
 $$
where  $\widetilde{D}(H_{\ov{\lambda}})$ is defined as in \eqref{eq:tildeD}. Noting that $T: \widetilde{D}(H_{\ov{\lambda}}) \to  D(H_\lambda) $, we can conclude as before that 
$\widetilde{H_{\ov{\lambda}}} \dashv H_\lambda$, i.e. $\widetilde{H\ha} \dashv H$, with the bounded intertwining operator $T=G^{-1}$.
The problem, of course, is that we don't know the operator $T$ explicitly. Being the inverse of the differential operator $G$, it is presumably an integral operator.  {Therefore, this second solution, albeit more standard,  is of little use.}

\section{Conclusion}

In this chapter, we have introduced  several generalizations of similarity between operators and we have obtained some results on the preservation of spectral properties under quasi-similarity, but only with a bounded metric operator with unbounded inverse. However, 
we have seen that the consideration of unbounded metric operators leads naturally to the formalism of \pip s.  
And indeed it turns out that exploiting the connection between metric operators and \pip s  does in certain cases improve the quasi-similarity of operators.
More precisely, given a symmetric operator $A=A^\times$ in a \pip\ with central \hs\ $\H$, one can apply the KLMN theorem, which may  yield a self-adjoint restriction  of $A$ in $\H$. Then additional quasi-similarity relations follow.

 Of course, these results are only a first step, many open problems subsist. In view of the applications, notably
in pseudo-Hermitian QM, the most crucial ones concern the behavior of spectral properties under some generalized similarity with an unbounded metric operator. In the same vein, there are few results about the spectral properties of self-adjoint operators derived in a \pip\ context from a symmetric operator via the KLMN theorem. Then, of course, one should investigate the connection between these two types of problems. In particular, one needs to investigate in more details the spectral properties of symmetric operators in a \pip, and in a LHS in the first place.
Research in this direction is in progress. Preliminary results may be found in \cite{pip-ops}, following the approach developed  for the case of a rigged Hilbert space  by Bellomonte, di Bella and one of us  \cite{bello-db-trap}.

\appendix{Partial inner product spaces}

\section{PIP-spaces    and   indexed  PIP-spaces}
\label{subsec-pip}

For the convenience of the reader, we have collected here the main features of partial inner product spaces and operators on them, keeping only what is needed for reading the chapter. 
Further information may be found in our monograph \cite{pip-book} or our review paper \cite{at-AMP}.

The general framework is that of a \pip\ $V$,  corresponding to the linear compatibility $\com$, that is,  
a symmetric binary relation $f \com g$  which preserves linearity.
We   call \emph{assaying subspace} of $V$ a  subspace $S$ such that $S^{\#\#} = S$ and 
we denote by ${\F}(V,\com)$   the family of all assaying subspaces of $V$, ordered  by inclusion. 
The assaying subspaces are denoted by $V_{r}, V_{q} , \ldots $ and the index set is $F$.  By definition, $q \leq r$ if and only if $V_{q} \subseteq V_{r}$.
 Thus we may write
\begin{equation}
\label{eq:gener2}
f\com g \; \Leftrightarrow \; \exists \;r \in F  \mbox{ such that } f \in V_{r}, g \in V_{\overline{r}}\,.
\en

General considerations \cite{birkhoff}  imply that the family   ${\F}(V,\com):= \{ V_r, r\in {F} \}$, ordered  by
inclusion, is  a complete involutive lattice, {i.e.}, it is stable under the following operations, arbitrarily iterated:
\medskip

\begin{tabular}{lccl}
. involution: &$V_r$                     & $\!\!\!\leftrightarrow\!\!\!$& $V_{\ov{r}}=(V_r)^{\#},$\\
. infimum:   & $V_{p \wedge q}$  &$\!\!\! :=\!\!\!$                & $V_p \wedge V_q = V_p \cap V_q,$   \qquad $(p,q,r \in {F})$\\
. supremum: & $V_{p \vee q}$     &$\!\!\! :=\!\!\!$               &$ V_p \vee V_q = (V_p + V_q)^{\#\#}$.
\end{tabular}
\\[2mm]
\noi The smallest element  of $ {\F }(V,\com)$ is  $V\co = \bigcap_r V_r $ and the greatest element   is $V = \bigcup_r V_r$.  

By definition, the index set  ${F}$    is  also a complete involutive  lattice; for instance,  
 $$
 (V_{p \wedge q})\co = V_{\ov{p \wedge q}} 
=  V_{\ov {p} \vee \ov {q}} = V_{\ov {p}} \vee V_{\ov {q}}.
$$

Given a vector space $V$ equipped with a linear compatibility $\com $, a \emph{partial inner product}   on   $(V, \,{\com})$ is a
   Hermitian form  $\ip{\cdot}{\cdot}$ defined exactly on compatible pairs of vectors. 
A \emph{partial inner product space}  (\pip)  is a  vector space $V$ equipped with a linear compatibility  and a partial inner product.
\index{{\sc pip} space (partial inner product space)}

From now on, we will assume that our \pip\  $(V, \com, \ip{\cdot}{\cdot})$ is \emph{nondegenerate}, 
that is,    $\ip{f}{g} = 0   $ for all $ f \in  V^{\#} $ implies $ g = 0$.  As a consequence,  $(V\co, V)$ and  
 every couple $(V_r , V_{\ov r} ), \,  r\in {F}, $  are a  dual pair in the sense of topological vector spaces \cite{kothe}. 
Next we assume that every $V_{r}$ carries  its Mackey topology $\tau(V_{r},V_{\ov{r}})$, so that its conjugate dual is $(V_r)^\times = V_{\ov {r}}, \; \forall\, r\in {F} $.
Then,   $r<s$ implies $V_r \subset V_s$, and the embedding operator $E_{sr}: V_r \to V_s$  is continuous and has dense range. In particular, $V\co$ is dense in every $V_{r}$.
 In the sequel, we also assume the partial inner product to be positive definite, $\ip{f}{f}>0$ whenever $f\neq0$.

As a matter of fact,   the whole structure can be reconstructed from a fairly
 small subset of $\F$, namely, a \emph{generating}    involutive sublattice $\J$   of $\F(V, \com)$, indexed by $J$, which means that
\be\label{eq:gener}
f\com g \; \Leftrightarrow \; \exists \;r \in J \mbox{ such that } f \in V_{r}, g \in V_{\overline{r}}\,.
\en
The resulting structure is called  an \emph{\ipip} and denoted simply by $V_{J} := (V, \J, \ip{\cdot}{\cdot}) $.

Then an \ipip\  $V_{J}$ is said to be:
\bei
\vspace*{-1mm}\item [(i)]
\emph{additive}, if $V_{p \vee q} = V_{p} + V_{q}, \; \forall \, p, q \in J$.

\vspace*{-2mm}\item [(ii)] \emph {projective}  if  
$V_{p \wedge q}|_{\tau} \simeq (V_{p} \cap V_{q})_{\mathrm{proj}}, \;\forall \, p, q \in J ;$
here $V_{p \wedge q}|_{\tau}$ denotes $V_{p \wedge q}$ equipped with the Mackey topology $\tau (V_{p \wedge q}, V_{\ov{p} \vee \ov{q}})$, 
the r.h.s. denotes $V_{p} \cap V_{q}$ with the topology of the projective limit from $V_{p}$ and $V_{q}$ and $\simeq $ denotes an isomorphism of locally convex
 topological spaces.

\eni
For practical applications, it is essentially sufficient to restrict oneself to the case of an \ipip\ satisfying the following conditions:
\bei
\vspace*{-2mm}\item [(i)]
 every $V_{r}, r\in J$, is a \hs\ or a reflexive Banach space, so that the Mackey topology $\tau(V_{r},V_{\ov{r}})$ coincides with the norm topology;

\vspace*{-2mm}\item [(ii)]   there is a unique self-dual, Hilbert,  assaying subspace $V_{o} =V_{\overline{o}}$.
\eni

\index{{\sc pip} space (partial inner product space)!indexed}
\index{{\sc pip} space (partial inner product space)!LHS}
\index{{\sc pip} space (partial inner product space)!LBS}
\index{lattice of Hilbert spaces (LHS)}
\index{lattice of Banach spaces (LBS)}
\noi In that case, the \emph{\ipip}  $V_{J} := (V, \J, \ip{\cdot}{\cdot}) $ is  called, respectively, 
a  \emph{lattice of \hs s} (LHS)  or a  \emph{lattice of Banach spaces} (LBS). 
 {We refer to \cite{pip-book} for more precise definitions, including explicit requirements on norms. In particular, the partial inner product $\ip{\cdot}{\cdot}$ coincides with the inner product of $V_o$ on the latter.}
The important facts here are that 
\bei
\vspace*{-2mm}\item [(i)] Every projective \ipip\  is additive.

\vspace*{-2mm}\item [(ii)]  A  LBS or a LHS is projective if and only if it is additive.
\eni
\vspace*{-2mm}
Note that $V\co, V $ themselves usually do \emph{not} belong to the family $\{V_{r}, \,r\in J\}$, but they can be recovered  as
$$
V\co = \bigcap _{r\in J}V_{r}, \quad V =\sum_{r\in J}V_{r}.
$$ 
 A standard, albeit trivial,  example is that of a Rigged Hilbert space (RHS) $\Phi \subset \H \subset \Phi\co$
(it is trivial because the lattice $\F$ contains only three elements). One should note that the construction of a RHS 
from a directed family of \hs s,  via projective and inductive limits, 
has been investigated recently by Bellomonte and Trapani \cite{bell-trap}. Similar constructions, in the language of categories, may be found in 
the work of Mityagin and Shvarts  \cite{mityagin} and that of Semadeni and Zidenberg \cite{semadeni}.
\medskip

Let us give some concrete examples.
\smallskip

\noi{\sl (i)   Sequence spaces }
\smallskip

Let  $V $ be  the space   $\omega$    of \emph{all} complex sequences $x = (x_n)$  and define on it (i)
a  compatibility relation by  $x {\com} y \Leftrightarrow \sum_{n=1}^\infty  |x_n \, y_n | < \infty$; (ii)
a partial inner product $ \ip{x}{y} = \sum_{n=1}^\infty  \overline{x_n} \, y_n $.
Then   $\omega ^{\#} =  \varphi $ , the space of   finite sequences, and
the complete lattice ${\F}(\omega,{\com})$ consists of K\"{o}the's perfect sequence spaces \cite[\S\,30]{kothe}.
  {Among these, a nice example is the lattice of  the so-called     $\ell_{\phi}$ spaces associated to symmetric norming functions or, more generally, Banach sequence ideals
discussed in \cite[Sec.4.3.2]{pip-book} and previously in \cite[\S\,6]{mityagin}
(in this example, the extreme spaces are, respectively, $\ell^1$ and $\ell^\infty$).}

\medskip

\noi{\sl (ii) Spaces of locally integrable functions}
\smallskip

Let  $V $ be $ L^1_{\rm loc}(\RN, \ud x)$, the space of Lebesgue measurable functions, integrable over compact subsets, and define 
a compatibility relation on it  by 
$$
f \com g\Leftrightarrow \int_{\RN} |f(x)g(x)|  \ud x < \infty
$$
and a partial inner product  given by $ \ip{f}{g} = \int_{\RN} \overline{f(x)} g(x)   \ud x$.
Then one gets $V^{\#} = $ \\
$L^\infty_{\rm c}(\RN, \ud x)$, the space of   bounded measurable functions of compact support.
The complete lattice ${\F }(L^1_{\rm loc},{\com})$ consists of the so-called K\"{o}the function spaces  \cite{dieu,goes}.
  {Here again, normed ideals of measurable functions in $L^1([0,1], \ud x)$ are described in \cite[\S\,8]{mityagin}.}
\medskip

\section{Operators on \ipip s}
\label{subsec:oper}

\index{operator!on {\sc pip}  space}
Let $V_{J}$ and $Y_{ K}$ be two nondegenerate \ipip s (in particular, two LHSs or LBSs). Then  an \emph{operator} from $V_J$  to $Y_{ K}$ is a map
from a subset $\D (A) \subset V$ into $Y$, such that
\smallskip

(i) $\D(A) = \bigcup_{q\in {\sf d}(A)} V_q$, where ${\sf d}(A)$ is a nonempty subset of $J$;
\smallskip

(ii)  For every $r \in  {\sf d}(A )$, there exists $u\in K$ such that the restriction of $A$ to $V_{r}$ is a continuous linear map into $Y_{u}$ (we denote this restriction by $A_{ur})$;
\smallskip

(iii) $A$ has no proper extension satisfying (i) and (ii).
\medskip

\noi We denote by Op$(V_J,Y_K)$  the set of all operators from  $V_J$ to $Y_{K}$ and, in particular, $ \mathrm{Op}(V_J) : = \mathrm{Op}(V_J,V_J)$.
 The continuous linear operator $A_{ur}: V_r \to Y_{u}$ is called a \emph{representative} of $A$.
The properties of $A$ are conveniently described
 by the set ${\sf j}(A)$ of all pairs $ (r,u )\in  J\times K$ such that $A$ maps $V_{r}$ continuously into $Y_{u}$
   Thus the operator $A$ may be identified with   the collection of its representatives,
\be\label{eqj(A)}
A \simeq \{ A_{ur}: V_{r} \to Y_{u} : (r,u) \in  {\sf j}(A)\}.
\en
 We will also need the following sets:
\vspace*{-2mm}\begin{align*}
{\sf d}(A) &= \{ r \in J : \mbox{there is a } \,   u \; \mbox{such that}\; A_{ur} \;\mbox{exists} \},
\\
{\sf i}(A) &= \{ u \in K : \mbox{there is a } \, r \; \mbox{such that}\; A_{ur} \;\mbox{exists} \}.
\end{align*}
\\[-4mm]
The following properties are immediate:
\bei
\vspace*{-2mm}
\item [{\bf .}]   
${\sf d}(A)$ is an initial subset of $J$:  if $r \in {\sf d}(A)$ and $r' < r$, then $r \in {\sf d}(A)$, and $A_{ur'} = A_{ur}E_{rr'}$,
 where  $E_{rr'}$ is a representative of the unit operator.  

\vspace*{-2mm}\item [{\bf .}]   
${\sf i}(A)$ is a final subset of $K$: if $u \in {\sf i}(A)$ and $u' > u$, then $u' \in {\sf i}(A)$ and $A_{u'r} = E_{u'u} A_{ur}$.
\eni
In the case of an operator $A\in \mathrm{Op}(V_J)$, the diagonal of $J\times J$ plays a special role. Hence the following set is useful
 \be\label{eqs(A)}
{\sf s}(A)=\{r \in J :  (r,r)\in {\sf j}(A) \}.
\en
This set can be conveniently used to describe operators similar or quasi-similar to some representative of $A$.
From the definitions \eqref{eq:lattice}, it is clear that  the set ${\sf s}(A)$ is invariant under the lattice operations $\cap$ and $+$.

Although an operator may be identified with a separately continous  sesquilinear form on $V^\# \times V^\#$,
it is more useful to keep also the \emph{algebraic operations} on operators, namely:
\index{operator!on {\sc pip}  space!adjoint}
 \bei
\vspace*{-1mm}\item[(i)] \emph{ Adjoint:}
every $A \in\mathrm{Op}(V_J,Y_K)$ has a {unique} adjoint $A\ta \in \mathrm {Op}(Y_K,V_J)$, defined by
\be\label{eq:adjoint}
\ip {A\ta y} {x} = \ip  {y} { Ax}   , \;\mathrm {for}\,  x \in V_r, \, r \in{\sf d}(A) \;\mathrm {and }\;\, y \in Y_{\ov{u}}, \, u \in{\sf i}(A),
\en
that is,    $(A\ta)_{\ov{r}\ov{u}} = (A_{ur})' $, where $(A_{ur})': Y_{\ov{u}} \to  V_{\ov{r}}$  is the  adjoint map  of $A_{ur}$.
Furthermore,  one has $A\ta{}\ta = A, $ for every $ A \in {\rm Op}(V_J,Y_K)$: no extension is allowed, by the maximality condition (iii)  of the definition.

\item[(ii)] 
\emph{Partial multiplication:}
Let $V_J$, $W_L$, and $Y_K$  be nondegenerate \ipip s (some, or all, may coincide).
 Given  two operators $A \in   \mathrm{Op}(V_J,W_L)$ and  $B \in   \mathrm{Op}(W_L,Y_K)$, we say that the product $BA$ is defined if and only  if
there is a $t \in{\sf i}(A) \cap{\sf d}(B)$, that is, if and only if   there is continuous factorization through some $W_t$:
\be\label{eq:mult}
V_r \; \stackrel{A}{\rightarrow} \; W_t \; \stackrel{B}{\rightarrow} \; Y_u , \,\mbox{{i.e.},} \quad  (BA)_{ur} = B_{ut} A_{tr}, \,\mbox{ for some } \;
r \in{\sf d}(A) , u\in {\sf i}(B).
\en
\eni
Concerning the adjoint, we note that  ${\sf j}(A\ta) = {\sf j}\ta(A):=\{(\ov{u},\ov{r}):  (r,u)\in {\sf j}(A)\} \subset J \times K$.
If $V=Y, \,{\sf j}(A\ta)$ is obtained by reflecting ${\sf j}(A)$  with respect to the anti-diagonal  $\{(r,\ov{r}), r \in J\}$.
In particular, if $(r,\ov{r})\in {\sf j}(A)$, then $(r,\ov{r})\in {\sf j}(A\ta)$ as well.
 Notice also that
${\sf s}(A^\times)=\{\ov{r} : r \in{\sf s}(A)\} .$

\subsection{Symmetric operators}
\label{subsubsec-sym}

\index{operator!on {\sc pip}  space!symmetric}
In Section \ref{subsect_71} we have discussed the problem of  showing that a given symmetric operator $H$  in a \hs\ $\H$ is self-adjoint.  
 The standard technique is to  use   quadratic forms 
(the Friedrichs extension) or  von Neumann's theory of self-adjoint extensions \cite{gitman,schm}.

However, since every operator $A \in \mathrm{Op}(V_{J}) $ satisfies the condition $A\taa = A $, there is no room for extensions. Thus the only notion we have at our disposal  is that of 
\emph{symmetric operator},  in the \pip\ sense, namely, an operator $A\in \mathrm{Op}(V_{J})$ such that $A=A\ta$.
If $A$ is symmetric, the set ${\sf j}(A)$  is symmetric with respect to the anti-diagonal  $\{(r,\ov{r}), r \in J\}$.
Now, if $r\in {\sf s}(A)$, then $\ov{r}\in {\sf s}(A)$ as well, hence by interpolation, $o\in {\sf s}(A)$, that is, $A$ has a bounded representative $A_{oo}: \H \to \H$
(\cite[Cor. 3.3.24]{pip-book}, reformulated as Proposition \ref{prop61}).

However, one may also  ask  whether $A$ has \emph{restrictions} 
 that are self-adjoint in $\H$.  The answer is given essentially by the KLMN theorem, which 
  can be extended to a \pip\ context \cite[Theor. 3.3.27-3.3.28]{pip-book}.
The most general version, adapted to a LHS, reads as follows.
\index{theorem!KLMN!{\sc pip} space version}
\begin{theo}[Generalized KLMN theorem] \label{theo:genKLMN2}
Let $V_{J}$ be a LHS with a central \hs\ $\H$ and  $A= A\ta\in \mathrm{Op} (V_{J})$.
Assume   there exists a $ \lambda \in \RN$ such that $A  - \lambda $ has 
a (boundedly)  invertible representative $A_{sr} - \lambda E_{sr} : V_{r} \to V_{s}$,  where   $V_{r} \subseteq V_{s}$.
Then there exists a unique restriction of $A_{sr}$
 to a self-adjoint operator  {\sf A} in the Hilbert space $\H$. The number $\lambda$  does not belong to the spectrum of  {\sf A}.
  The domain of  {\sf A}  is obtained by eliminating from $V_{r}  $ exactly the vectors $f$ that are mapped by $A_{sr}$ beyond $\H$ 
  (i.e., satisfy $A_{sr} f \notin \H $).
The resolvent $({\sf A}  - \lambda)^{-1}$ is  compact (trace class, etc.) if and only if the natural embedding $ E_{sr} : V_{r} \to V_{s} $ is compact (trace class, etc.).

\entheo

\subsection{Regular operators, morphisms and projections}
\label{subsubsec:regops}

\index{operator!on {\sc pip}  space!regular}
Besides symmetric operators, other classes of operators on \pip s may be defined.
First, an operator $A\in   \mathrm{Op}(V_{J},Y_{K}) $ is called \emph{regular} if ${\sf d}(A)= J$ and   ${\sf i}(A)= K$ or, equivalently, if  $ A:V\co\to Y\co \mbox{ and } A:V\to Y$ continuously for the respective Mackey topologies.
This notion depends only on the pairs $(V\co,V)$ and $(Y\co,Y)$, \emph{not} on the particular compatibilities on them.
In the case $V_{J}=Y_{K}$, an operator  $A\in \mathrm{Op}(V_{J})$ is regular if and only if  $A\ta$ is.

Next, an operator  $A\in \mathrm{Op}(V_{J})$ is called \emph{totally regular} if  ${\sf j}(A) $ contains the diagonal of   $J\times J$, i.e.,  $A_{rr}$ exists for every  $  r\in J$ or $A$ maps every $V_{r}$ into itself continuously, in other words,  ${\sf s}(A) =J$. This class leads to the identification of *-algebras of operators in $\mathrm{Op}(V_{J})$
\cite[Sec. 3.3.3]{pip-book}.

\index{operator!on {\sc pip}  space!homomorphism}
Among operators on \ipip s, a special role is played by morphisms.
An operator     $A\in \mathrm{Op}(V_{J},Y_{K})$ is called a \emph{homomorphism} if
\bei
\vspace*{-2mm}\item [(i)] for every $r\in J$, there exists $u\in K$ such that both $A_{ur}$ and $A_{\ov u \ov r}$  exist;

\vspace*{-2mm}\item [(ii)] for every $u\in K$, there exists $r\in J$ such that both $A_{ur}$ and $A_{\ov u \ov r}$  exist.
\eni
\vspace*{-2mm}

We denote by Hom($V_{J},Y_{K}$) the set of all homomorphisms from $V_{J}$ into $Y_{K}$ and by $\mathrm{Hom}(V_{J})$ those from $V_{J}$ into itself.
The following properties are immediate.

\beprop \label{prop:homom} Let $ V_{J},Y_{K},\ldots$ be indexed  \pip s. Then:\vspace*{-1mm}
\begin{itemize}
\item [(i)]
$A\in \mathrm{Hom}(V_{J}, Y_{K})$ if and only if  $A\ta\in \mathrm{Hom}(Y_{K},V_{J})$.
\vspace*{-2mm}
\item [(ii)]   The product of any number of homomorphisms (between successive \pip s) is defined and is a homomorphism.
\vspace*{-2mm}
\item [(iii)]    If $A\in\mathrm{Hom}(V_{J}, Y_{K})$, then $f\com g$ implies  $Af\com Ag$.
\vspace*{-2mm}
\item [(iv)]    If  $A\in \mathrm{Hom}(V_{J}, Y_{K}) $, then ${\sf j}(A\ta A)$ contains the diagonal of $J\times J$ 
and  ${\sf j}(A A\ta)$ contains the diagonal of $K\times K$ .
 \end{itemize}
\enprop
Note that an arbitrary homomorphism $A\in \mathrm{Hom}(V_{J}) $ need not be totally regular,  but both $A\ta A$ and $A A\ta$ are.

The definition of homomorphisms just given is tailored  in such a way that one may consider the category of all  \ipip s, with the 
homomorphisms as morphisms (arrows), as we have done in \cite{alt-cat}. 
In the same  language, we may define particular classes of morphisms, such as monomorphisms, epimorphisms and isomorphisms.
In particular, unitary isomorphisms are the proper tool for defining representations of Lie groups and Lie algebras in \pip s.
Examples and further properties of morphisms may be found in  \cite[Sec.3.3]{pip-book} and in \cite{at-iwota}.

\index{operator!on {\sc pip}  space!orthogonal projection}
Finally,  an \emph{orthogonal projection} on a nondegenerate \ipip\ $V_J$, in particular,  a LBS or a LHS,  is a homomorphism 
 $P\in \mathrm{Hom}(V_J)$ such that $P^2 = P\ta = P$.

A \spip\ $W$ of a \pip\ $V$ is defined  in \cite[Sec.3.4.2]{pip-book}  as an \emph{orthocomplemented} subspace of $V$, that is,
 a vector subspace $W$ for which there exists a vector subspace $Z \subseteq V$ such that  $V = W \oplus Z$ and
\bei
\vspace*{-2mm}\item [(i)] 
 $\{f\}\co = \{f_{W}\}\co \cap \{f_{Z}\}\co \; $ for every $f \in V$, where$f= f_{W} + f_{Z}, \, f_{W}\in W, f_{Z}\in Z$; 
\vspace*{-2mm}\item [(ii)]
 if $f\in W, g\in Z$ and  $f\# g$, then $\ip{f}{g} = 0$. 
\eni
\vspace*{-2mm} Condition (i) means that the compatibility $\#$ can be recovered  from its restriction to $W$ and $Z$.
\noi
In the same Section 3.4.2 of \cite{pip-book}, it is shown that a vector subspace $W$ of a nondegenerate \pip\ is orthocomplemented if and only if it is
\emph {topologically regular}, which means that   it satisfies the following two conditions:
\bei
\vspace*{-2mm}\item [(i)]   
for every assaying subset $V_{r}\subseteq  V$, the intersections $W_{r} = W \cap V_{r}$  and $W_{\ov{r}} = W\cap V_{\ov{r}} $ are a dual pair in $V$;

\vspace*{-2mm}\item [(ii)] 
the intrinsic Mackey topology $\tau (W_{r}, W_{\ov{r}})$ coincides with the Mackey topology  
 $\tau(V_{r}, V_{\ov{r}}) |_{W_{r}} $ induced by $V_{r}$.
\eni

Then the fundamental result, which is the analogue to the similar statement for a \hs, says that
a vector subspace $W$ of the nondegenerate \pip\ $V$ is orthocomplemented if and only if it is the range of an orthogonal projection :
$$
W = PV \mbox{ and } V = W \oplus W^\bot = PV \oplus (1-P)V.
$$ 
Clearly, this raises the question of identifying the subobjects of any category consisting of \pip s. 
  {For instance, in a category consisting of LHSs/LBSs   only, a subspace is a LHS/LBS if and only if it is topologically regular, thus 
orthocomplemented. In that case, the subobjects are precisely the orthocomplemented subspaces. But, for more general  \ipip s, this need not be true.
Orthocomplemented subspaces are subobjects, but there might be other ones. A simple example is that of a noncomplete prehilbert space (i.e.,
$V=V\ta$: then every subspace is a subobject, but need not be orthocomplemented. Further details may be found in \cite{at-iwota}.}

\end{document}